\documentclass[12pt]{article}
\usepackage{titlesec}
\usepackage[T1]{fontenc}
\usepackage[utf8]{inputenc}
\usepackage{authblk}

    \usepackage{amsfonts}
    \usepackage{amssymb}
    \usepackage{amsbsy}
    \usepackage{amsmath}
    \usepackage{amsthm}
    \usepackage{bbm}
    \usepackage{bm}
    \usepackage{mathtools}

    \newcounter{subremark}[remark]
	
	\makeatletter
	\renewcommand{\p@subremark}{\theremark}
	\makeatother

    \newtheorem{assum}{Assumption}
	\newcounter{subassumption}[assum]
	
	\makeatletter
	\renewcommand{\p@subassumption}{\theassum}
	\makeatother

    \newcounter{subdefinition}[definition]
	
	\makeatletter
	\renewcommand{\p@subdefinition}{\thedefinition}
	\makeatother

	\newtheorem*{proof*}{Proof}
	\newtheorem*{obs*}{Observation}

    \usepackage[flushleft,online,para]{threeparttable}
	\usepackage{graphics}
\usepackage[vlined,linesnumbered,ruled,resetcount]{algorithm2e}
\SetKwInOut{Initialization}{Initialization}

    \usepackage{subfig}
    \graphicspath{{Graphs/}}
    \usepackage{multirow}

    \usepackage{tikz}
    \usetikzlibrary{arrows.meta}
    \usetikzlibrary{arrows}
    \usetikzlibrary{decorations.pathreplacing}
    \usepackage{pgfplots}
\usepackage{indentfirst}             
\usepackage[capposition=top]{floatrow}
    \usepackage{tabularx}
        \newcolumntype{Z}{>{\centering\arraybackslash}X}
        \newcolumntype{L}{>{\raggedright\arraybackslash}X}
    \usepackage{dcolumn}
        \newcolumntype{d}[1]{D{.}{.}{#1}}
    \usepackage{rotating}                     
    \usepackage{lscape}
    \usepackage{pdflscape}   
\usepackage{authblk}

	\usepackage{csquotes}
	\usepackage[round]{natbib}
	\usepackage{url}
	\let\OLDthebibliography\thebibliography
\renewcommand\thebibliography[1]{
  \OLDthebibliography{#1}
  \setlength{\parskip}{0pt}
  \setlength{\itemsep}{0pt plus 0.3ex}
}

    \usepackage{xcolor}
        \definecolor{darkblue}{rgb}{0,0,0.4}
    \usepackage{hyperref}
        \hypersetup{
            colorlinks = true,
            linkcolor = darkblue,
            citecolor = darkblue,
            pdfborder = 0 0 0,
            pdfdisplaydoctitle = true,
            pdfhighlight = /N,
            pdfpagelayout = OneColumn,
            pdfpagemode = UseNone,
            pdfstartview = {FitH},
            pdfauthor = {{AB, CF \& JR}},
            pdftitle = {{par_exp}},
            pdfsubject = {{}}
        }

    \usepackage[textsize=footnotesize, colorinlistoftodos, textwidth=4cm, obeyDraft]{todonotes}
    \usepackage{geometry}
    \usepackage{setspace}
    \usepackage[bottom, multiple]{footmisc}  
    \usepackage{verbatim}
    \usepackage[normalem]{ulem}     
    \usepackage{mathpazo}
     \titlespacing*{\section}
{0pt}{0.1\baselineskip}{0.1\baselineskip}
    \titlespacing*{\subsection}
{0pt}{0.1\baselineskip}{0.1\baselineskip}
    \titlespacing*{\subsubsection}
{0pt}{0.1\baselineskip}{0.1\baselineskip}
    \titlespacing*{\paragraph}
{0pt}{0.1\baselineskip}{0.1\baselineskip}
    \titlespacing*{\assum}
{0pt}{0.1\baselineskip}{0.1\baselineskip}
\titlespacing*{\table}
{0pt}{0.1\baselineskip}{0.1\baselineskip}
\titlespacing*{\align}
{0pt}{0.1\baselineskip}{0.1\baselineskip}

    \usepackage[toc,page]{appendix}
	\usepackage{lipsum}

    \newcommand{\ubar}{\overline}

\begin{document}

\begin{spacing}{0}

\title{Parallel Experimentation and Competitive Interference on Online Advertising Platforms\thanks{This article was previously circulated under the titles ``Parallel Experimentation in a Competitive Advertising Marketplace'' and ``Parellel Experimentation on Advertising Platforms.'' Waisman, Nair, and Lin were part of \texttt{JD.com} when this research was initiated. The views represent that of the authors and not \texttt{JD.com}. Some aspects of the data, institutional context, and implementation are masked to address business confidentiality. We thank Pankhuri Saxena and Lijing Wang for research assistance, and Jun Hao, Jack Lin, Lei Wu, and Paul Yan for their support, collegiality and collaboration during the project. Thanks to Carlos Carrion, Dean Eckles, G\"{u}nter Hitsch, Carl Mela, Duncan Simester, Stefan Wager; seminar participants at Amazon, Carlson-UMinn, Facebook, Fuqua-Duke, Haas-Berkeley, Kellogg-Northwestern, Kenan-Flagan-UNC, Tepper-CMU, and at the 2020 Joint Statistical Meetings, Marketing Science, Virtual Quant Marketing Seminar, and the 2019 Choice Symposium conferences for thoughtful comments and suggestions. Please contact the authors at \texttt{caio.waisman@kellogg.northwestern.edu} (Waisman), \texttt{navdeep.sahni@stanford.edu} (Sahni), \texttt{harikesh.nair@stanford.edu} (Nair) or \texttt{xilianglin@gmail.com} (Lin) for correspondence.} }

\end{spacing}

\author{Caio Waisman \quad Navdeep S. Sahni  \quad Harikesh S. Nair \quad  Xiliang Lin}

\date{This draft: \today}

\maketitle

\begin{abstract}
\begin{singlespace}
\noindent This paper studies the measurement of advertising effects on online platforms when parallel experimentation occurs, that is, when multiple advertisers experiment concurrently. It provides a framework that makes precise how parallel experimentation affects the experiment's value: while ignoring parallel experimentation yields an estimate of the average effect of advertising in-place, which has limited value in decision-making in an environment with variable advertising competition, accounting for parallel experimentation captures the actual uncertainty advertisers face due to competitive actions. It then implements an experimental design that enables the estimation of these effects on \texttt{JD.com}, a large e-commerce platform that is also a publisher of digital ads. Using traditional and kernel-based estimators, it shows that not accounting for competitive actions can result in the advertiser inaccurately estimating the advertising lift by a factor of two or higher, which can be consequential for decision-making. \hfill\break

\end{singlespace}

\begin{singlespace}
\end{singlespace}

\end{abstract}

\pagebreak{}

\begin{section}{Introduction}

\noindent Experimentation is a fixture of the digital age. Companies routinely run numerous randomized experiments daily on their digital platforms to measure the effect of various interventions, such as advertising, and to formulate business strategy. Several digital platforms now offer ``experimentation as a service'' to their participating clients, whereby the platform experiments on behalf of firms, acting as their agents to facilitate measurement.\footnote{Examples include \texttt{Adobe}'s \textit{Target A/B Tests}, and \texttt{Google} and \texttt{Facebook}'s \textit{Optimize} for webpages and \textit{Conversion Lift} and \textit{Brand Lift} for advertising.}

The proliferation of experimentation has created pervasive situations in which competing firms experiment concurrently. Our objective is to assess the real-world impact of competitive advertising interference. We aim to underscore the decision-making challenges posed by competition-driven parallel experimentation and to enable the development of practical solutions to overcome these challenges.

While the literature on experimentation for measuring the effects of digital advertising has grown (e.g., \citealp{GordonetalInefficienciesdigitalAds21}), the impact of parallel experimentation on platforms by competing firms has received limited attention. The typical approach randomizes users independently across experiments and estimates effects from each experiment separately, ignoring multiple experimentation (\citealp{Kohavi2009}). Implicitly, this approach assumes that cross-experimental interaction effects are inconsequential in practice.\footnote{For example, speaking on online experiments, \cite{Kohavi2009} state on page 158: ``Parallel experiments. Our experience is that strong interactions are rare in practice \citep{vanbelle2002}, and we believe this concern is overrated.'' This appears to reflect a broader informal understanding that only lower order interactions matter. While recognizing this as an assumption, on page 169, \cite{vanbelle2002} states: ``High-order interactions occur rarely, therefore it is not necessary to design experiments that incorporate tests for higher-order interactions.''}

However, ignoring competitive advertising interactions this way is at odds with consumer behavior research and with how competitive marketplaces operate. First, when firms are competing for the same advertising slot, for example, on platform front-pages, a competitor targeting the same audience will affect who an advertiser is able to display their ad to. Second, the effect of a firm's advertising fundamentally depends on its competitors' actions, which may increase or decrease the advertising's effect.\footnote{The possibility of such competitive interactions has long been recognized in the marketing literature. Examples of such work span over five decades and include \cite{clarke1973sales}, \cite{keller1987memory}, \cite{burke1988competitive}, \cite{dm2005}, and, more recently, \cite{andersonsimester_2013}, \cite{sahni2016}, \cite{shapiro2018}, \cite{snr_2018}, and \cite{sh_2020} to quote a non-exhaustive list.} Consequently, findings from a firm's experiment measuring its ad effectiveness depend on which competitors advertised and experimented during this firm's experiment. 

Given the limited consideration of competitor behavior in ad-experiments, we address the following fundamental questions:
\begin{itemize}

    \item What is the nature of competitive interactions that could occur on platforms where competing firms experiment in parallel so that many users are simultaneously in multiple experiments? What is the quantitative significance of such interactions in modern, real-world marketplaces? What are the consequences of ignoring such parallel experimentation for experiment-based decision-making and causal inference on the effect of interventions? 
    
    \item What internally consistent causal estimands can be defined and estimated in environments where competitive interactions are salient? 
    
\end{itemize}

\paragraph{Contributions} \quad 
The main contribution of this paper is to show that parallel experimentation can indeed create an economically significant challenge that deserves attention from advertisers and advertising platforms that provide experimentation services to advertisers. Another firm experimenting in parallel can affect the experimental readout significantly, which can undermine the experiment's ability to improve decision-making. 

We believe that competitive interactions, while conceivable, are often ignored in practice because of a lack of rigorous empirical evidence on their significance, which is likely due to the difficulty of merging cross-competitor data. This paper fills this gap by bringing to bear data on large scale parallel experimentation occurring naturally at \texttt{JD.com}, an e-commerce platform that also is a large publisher of ads in China. Our setting and data uniquely enable us to analyze experimentally-varied competitive advertising, which is rarely observed outside a lab but necessary to assess the significance of competitive interference.

To analyze this problem, we first outline a formal framework that characterizes the dependence of a focal firm's experiment-based advertising decision on its competitors' advertising and experimentation. This framework shows the sources of competitive interference effects, which we refer to as \textit{ad allocation change} and \textit{cross-campaign externalities}. Ad allocation change refers to the impact of competitor actions on the likelihood of the focal firm's ad being shown. Cross-campaign externalities are changes in the focal firm's payoff from advertising caused by the presence of a competitor's ads.\footnote{The more competitive a platform is for advertising, the more likely these effects may manifest. For instance, in a more competitive platform a competing experimenting campaign may be more likely to affect an  experimenting campaign's ability to show its ad, driving ad allocation change. In a more competitive platform, it may also be more likely that a competing experimenting campaign's ad is shown to users in an experimenting campaign's control group, driving cross-campaign externalities.}

To enable a path towards a rigorous solution, we outline a full factorial experimental design and specify estimands that account for competitive interference, and discuss the transportability of experimental results for post experiment decision-making. Such experimental design can be implemented by the advertising platform, in line with the prevalent ``experimentation as a service'' phenomenon, where the platform runs the experiment and reports to the advertiser the effects estimated from the experimental data. We further present two estimators that can recover these effects. The first is a difference in means estimator, which is conceptually simple, but not scalable with the number of factor combinations.\footnote{In our empirical application, we consider an experiment with 16 firms, which yields $2^{16}=65,536$ such combinations, implying that for each firm we would need to assess how its campaign's effect varies over $2^{15}=32,768$ possible scenarios depending on each of its competitors' presence.} To overcome this problem, we build upon recent advances in nonparametric econometrics and specify a second estimator, which pools observations from different factor combinations and borrows information across them instead of treating each combination in isolation. We show that this kernel-based estimator, proposed by \cite{lor2013}, is effective even in high dimensions and has interpretable bandwidths. To our knowledge, this is the first application of this estimator to handle data obtained from a factorial design, which might be of separate interest for researchers given the popularity of such experimental designs.

Finally, we incorporate our solution into an advertising experimentation platform we helped build for \texttt{JD.com}. We show how the experimental design can be engineered into an auction-driven marketplace, which may be useful for researchers and firms interested in practical solutions.

\paragraph{Empirical application and results} \quad Our field experiment involves 16 experimental ad campaigns that ran parallel experiments for a three-day period in September 2018 on \texttt{JD.com}. The experiments include approximately 22 million users.

Our analysis of these data shows that advertising significantly affects our outcome measure---visits to the advertised product's pages---on average. However, competitive interactions play a crucial role across the experimental campaigns as the presence or absence of a competitor's advertising can have a substantial effect on the effectiveness of advertising, with the potential to alter its impact by up to 67\%. This quantitative evidence highlights the significant influence competitor behavior can have on the impact of a campaign. Unpacking the sources of such interactions, we find evidence for the presence of both ad allocation change and cross-campaign externalities, which indicates that both channels of impact are relevant on this marketplace. Implementing the kernel-based nonparametric estimator for a focal campaign, we find that it can recover treatment effect parameters precisely (35\% of the $2^{15}=32,768$ competitive interaction effects are statistically significant at the 5\% level). The recovered parameters show significant dispersion (the standard deviation is approximately 128\% of the mean parameter), indicating that the effect of advertising can be substantively impacted by competitors' presence or absence.

The analysis of dispersion in the distribution of treatment effects conditional on competitors' advertising ($CATE$s) is relevant to assess the importance of what we refer to as “competitive environmental uncertainty”: it is a reflection of a firm’s uncertainty over which of its competitors is advertising. Because $CATE$s are estimated to aid firms' decision-making, high dispersion across them indicates that accounting for this environmental uncertainty is important, even though typical practice ignores it. 

Firms often use the advertising campaign ``lift'' metric, the incremental outcome relative to the baseline of not having the campaign, as a reference for decision-making. The dispersion in $CATE$s we detect translates into dispersion in lifts. In our analysis, we discover that if a firm fails to consider competitive actions, the actual lift it experiences could be inaccurately estimated by a factor of two or more compared to the lift it might expect without taking into account competitive interactions.  

In addition, we simulate a scenario where a firm chooses to advertise only if the expected effect outweighs the cost to assess how considering competitive actions affects a firm's decisions. Our data indicate that by neglecting competitive factors, a firm might wrongly decide against advertising when it would be beneficial to do so. Depending on the advertising cost, this oversight could lead to a loss of more than 30\% of the firm's potential incremental profits.

The rest of the paper proceeds as follows. First, we discuss its relationship to the extant literature. Section \ref{sec:example} outlines our framework using a simple example. Sections \ref{sec:experiment} and \ref{sec:estimation} describe our experiment implementation on \texttt{JD.com} and estimation, respectively. Section \ref{sec:data} describes our data and Section \ref{sec:results} shows the results. Section \ref{sec:conclusion} concludes.

\subsection{Related literature}

\noindent This paper relates to several sub-literatures on experimentation, causal inference, and digital advertising. It also relates to the broad literature that studies advertising competition. We now discuss these relationships in more detail. 

Our setting of interest is one in which experiments are conducted in environments with competitive interactions. Hence, our work relates to studies that consider the impact of network externalities in learning from experiments and to ideas regarding the measurement of global effects of interventions from local experiments, such as \cite{heckman2000substitution}, \cite{acemoglu2010}, \cite{mn2017}, \cite{msw_2021} and \cite{sw_2021}. 

This paper addresses a conceptually different problem from this literature. Broadly speaking, this literature addresses the vexing problem of extrapolating from effects learned from small-scale \textit{local} experiments to \textit{global} (market or platform) level effects, which could be different because local experiments reflect the best equilibrium responses to the intervention. To address this problem, this literature, which we refer to as the \textit{local-global-gap} literature,  presents innovative solutions to extrapolate from local effects to global interventions. 

Unlike this literature, we focus on understanding how the presence of \textit{simultaneous} local experiments affects what can be learned from experimentation. Further, the current study considers how what can be learned in one local experiment depends on the experimentation policy of competitors. We show that this dependence affects the transportability of the effects to the post-experimental period when experimentation ends and policies have to be implemented. By considering both these problems that are not addressed in the current local-global-gap literature, our paper complements this literature and outlines a new interaction arising from competition that is relevant for platform experimentation.

Our paper is also related to the recent literature on experimentation with interference, such as on social networks and due to spillovers (e.g., \citealp{hh2008,aei2018,ijm2021}). Broadly speaking, the thrust of this literature is to accommodate situations where there are interactions between \textit{experimental units} and so the stable unit treatment value assumption (SUTVA) is violated. In the settings considered in this paper, interference occurs not due to interactions between units, but due to the fact that they are simultaneously in multiple experiments.

This paper is also related to the literature on measuring digital advertising effects via experiments and to the empirical literature on measuring advertising effects in competition. Apart from the papers cited in the introduction, examples at this intersection include \cite{gim2009}, \cite{lt2013}, and \cite{ln_2015}. Unlike our paper, this literature does not focus on the challenges induced by parallel experimentation that we focus on. We contribute to this literature by showing the ``ad allocation effect'' of competitive advertising, which we are able to do using uniquely detailed data.\footnote{\cite{sahni2019experimental} show empirical patterns consistent with an ad allocation effect in the context of retargeted advertising, but are unable to observe this effect due to lack of competitive advertising data.} Broadly, we also contribute to the copious literature on digital ad-experimentation. We keep the review of this literature short for brevity and point the reader to \cite{GordonetalInefficienciesdigitalAds21} and the references therein for a recent comprehensive overview.

From an experimental design perspective, we rely on advances that have been made in developing infrastructure for implementing overlapping experiments at scale (e.g., \citealp{taom2010}). Our specific design is a full factorial design, which aims to enable the estimation of interaction effects between multiple factors.\footnote{For a textbook treatment of full factorial designs, see, for instance, \cite{montgomery2017design}.} Full factorial designs have long been used in marketing, dating back at least to \cite{barclay1969factorial}. We leverage such infrastructure and design to recover new estimands with economic content in real marketplaces. 

In a recent contribution on this front, \cite{yzzzz2023} propose a method to estimate the effects of multiple treatments and their combinations from data obtained through partial factorial designs that do not implement all factor combinations. \cite{yzzzz2023} accomplish this task by imposing a specific structure on treatment interactions and proposes a deep learning estimator. In turn, since advertising theories and our data do not comply with such structure, our objective is to estimate all interaction effects without imposing restrictions on them.

To the extent that we leverage counterfactual policy logging to improve the precision of our estimates, our work is also related to the recent literature that has suggested such strategies for improving statistical efficiency (e.g., \citealp{johnson2017ghost, smt2020}). One contribution of this paper is to step toward combining all these aspects in a coherent practical system that can be implemented at scale on complex auction-driven ad platforms.

Beyond the methodological literature on experimentation, this paper is related to research studying advertising competition, with various objectives ranging from understanding overall advertising patterns under competition (e.g., \citealp{dube2005empirical}) to estimating advertising in different counterfactual scenarios (e.g. \citealp{vilcassim1999investigating}). Several theories have been proposed on how exposure to competitors' ads can impact the effects of a company's ads (e.g., \citealp{keller1991memory,unnava1994reducing,danaher2008effect}). The ad allocation effect has received less attention in the past, possibly due to the limited availability of competitor behavior data and the ambiguity around the identity of the counterfactual ad, especially for traditional media. However, research on search advertising has considered position effects that bear similarity, since a firm's decision not to advertise results in higher placement for others (e.g., \citealp{yang2014modeling,snr_2018}). We build on this literature by focusing on consequences of such competition for parallel experimentation, and the subsequent inference and decision-making. 

\end{section}

\begin{section}{Advertising environment}\label{sec:example}

We introduce our advertising environment through a simple example with two advertisers, one target audience, and one impression opportunity per user. We first establish some simple notation to define our estimands of interest and how they fit in an online advertising marketplace where firms experiment. Second, we introduce a simple advertising decision problem that highlights the double-edged sword nature of parallel experimentation: exploiting it allows for the estimation of a set of policy relevant treatment effect parameters, but ignoring it only allows for the estimation of the $ATE$ that held during the experiment. The decision problem we consider shows that this $ATE$ can be of limited use after the experiment ends. Finally, we describe how this example can be generalized to a complex setting that matches real-world online advertising marketplaces.

\subsection{Objects of interest}\label{sec:objects}

\noindent Consider an advertising marketplace where two firms, $f$ and $g$, compete to show their ads to users and have only one opportunity to expose each user to their ad. Firm $f$ seeks to estimate the expected effect of its advertising with the intention of using this estimate to better inform its decisions, such as whether to advertise in the first place. We now establish some notation to define this effect and how it can vary as a function of $g$'s advertising.

Denote firm $f$'s potential outcomes from user $i$ as $Y_{if}(f)$ if its ad is displayed, $Y_{if}(g)$ if $g$'s ad is displayed, and $Y_{if}(0)$ if no ad is displayed. Let $A_f$ and $A_g$ be indicators for whether $f$ and $g$ are advertising, respectively, and $W_i$ be the identity of the ad to which user $i$ is exposed. Define $p_w(A_f,A_g)\equiv \Pr \left ( W_i=w \middle \vert A_f, A_g \right )$, where $w\in\{f,g,0\}$.\footnote{Note that users may not be exposed to the ad even when they are targeted by the advertiser, that is, $\Pr(W_i=f|A_f=1, A_g=0) \leq 1$. For instance, this can happen because the platform allocates ads through an auction and the advertiser's bid may fall short of the reserve price.} 
Let $\mathbb{W} \left (A_f, A_g \right )$ be the set of ads users can be exposed to as a function of the firms' advertising policies.\footnote{For example, if $A_f=A_g=1$, then $\mathbb{W} \left (1, 1 \right )=\left \{f,g,0 \right \}$, while if $A_f=1$ and $A_g=0$, then $\mathbb{W} \left (1, 0 \right )=\left \{f,0 \right \}$. Notice that 0 is always in $\mathbb{W}\left(A_f,A_g\right)$.} The observed outcome from user $i$ to firm $f$ is $Y_{if}=\sum_{w \in \mathbb{W}(A_f,A_g)}\mathbbm{1}\{W_i=w\}\times Y_{if}(W_i)$. Firm $f$'s \textit{conditional} $ATE$ on $g$'s advertising policy, $CATE$, denoted by $\tau_f(A_g)$, is:
\begin{align}\label{eq:simp_ate}
\tau_f(A_g)&\equiv\mathbb{E} \left [ Y_{if} \middle \vert 1, A_g \right ] - \mathbb{E} \left [ Y_{if} \middle \vert 0, A_g \right ] \nonumber \\
	&= \sum_{w\in\{f,g,0\}} p_w \left (1,A_g \right )\times\mathbb{E} \left [Y_{if}(w) \right ]  - \sum_{w\in\{g,0\}} p_w \left ( 0,A_g \right )\times\mathbb{E} \left [Y_{if}(w)\right ].
\end{align}

Equation (\ref{eq:simp_ate}) highlights two ways in which interference can arise, each through one of the components within the sums in the right-hand side . First, it can arise if $g$'s presence affects the probability with which $f$'s ad is shown, that is, if $p_f(1,1)\neq p_f(1,0)$. We refer to this effect as \textit{ad allocation change}. Second, interference can arise if the outcome for $f$ depends on $g$'s presence when $f$ does not show its ad, that is, if $Y_{if}(g)\neq Y_{if}(0)$. We refer to this effect as \textit{cross-campaign externalities}. Finally, notice that the $CATE$ we defined conditions on $f$'s competitor's actions. This stands in contrast to the way the literature usually defines $CATE$s, which condition on unit-level attributes.

\subsection{Experimentation}\label{sec:exper}

\noindent We now incorporate experimentation into this environment. Let $E_g$ be an indicator for whether $g$ is experimenting and $\sigma_g$ be the probability with which user $i$ is randomly allocated to $g$'s treatment group, in which case $i$ is eligible to see $g$'s ad. Hence, $f$'s $ATE$, conditional on the user being in $g$'s treatment group, is $\tau_f(1)$, while the corresponding $CATE$ from users in $g$'s control group is $\tau_f(0)$. Firm $f$'s  $ATE$ during the experiment is thus a convex combination between $\tau_f(1)$ and $\tau_f(0)$, with weights determined by $\sigma_g$. Formally, we can define this $ATE$, which we denote by $\xi_f(\sigma_g)$, as
\begin{align}\label{eq:simp_ate_int}
\xi_f(\sigma_g)&\equiv \mathbb{E} \left [Y_{if} \middle \vert 1,\sigma_g \right ] - \mathbb{E} \left [Y_{if} \middle \vert 0,\sigma_g \right ] 
	=\sigma_g\times\tau_f(1) + (1-\sigma_g)\times\tau_f(0).
\end{align}

Notice that $\xi_f(\cdot)$ can be seen as a generalized version of $\tau_f(\cdot)$. When $\sigma_g=1$, $g$ allocates all users to its treatment group, which corresponds to a situation in which $A_g=1$ and $E_g=0$, that is, where $g$ is advertising but not experimenting, so that $\xi_f(1)=\tau_f(1)$. Analogously, if $\sigma_g=0$, then firm $g$ allocates all users to its control group, which corresponds to a situation where $g$ is not advertising, so that $A_g=E_g=0$ and $\xi_f(0)=\tau_f(0)$. Finally, when $\sigma_g\in(0,1)$, $A_g=E_g=1$ and therefore $g$ is advertising \textit{and} experimenting, so that $f$'s $ATE$ during the experiment becomes a function of $\sigma_g$. This is summarized in Figure \ref{fig:ex_ATEs_f}.

\begin{figure}[H]
\begin{tikzpicture}[scale=0.8,every node/.style={scale=0.8}]
\draw[fill] (0,0) circle [radius=2.5pt];
\node[align=center, above] at (0,0)
{Firm $g$};
\draw [ultra thick][-] (0,0) -- (3,-2);
\node[align=center, right] at (2,-0.75)
{$A_g=0$};
\node[align=right, below] at (3,-2)
{ $\xi_f(0)=\tau_f(0)$};
\draw [ultra thick][-] (0,0) -- (-3,-2);
\node[align=center, left] at (-2,-0.75)
{ $A_g=1$};
\draw[fill] (-3,-2) circle [radius=2.5pt];
\draw[fill] (3,-2) circle [radius=2.5pt];
\draw [ultra thick][-] (-3,-2) -- (-6,-4);
\node[align=center, left] at (-5,-3)
{ $E_g=1$ };
\draw[fill] (-6,-4) circle [radius=2.5pt];
\node[align=left, below] at (-6,-4)
{ $\xi_f(\sigma_g)=\sigma_g\times\tau_f(1) +(1-\sigma_g)\times \tau_f(0)$ };
\draw [ultra thick][-] (-3,-2) -- (0,-4);
\node[align=center, right] at (-1,-3)
{ $E_g=0$ };
\draw[fill] (0,-4) circle [radius=2.5pt];
\node[align=right, below] at (0,-4)
{ $\xi_f(1)=\tau_f(1)$ };
\end{tikzpicture}
\caption{Illustration of $f$'s $ATE$ during the experiment as a function of $g$'s policies}
\label{fig:ex_ATEs_f}
\end{figure}

\subsection{Estimating advertising effects} \label{sec:meas_int}

\noindent Assume that the platform runs an experiment on $f$'s behalf to obtain data to estimate $f$'s advertising effectiveness and that it runs an experiment on $g$'s behalf in parallel. Let $D_{if}$ and $D_{ig}$ be indicators for whether $i$ is in $f$'s and $g$'s treatment groups, respectively. We assume that the platform allocates users to $f$'s and $g$'s treatment groups independently. This experiment follows a $2^2$ factorial design, which we depict in Figure \ref{fig:ex_treat_complete}.

\begin{figure}[htp]
\begin{tikzpicture}[scale=0.8,every node/.style={scale=0.8}]
\draw [fill=gray!100] (-4,0) to (0,0) to (0,-4) to (-4,-4);
\draw [fill=gray!66.6] (0,0) to (4,0) to (4,-4) to (0,-4);
\draw [fill=gray!33.3] (0,-4) to (4,-4) to (4,-8) to (0,-8);
\draw [fill=gray!0] (-4,-4) to (0,-4) to (0,-8) to (-4,-8);
\node[align=center, above] at (-2,0)
{$D_{ig}=1$};
\node[align=center, left] at (-4,-2)
{$D_{if}=1$};
\node[align=center] at (-2,-2)
{$\mathbb{E} \left [ Y_{if} \middle \vert 1,1 \right]$};
\node[align=center] at (2,-2)
{$\mathbb{E} \left [ Y_{if} \middle \vert 1,0  \right]$};
\node[align=center] at (-2,-6)
{$\mathbb{E} \left [ Y_{if} \middle \vert  0,1  \right]$};
\node[align=center] at (2,-6)
{$\mathbb{E} \left [ Y_{if} \middle \vert  0,0  \right]$};
\node[align=center, above] at (2,0)
{$D_{ig}=0$};
\node[align=center, left] at (-4,-6)
{$D_{if}=0$};
\draw [ultra thick][-] (-4,0) -- (4,0);
\draw [ultra thick][-] (-4,-4) -- (4,-4);
\draw [ultra thick][-] (-4,-8) -- (4,-8);
\draw [ultra thick][-] (-4,0) -- (-4,-8);
\draw [ultra thick][-] (0,0) -- (0,-8);
\draw [ultra thick][-] (4,0) -- (4,-8);
\end{tikzpicture}
\caption{Illustration of $2^2$ factorial design and identified objects}
\label{fig:ex_treat_complete}
\end{figure}
 
Identification of $f$'s $CATE$s based on data collected from such experiment is straightforward because randomization implies that $\mathbb{E} \left [Y_{if} \middle \vert D_{if}, D_{ig} \right ] = \mathbb{E} \left [Y_{if} \middle \vert A_{f}, A_{g} \right ]$. By focusing on the average outcome of users who belong to $f$'s and $g$'s treatment groups, the platform can estimate $\mathbb{E} \left [ Y_{if} \middle \vert  1,1  \right]$, as illustrated in the northwest quadrant of Figure \ref{fig:ex_treat_complete}. The same logic applies to $\mathbb{E} \left [ Y_{if} \middle \vert  1,0  \right]$, $\mathbb{E} \left [ Y_{if} \middle \vert  0,1  \right]$ and $\mathbb{E} \left [ Y_{if}\middle \vert  0,0  \right]$ using the appropriate subgroups of users. With these four estimates, it is straightforward to recover $\tau_f(1)$ and $\tau_f(0)$. 

This approach presupposes that $g$'s treatment assignment is taken into account, which is often not the case. If the platform ignores $g$'s experiment when focusing on $f$'s treatment group, it considers users who were eligible to see $g$'s ads with probability $\sigma_g$ and ineligible with probability $1-\sigma_g$. Hence, the average outcome from these users corresponds to a convex combination of $\mathbb{E} \left [ Y_{if} \middle \vert  1,1  \right]$ and $\mathbb{E} \left [ Y_{if} \middle \vert  1,0  \right]$, as illustrated in the upper rectangle of Figure \ref{fig:ex_treat_simple}. An analogous result is obtained for users in $f$'s control group. By following the typical approach of taking the difference in mean outcomes between users in $f$'s treatment and control groups to estimate an $ATE$, the object this estimator recovers is $\xi_f(\sigma_g)$, as given in equation (\ref{eq:simp_ate_int}), that is, the $ATE$ that held during the experiment.

\begin{figure}[htp]
\begin{tikzpicture}[scale=0.8,every node/.style={scale=0.8}]
\draw [fill=gray!100] (-4,0) to (4,0) to (4,-4) to (-4,-4);
\draw [fill=gray!0] (-4,-4) to (4,-4) to (4,-8) to (-4,-8);
\node[align=center, above] at (-2,0)
{$D_g=1$};
\node[align=center, left] at (-4,-2)
{$D_f=1$};
\node[align=center] at (0,-2)
{$\sigma_g \times \mathbb{E} \left [ Y_{if}\middle \vert  1,1  \right] + (1-\sigma_g) \times  \mathbb{E} \left [ Y_{if} \middle \vert  1,0  \right]$};
\node[align=center] at (0,-6)
{$\sigma_g\times \mathbb{E} \left [ Y_{if} \middle \vert  0,1  \right] + (1-\sigma_g) \times  \mathbb{E} \left [ Y_{if} \middle \vert  0,0  \right]$};
\node[align=center, above] at (2,0)
{$D_g=0$};
\node[align=center, left] at (-4,-6)
{$D_f=0$};
\draw [ultra thick][-] (-4,0) -- (4,0);
\draw [ultra thick][-] (-4,-4) -- (4,-4);
\draw [ultra thick][-] (-4,-8) -- (4,-8);
\draw [ultra thick][-] (-4,0) -- (-4,-8);
\draw [ultra thick][-] (4,0) -- (4,-8);
\draw [dotted, thick][-] (0,0) -- (0,-1.75);
\draw [dotted, thick][-] (0,-2.25) -- (0,-5.75);
\draw [dotted, thick][-] (0,-6.25) -- (0,-8);
\end{tikzpicture}
\caption{Illustration of typical experimental approach and identified objects}
\label{fig:ex_treat_simple}
\end{figure}

The differences between Figures \ref{fig:ex_treat_complete} and \ref{fig:ex_treat_simple} highlight why parallel experimentation can act as a double-edged sword. On the one hand, when taken into account parallel experimentation enables the measurement of a set of treatment effect parameters that show if and how much one's advertising effects are impacted by one's competitors' advertising policies---$\tau_f(1)$ and $\tau_g(0)$. These $CATE$s can then be used to construct policy relevant treatment effect parameters for the post-experimentation period.

On the other hand, when ignored parallel experimentation leads to the measurement only of the $ATE$ that held during the experiment, $\xi_f(\sigma_g)$. Even though this object does convey information about treatment effects, its usefulness can be limited for decision-making after experiments end. We discuss this next.

\subsection{Decision-making}\label{sec:ex_post}

\noindent We now consider how the objects recovered from the experiment can be used by $f$ to make decisions after the experiment  ends. While there are many possible ways $f$ can act on this information, we consider one we believe is simple, practical, and reasonable: integrating $f$'s $CATE$s using its beliefs to obtain a weighted $ATE$ to be used for decision-making.

Assume that the platform discloses to $f$ its $CATE$s, $\tau_f(1)$ and $\tau_f(0)$, but nothing more.\footnote{This presupposes that the platform accounts for parallel experimentation when measuring advertising effects and that it can commit to report the estimates it obtains truthfully.} We further assume that firm $f$ has beliefs, $\pi_f(\cdot)$, which correspond to $\Pr_f(A_g=1)$. We do not take a stand on how these beliefs are formed.\footnote{Beliefs could be formed based on historical patterns, for example, through competitive data (e.g., Comscore media metrix data in the US). As another possibility, the platform could compute, over a period of time, the frequency with which firm $f$'s competitors were advertising and report it to $f$. The beliefs would then match these empirical frequencies.} Given the estimates provided by the platform and $\pi_f(\cdot)$, firm $f$'s projected $ATE$ is given by 
\begin{align}\label{eq:simp_pol}
\xi_f \left [ \pi_f \left (\cdot \right ) \right ]= \sum_{A_g \in \{0,1\}} \pi_f (A_g)\times \tau_f (A_g),
\end{align} 
which can be computed for any $\pi_f(\cdot)$ if $\tau_f(1)$ and $\tau_f(0)$ are known. Assuming that firm $f$ maximizes its expected payoffs from advertising, it then follows that $A_f=1$ if and only if $\xi_f \left [ \pi_f \left (\cdot \right ) \right ] \geq \kappa$, where $\kappa$ is the cost of advertising, which we assume $f$ knows. 

Firm $f$ can only follow this course of action if it acquires knowledge of its $CATE$s. This does \textit{not} require $f$ to know $g$'s $CATE$s, which, in practice, the platform probably would not disclose to $f$ to preserve $g$'s privacy. The relevant information can only be acquired through an experiment using a full factorial design, which, in turn, can only be performed by the platform and therefore demonstrates the critical role it can play in facilitating decision-making by firms.

Consider now the consequences of ignoring parallel experimentation in the context of this decision problem. When parallel experimentation is ignored, the only object that can be recovered is the $ATE$ that held during the experiment, shown in equation (\ref{eq:simp_ate_int}). This object has a direct relationship with the projected $ATE$ from equation (\ref{eq:simp_pol}): they become equal when $\pi_f(1)=\sigma_g$. In other words, the $ATE$ that held during the experiment is the policy relevant parameter as long as the conditions that also held during the experiment do not change. However, because firms utilize the information they obtain from experiments to make decisions, it is likely that such conditions will change. It is unclear how, without additional assumptions, the $ATE$ that held during the experiment can be used for decision-making once the experiment ends.\footnote{This will trivially be the case when interference effects are ruled out, that is, when $\tau_f(1)=\tau_f(0)=\tau_f$, in which case the $ATE$ from equation (\ref{eq:simp_ate_int}) and the projected $ATE$ from equation (\ref{eq:simp_pol}) also equal $\tau_f$.}    

\subsection{Generalization}

In practice, online advertising environments are more complex: there are more firms competing to show their ads, users belong to different target audiences, and each user can be exposed to ads multiple times. It is straightforward to generalize the expressions above to capture these additional complexities.
This generalization consists of first expressing treatments at the impression level and then aggregating them to sequences of ads. In turn, the sets of possible sequences of ads are determined by which firms are advertising, leading to $CATE$s that are generalized versions of those from equation (\ref{eq:simp_ate}). When there are $F$ firms advertising, for each firm $f$ this characterizes a family of $2^{F-1}$ $CATE$s.

Identification of the $CATE$s follows the same logic as the one presented above: by focusing on specific treatment-control combinations, it is possible to recover firm $f$'s expected outcomes as a function of its competitors' advertising policies, akin to what is displayed in Figure \ref{fig:ex_treat_complete}. However, the higher complexity of this new environment creates additional difficulties in implementing the full factorial design necessary so that the $CATE$s can be recovered. We discuss these difficulties and how they can be overcome in the next section.

\end{section}

\begin{section}{Implementing the experiment}\label{sec:experiment}

We implemented the full factorial design  we presented in Section \ref{sec:exper} on \texttt{JD.com}'s \textit{Conversion Lift System} \citep{JD}. \textit{Conversion Lift System} is an on-demand product we designed that allows advertisers to run experiments to assess the performance of their campaigns on \texttt{JD.com}'s ad inventory. There are many particularities and details that need to be handled in translating the factorial design and the statistical framework we outlined to a practical functioning system. The ranking of ads is determined by the auction-driven allocation system used to serve ads. In addition, scalability requires reducing latency in serving ads while experimenting and improving statistical precision in the analysis of the data from the experiment. These particularities are common in large-scale ad auctions environments, so that these implementation details may be of more general interest in and of themselves. We describe each of them next.

\subsection{Front-focused mobile ads}\label{sec:front_focus}

\noindent Within the system, the specific experiments reported in this paper pertain to ``front-focused'' ad positions on \texttt{JD.com}'s app or mobile page. These ads are served at the top of the app or mobile home page. Panel (a) of Figure \ref{fig:jed_ex} shows a screenshot of the app. There are eight front-focused ad positions and each position can show a different ad in the upper rectangle. A user can only see one front-focused ad at a time. 

\begin{figure}[htb]
    \centering
    \begin{subfloat}[Front-focused ad (enclosed in green box)\label{fig:jed_ex1}]
        {\includegraphics[width=0.30\textwidth]{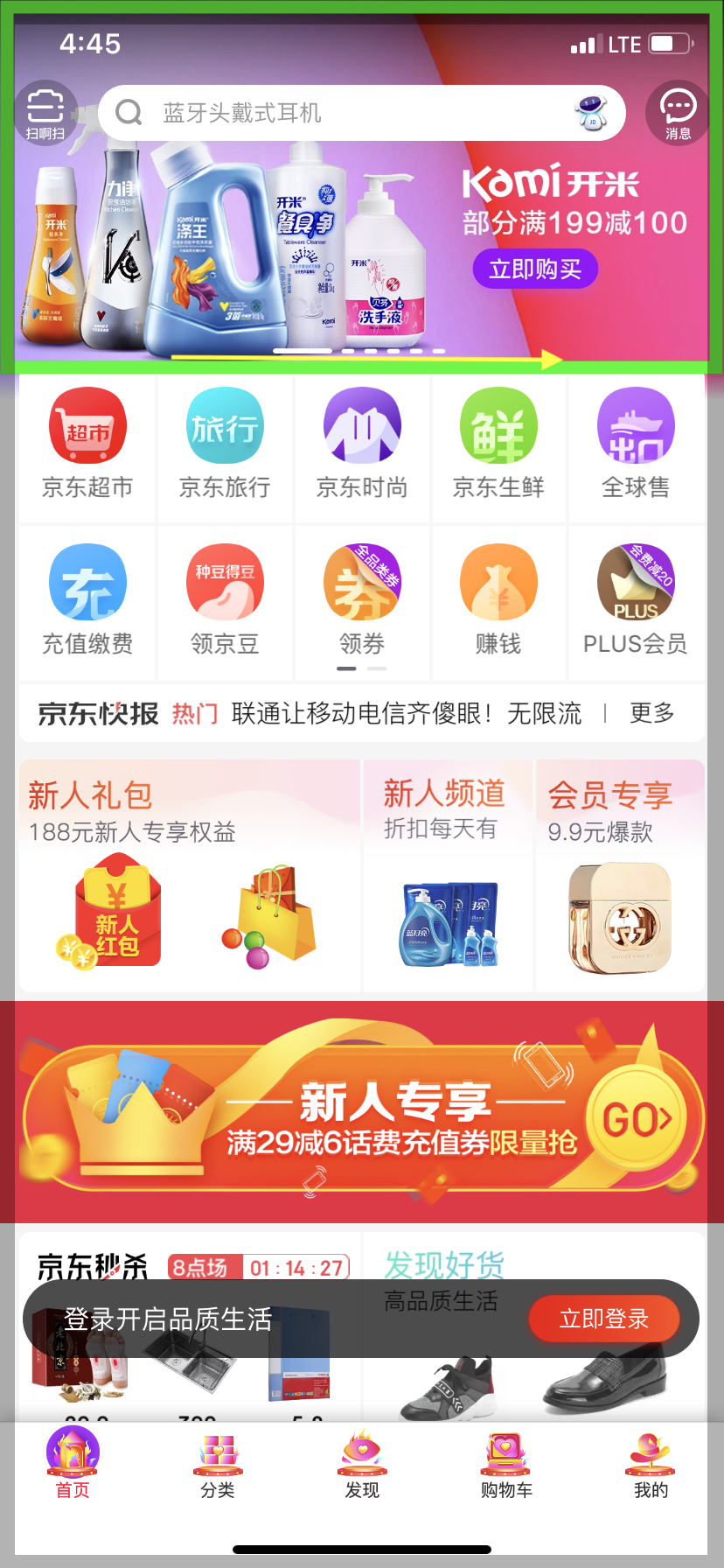}}
    \end{subfloat}
    \qquad \qquad
    \begin{subfloat}[Landing page (reached by clicking on ad)\label{fig:jed_Ex2}]
        {\includegraphics[width=0.30\textwidth]{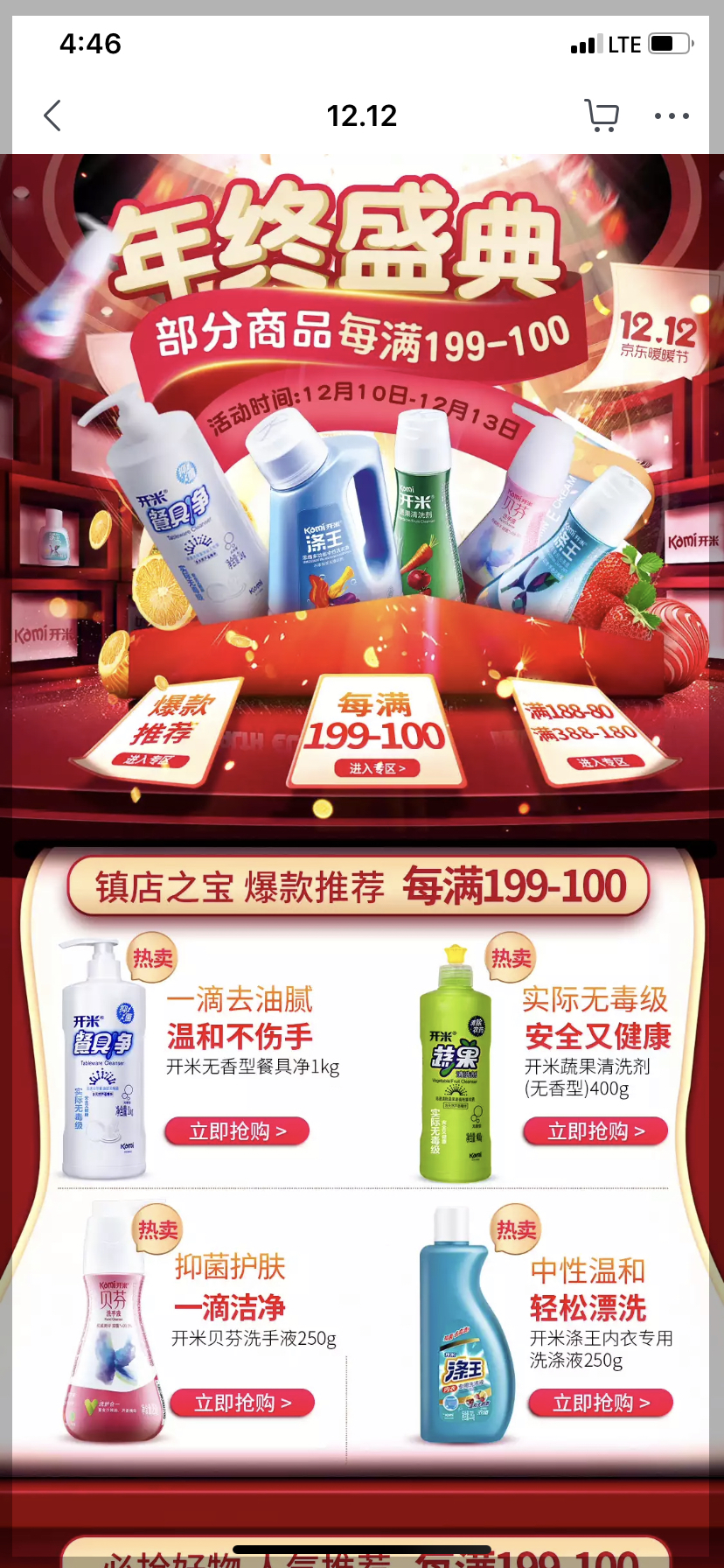}}
    \end{subfloat}
    \caption{Front-focused ad position on \texttt{JD.com}' app}
    \label{fig:jed_ex}%
\end{figure}

When a user arrives at \texttt{JD.com}'s app or mobile home page, they are served ad number one. After a few seconds, the banner automatically rotates to the next ad. The user can also manually rotate the ads. Once the user clicks on an ad, the app directs them to a ``landing page'' that shows more information including larger images of the featured products, promotional coupons, and a collection of relevant products. Panel (b) of Figure \ref{fig:jed_ex} shows an example. If a user clicks on a product on the landing page, they arrive at a product detail page featuring detailed information about the product including price, coupons, and reviews. They can then add the product to their shopping cart and eventually purchase it.

\subsection{Auction-driven ads marketplace}

\noindent Like most digital ad platforms, a large amount of ad inventory on \texttt{JD.com} is allocated via real-time bidding (RTB) auctions. The process is as follows. Advertisers first set up campaigns on \texttt{JD.com}. 
Each campaign specifies a target audience, a set of ad positions where users belonging to these target audiences can be exposed to ads, a creative or set of creatives (i.e., images) to be shown when an ad is served, and a set of rules specifying the bidding policy.

When a user arrives at an ad position on \texttt{JD.com}, the system retrieves in real time a queue containing the list of advertisers that are eligible to show their ads to this user. The list is then sorted on the basis of a proprietary ad quality score that comprises several variables, including bids, and which generates the rankings that determine which ad is served. The use of this quality score in addition to just the bids reflects the desire of the publisher to ensure an experience that reduces user annoyance and to generate long-term value to all players in the ecosystem. Programmatic ads on most digital platforms are sold this way (e.g., \citealp{nk2015}).

\subsection{Details behind ad serving mechanism}

\noindent The front-focused ad positions are considered premium ad inventory by advertisers because of their prominence and their large number of exposures. They are key to drive visits to product detail pages, which, in turn, are key for conversions. Front-focused ad auctions feature specific characteristics that are relevant for engineering the experiment.

When setting up campaigns for such inventory, advertisers must specify a base level bid that applies to all users and then select specific target audiences to which premium bids may apply. No user is explicitly excluded: all exclusions and inclusions are induced by the bids the advertiser specifies. Advertisers also cannot specify a particular ad position to bid for. They can only specify they wish to show an ad at any of the available ad positions.

Upon a user's arrival, advertisers targeting this user participate in an RTB auction to show their ad at an available ad position. 
Auction queues are generated independently for each ad position and they can differ due to technical specifications. First, queue rankings are based on the aforementioned quality scores. Second, a user experience control system filters the ranked queues to avoid repeating ads from the same advertiser across the various ad positions at a given impression opportunity. Finally, additional user experience controls, such as pacing, feed into the ad serving rules. Once the queue clears this system's filters, the top ranked ad at the corresponding ad position is served. 

Importantly, there is a distinction between the ad that is served and whether this ad is actually seen. Serving refers to the process by which ads are sent from the server to the user's app. To reduce latency, the ads for all eight positions are served to the user in one shot. A user may choose not to look or navigate away from the front-focused space before seeing the served ad at one of the ad positions. The ``compliance'' with the served ad is therefore a user's decision. 
Generally speaking, ad positions with lower index have higher chances of being seen by users because they are shown earlier. 

The complexity of this mechanism implies that the relevant counterfactual to showing a particular ad is  not trivial to obtain. If an advertiser chose not to show their ad to a user, the next ad in the ranking queue would be served instead subject to this user's experience control system filter. Since only the platform can retrieve this in real time, only the platform can utilize this experiment with a precisely constituted counterfactual. This is a motivation for developing a product that delivers ``experiments as a service.'' The challenge is to engineer the experiment in the context of this complex system. 

\subsection{Engineering the experiment}

\noindent Engineering this experiment requires working out strategies for randomization, for adapting the ad serving mechanism to deliver the right factual and counterfactual ads for users in the experiment, and for logging the data obtained through this experiment. We now discuss each of these components.

\subsubsection{Randomization}

\noindent At the beginning of the experiment, the engineered system first retrieves all advertisers and assigns a campaign index to each of them. It then uses the \texttt{hash MD5} quasi-randomization method \citep{rivest1992} to assign each user $i$ and advertiser $f$ to a common randomization seed, so that $D_{if}=\text{hash split(user id, campaign index, seed)}$, where $D_{if}$ is user $i$'s treatment assignment for advertiser $f$. This approach effectively stores a consistent randomization method instead of storing the full assignment vector, $D_i$, for each user $i$ and is helpful for reducing latency in the online system.\footnote{Otherwise, it would  have to store tens of  billions of treatment assignments and impose a heavy cost to the online ad serving system.} The hash method only takes user id, campaign index, and a seed as independent, and, therefore, users are independently assigned to treatment or control groups across advertisers.

\subsubsection{Serving ads}

\noindent At ad serving time, the queue for each ad impression opportunity is generated as described above and passed through the auction and user experience control systems. To induce the experiment into the systems, we first remove all $f$ such that $D_{if}=0$ from the auction queue, which ensures that user $i$ is eligible to see $f$'s ad only if $D_{if}=1$. The remaining ads in the ranked queue are passed through the user experience control system again and the top ad is served to the user. Thus, we allocate the ad position to the top eligible ad, which is economically efficient to the platform while ensuring that the user experience is not degraded during the experiment. 

To illustrate this system, Figure \ref{fig:ex_system} shows an example with one ad position and three advertisers. Advertisers 1 and 2 are experimenting but advertiser 3 is not. User $i$ is independently assigned to the treatment and control groups of 1 and 2, which determine $i$'s full treatment assignment. Figure \ref{fig:ex_system} illustrates which ad would be served under each possible full treatment assignment. For example, the first column shows an auction queue with Ad 1 at the top, followed by Ad 2 and then Ad 3. If the user was in advertiser 1's treatment group, they would be served Ad 1 regardless of whether they were in advertiser 2's treatment group or not. If the user was in advertiser 1's control group but in 2's treatment group, they would be served Ad 2. Finally, if the user was in 1's and 2's control groups, they would be served Ad 3.

\begin{figure}[htp]

\begin{tikzpicture}[scale=0.7,every node/.style={scale=0.7}]

\draw [very thick][-] (-16,0.5) -- (-14,0.5);
\draw [very thick][-] (-16,-0.5) -- (-14,-0.5);
\draw [very thick][-] (-16,0.5) -- (-16,-0.5);
\draw [very thick][-] (-14,0.5) -- (-14,-0.5);
\node[align=center] at (-15,0) { User $i$};

\draw [very thick][-Triangle] (-14,0) -- (-10,3);
\draw [very thick][-Triangle] (-14,0) -- (-10,1);
\draw [very thick][-Triangle] (-14,0) -- (-10,-1);
\draw [very thick][-Triangle] (-14,0) -- (-10,-3);

\node[above] at (-8,4) {Possible treatment assignments};
\node[above] at (1.6,4) {Ad served to user $i$ in the experiment};
\node[above] at (1.6,7.2) {Possible auction queues};

\draw [very thick][-] (-10,3.5) -- (-10,2.5);
\draw [very thick][-] (-6.3,3.5) -- (-6.3,2.5);
\draw [very thick][-] (-10,3.5) -- (-6.3,3.5);
\draw [very thick][-] (-10,2.5) -- (-6.3,2.5);
\node[align=center] at (-9,3) {$\qquad \qquad D_{i1}=1, D_{i2}=1$};

\draw [very thick][-] (-10,1.5) -- (-10,0.5);
\draw [very thick][-] (-6.3,1.5) -- (-6.3,0.5);
\draw [very thick][-] (-10,1.5) -- (-6.3,1.5);
\draw [very thick][-] (-10,0.5) -- (-6.3,0.5);
\node[align=center] at (-9,1) {$\qquad \qquad D_{i1}=1, D_{i2}=0$};

\draw [very thick][-] (-10,-0.5) -- (-10,-1.5);
\draw [very thick][-] (-6.3,-0.5) -- (-6.3,-1.5);
\draw [very thick][-] (-10,-1.5) -- (-6.3,-1.5);
\draw [very thick][-] (-10,-0.5) -- (-6.3,-0.5);
\node[align=center] at (-9,-1) {$\qquad \qquad D_{i1}=0, D_{i2}=1$};

\draw [very thick][-] (-10,-2.5) -- (-10,-3.5);
\draw [very thick][-] (-6.3,-2.5) -- (-6.3,-3.5);
\draw [very thick][-] (-10,-3.5) -- (-6.3,-3.5);
\draw [very thick][-] (-10,-2.5) -- (-6.3,-2.5);
\node[align=center] at (-9,-3) {$\qquad \qquad D_{i1}=0, D_{i2}=0$};

\draw [very thick][-] (-3.2,4) -- (-3.2,-4);
\draw [very thick][-] (-1.6,4) -- (-1.6,-4);
\draw [very thick][-] (0,4) -- (0,-4);
\draw [very thick][-] (1.6,4) -- (1.6,-4);
\draw [very thick][-] (3.2,4) -- (3.2,-4);
\draw [very thick][-] (4.8,4) -- (4.8,-4);
\draw [very thick][-] (6.4,4) -- (6.4,-4);

\draw [very thick][-Triangle] (-6.3,3) -- (-3.2,3);
\draw [very thick][-Triangle] (-6.3,1) -- (-3.2,1);
\draw [very thick][-Triangle] (-6.3,-1) -- (-3.2,-1);
\draw [very thick][-Triangle] (-6.3,-3) -- (-3.2,-3);

\draw [very thick][-] (-3.2,4) -- (-3.2,8);
\draw [very thick][-] (6.4,4) -- (6.4,8);
\draw [very thick][-] (-3.2,8) -- (6.4,8);
\draw [very thick][-] (-3.2,5.6) -- (6.4,5.6);
\draw [very thick][-] (-3.2,6.4) -- (6.4,6.4);
\draw [very thick][-] (-3.2,4.8) -- (6.4,4.8);
\draw [very thick][-] (-3.2,7.2) -- (6.4,7.2);

\draw [very thick][-] (-1.6,7.2) -- (-1.6,4.8);
\draw [very thick][-] (0,7.2) -- (0,4.8);
\draw [very thick][-] (1.6,7.2) -- (1.6,4.8);
\draw [very thick][-] (3.2,7.2) -- (3.2,4.8);
\draw [very thick][-] (4.8,7.2) -- (4.8,4.8);

\node[align=center] at (-2.4,8.6) {};

\node[align=center] at (-2.4,6.8) {Ad 1};
\node[align=center] at (-2.4,6) {Ad 2};
\node[align=center] at (-2.4,5.2) {Ad 3};

\node[align=center] at (-0.8,6.8) {Ad 1};
\node[align=center] at (-0.8,6) {Ad 3};
\node[align=center] at (-0.8,5.2) {Ad 2};

\node[align=center] at (0.8,6.8) {Ad 2};
\node[align=center] at (0.8,6) {Ad 1};
\node[align=center] at (0.8,5.2) {Ad 3};

\node[align=center] at (2.4,6.8) {Ad 2};
\node[align=center] at (2.4,6) {Ad 3};
\node[align=center] at (2.4,5.2) {Ad 1};

\node[align=center] at (4,6.8) {Ad 3};
\node[align=center] at (4,6) {Ad 1};
\node[align=center] at (4,5.2) {Ad 2};

\node[align=center] at (5.6,6.8) {Ad 3};
\node[align=center] at (5.6,6) {Ad 2};
\node[align=center] at (5.6,5.2) {Ad 1};

\draw [very thick][-] (-3.2,4) -- (6.4,4);
\draw [very thick][-] (-3.2,2) -- (6.4,2);
\draw [very thick][-] (-3.2,0) -- (6.4,0);
\draw [very thick][-] (-3.2,-2) -- (6.4,-2);
\draw [very thick][-] (-3.2,-4) -- (6.4,-4);

\node[align=center] at (-2.4,3) {Ad 1};
\node[align=center] at (-0.8,3) {Ad 1};
\node[align=center] at (0.8,3) {Ad 2};
\node[align=center] at (2.4,3) {Ad 2};
\node[align=center] at (4,3) {Ad 3};
\node[align=center] at (5.6,3) {Ad 3};

\node[align=center] at (-2.4,1) {Ad 1};
\node[align=center] at (-0.8,1) {Ad 1};
\node[align=center] at (0.8,1) {Ad 1};
\node[align=center] at (2.4,1) {Ad 3};
\node[align=center] at (4,1) {Ad 3};
\node[align=center] at (5.6,1) {Ad 3};

\node[align=center] at (-2.4,-1) {Ad 2};
\node[align=center] at (-0.8,-1) {Ad 3};
\node[align=center] at (0.8,-1) {Ad 2};
\node[align=center] at (2.4,-1) {Ad 2};
\node[align=center] at (4,-1) {Ad 3};
\node[align=center] at (5.6,-1) {Ad 3};

\node[align=center] at (-2.4,-3) {Ad 3};
\node[align=center] at (-0.8,-3) {Ad 3};
\node[align=center] at (0.8,-3) {Ad 3};
\node[align=center] at (2.4,-3) {Ad 3};
\node[align=center] at (4,-3) {Ad 3};
\node[align=center] at (5.6,-3) {Ad 3};

\end{tikzpicture}

\caption{Treatment assignments and served ads under different rankings}

\label{fig:ex_system}

\end{figure}

\paragraph{Multiple ad exposures} \quad An experiment can last for multiple days and pertains to multiple ad positions. During the course of the experiment, users may visit \texttt{JD.com}'s app multiple times. Hence, the engineered system needs to handle multiple auctions for the same user. It is critical that the assignment of a user to treatment or control groups is consistent during the entire experiment, which the hash split presented above guarantees. 

\subsubsection{Counterfactual policy logging and interpretation}

The effects of digital ads may be small and require data on hundreds of thousands of users to be estimated with precision. To improve precision, as part of the engineered system we log the auction queue before control ads are dropped, which represents the counterfactual ad queue if none of the firms were experimenting. In this counterfactual scenario, the top ad on the logged queue would be served to the user, so we refer to it as ``auction-specific counterfactual ad.'' For each auction, we also log which ad is actually served to the user. We call this ``auction-specific factual ad.'' Our data only contain users who factually or counterfactually are served the focal experimental ad in at least one auction. This ensures that we implement our analysis on a set of users in the treatment group who are most likely to be factually served the focal ad and on a set of users in the control group who are most likely to be counterfactually served the focal ad. By removing users with low propensity to be served the focal ad from the analysis, we improve precision.

The treatment effects we measure capture the effect of being assigned to a treatment group compared to the control group. Assignment to treatment implies a user is eligible to be served ads of the focal advertiser, while assignment to the control group implies they are ineligible. Hence, the effects we measure should be interpreted as a local average treatment effect ($LATE$), local to the subpopulation of users the publisher would serve the advertiser's ad on its platform. 

\end{section}

\begin{section}{Estimation} \label{sec:estimation}

We now briefly discuss estimation of the $CATE$s analogous to $\tau_f(A_g)$ as defined in equation (\ref{eq:simp_ate}), which are conditional average treatment effect parameters with respect to firm $f$'s competitors' advertising policies. We discuss standard estimation here and revisit this issue in Section \ref{sec:kernel}, where we present an alternative estimator better-suited to accurately estimate all the $CATE$s. Further details on other estimators can be found in the Appendix.

To define the estimators, we introduce some additional notation. Assume there are $F$ firms experimenting. We define the $F$-dimensional vector $D_i \equiv \left [D_{i1},\dots,D_{iF} \right ]'$ and the $(F-1)$-dimensional vector $D_{i,-f} \equiv \left [D_{i1},\dots, D_{i,f-1}, D_{i,f+1},\dots,D_{iF} \right ]'$,
which we refer to as \textit{full} and \textit{partial treatment assignment} vectors, respectively. Analogously, we define vectors $A$ and $A_{-f}$ that indicate which advertisers and which of $f$'s competitors are advertising, respectively. Notice that the experimental design we described above guarantees that for each vector $A$ there is a vector $D_i$ that is its exact analogous.

We assume we have data that come from the full factorial design described above. The observed data for firm $f$ are $\left \{Y_{if}, D_{if}, D_{i,-f} \right \}_{i=1}^n$, where $n$ is the number of observations. Finally, we assume that all partial treatment assignments $D_{i,-f}$ are observed in the data, so that all $CATE$s are estimable.

The simplest estimator for $\tau_{f} \left ( A_{-f} \right )$  is given by the usual difference in means. Let $D_{i,-f}=d$ be the partial treatment assignment that is equivalent to $A_{-f}$. This estimator is:
\begin{align}\label{eq:est_tauF}
    \hat{\tau}_{f} \left ( d \right ) &= \frac{\mathlarger{\sum}\limits_{i=1;D_{i,-f}=d}^n D_{if}Y_{if}}{\mathlarger{\sum}\limits_{i=1;D_{i,-f}=d}^n D_{if}} - \frac{\mathlarger{\sum}\limits_{i=1;D_{i,-f}=d}^n (1-D_{if})Y_{if}}{\mathlarger{\sum}\limits_{i=1;D_{i,-f}=d}^n (1-D_{if})}    
\end{align}

Under standard conditions, $\hat{\tau}_{f} \left ( d \right )$ is a consistent estimator for $\tau_{f} \left (A_{-f} \right )$ and is asymptotically normal at the usual square root rate. The estimator in (\ref{eq:est_tauF}) corresponds to the OLS estimator of the slope coefficient of a linear regression of $Y_{if}$ on $D_{if}$ using only observations such that $D_{i,-f}=d$. Hence, it can be computed through the saturated regression:
\begin{align}\label{eq:est_tau_f_regr}
	Y_{if}&=  \sum_{d \in \mathbb{D} } \mathbbm{1} \left \{ D_{i,-f} = d \right \} \left [  \alpha_{f,d} + \tau_{f,d} D_{if}\right ] +\epsilon_{if},
\end{align}
where $\mathbb{D}$ corresponds to the set of all partial treatment assignments and the $\epsilon_{if}$s are error terms.

The number of coefficients to be estimated in (\ref{eq:est_tau_f_regr}) grows exponentially with the number of partial treatment assignments. Hence, in practice OLS may not be the best estimator for this problem \citep{stein1956}. When we tackle the estimation of all $CATE$s from our experiment, which we describe below, we use a different estimator. We provide details immediately before this exercise in Section \ref{sec:full_analysis}.

\end{section}

\begin{section}{Data} \label{sec:data}

Our data originally correspond to parallel experiments on front-focused ads for 27 campaigns that ran for a three-day period in September 2018. In this experiment, 70\% of users are independently assigned to each campaign's treatment group and the remaining 30\% are assigned to each campaign's control group. Out of the 27 campaigns, we only keep the 16 campaigns that had at least 200,000 experimental users. We track user visits to the product detail page through searches, other ads, and organic recommendations. Advertisers value visits to the product detail page because it is an antecedent to actual conversion and an explicit role of front-focused ads is to drive such visits. They are also a key outcome metric reported to advertisers as part of campaign performance. This forms our dependent variable in the analysis reported below. 

We describe our sample focusing on the overlap across campaigns because this is the key feature that generates interference between experimenting firms. We display a series of randomization checks in the Appendix that show that the experiment was implemented successfully.

Our sample comprises approximately 22 million users who are exposed to at least one of the 16 experimental ad campaigns. We say an individual is ``exposed'' to a campaign if they would have been served the ad absent experimentation. About 27\% of users in our sample are exposed to more than one of the 16 campaigns. Table \ref{tab:overlap} shows the distribution of exposures across users.

\begin{table}[htp]
\begin{threeparttable}
\caption{Distribution of users across campaign exposures}
\label{tab:overlap}
\begin{tabular}{c|cc}
    \hline \hline 
\# of campaigns exposed to & \# of users & \% of sample   \\
\hline
  1 & 15,830,998 & 72.86  \\ 
  2 & 4,629,025  & 21.31  \\ 
  3 & 1,040,805 & 4.79  \\ 
  4 & 193,192 & 0.89  \\ 
  5 & 29,007 & 0.13  \\ 
  6 or more & 3,854 & 0.02  \\ 
 \hline \hline
    \end{tabular}
  \end{threeparttable}
\end{table} 

The percentage of a campaign's experimental users that are also exposed to other campaigns affects the potential for competitive interference between them. Table \ref{tab:overlap2} shows that 50\% of users in the median campaign are also exposed to at least one other campaign. It further shows that substantial overlap can be due to a single competing campaign.

\begin{table}[htp]
\resizebox{\textwidth}{!}{\begin{threeparttable}
\caption{Overlap of users across campaigns}
\label{tab:overlap2}
\begin{tabular}{c|cc||c|cc}
    \hline \hline 
\multirow{3}{*}{Campaign number} & \% of users exposed & \% of users exposed & \multirow{3}{*}{Campaign number} & \% of users exposed & \% of users exposed   \\
& to at least one other & to campaign with & & to at least one other & to campaign with   \\
& campaign & highest overlap & & campaign & highest overlap     \\
\hline
  1 & 74\% & 53\%  & 9 & 51\% & 17\%   \\
  2 & 38\% & 10\%  & 10 & 59\% & 19\%  \\ 
  3 & 50\% & 18\%   & 11 & 51\% & 11\%  \\ 
  4 & 46\% & 11\%   & 12 & 50\% & 14\%  \\ 
  5 & 52\% & 12\%   & 13 & 58\% & 22\%  \\ 
  6 & 37\% & 13\%   & 14 & 49\% & 21\%  \\
  7 & 43\% & 11\%  & 15 & 52\% & 21\%  \\ 
  8 & 45\% & 20\%   & 16 & 60\% & 27\%  \\
 \hline \hline
    \end{tabular}
  \end{threeparttable}}
\end{table} 

We will refer to the advertiser corresponding to the second column of Table \ref{tab:overlap2} as the focal campaign's \textit{main competitor}. Part of the analysis below focuses solely on the focal campaign and its main competitor, which we denote by $f$ and $g$, respectively.\footnote{Note that such overlap is only a crude measure of which other firm is the focal firm's main competitor. For instance, a firm $g$ may not be a relevant competitor of firm $f$ for all overlapping users, and the extent to which $g$'s presence affects $f$ may depend on another firm, $h$.}

\end{section}

\begin{section}{Results}\label{sec:results}

\noindent We now present our main results. First, we display the baseline results obtained from the typical approach in experimental studies, that is, a simple difference in means between treatment and control observations. We refer to these as the difference in means estimates of the unconditional $ATE$. They correspond to estimates of objects analogous to $\xi_{f}(A_g)$ from equation (\ref{eq:simp_ate_int}) for each of the 16 campaigns. Second, we assess and find evidence of the existence of competitive interference effects, and then investigate whether these effects are associated with the previously defined ad allocation change and cross-campaign externalities. Finally, we dissect the heterogeneity in $CATE$s that is induced by this interference and outline implications for advertiser's decision-making.

\subsection{Baseline results}\label{sec:baseline}

The difference in means estimates of the unconditional \textit{ATE} for each of the 16 campaigns are shown in Table \ref{tab:base1}. We only use observations that, absent the experiment, would have been exposed to the focal campaign, $f$, and to its main competitor, $g$. We choose this sample for the purposes of comparability to results we find when we investigate the existence of competitive interference effects.

\begin{table}[htp]
\begin{threeparttable}
\caption{Baseline results}
\label{tab:base1}
\begin{tabular}{c|cc||c|cc}
    \hline \hline 
Campaign & Treatment & Constant  & Campaign & Treatment & Constant \\
\hline 
 1 & 0.0293 & 0.1015  & 9  & 0.0286 & 0.1212  \\
($n=128,576$) & (0.0093)*** & (0.0078)*** & ($n=677,576$) & (0.0079)*** & (0.0066) *** \\ 
 2 & -0.0018 & 0.0268  & 10  & 0.0297 & 0.1463  \\
($n=535,426$) & (0.0031) & (0.0036)*** & ($n=68,454$) & (0.0251) & (0.0233)*** \\ 
 3 & 0.0436 & 0.0473  & 11  & 0.0187 & 0.0177  \\
($n=24,109$) & (0.0127)*** & (0.0080)*** & ($n=161,701$) & (0.0030)*** & (0.0020)*** \\
 4 & -0.0036 & 0.0141  & 12  & -0.4537 & 1.6134  \\
($n=27,883$) & (0.0087) & (0.0085)* & ($n=17,347$) & (0.4060) & (0.4027)*** \\ 
 5 & 0.0090 & 0.0395  & 13  & 0.0284 & 1.4218  \\
($n=231,047$) & (0.0044)** & (0.0038)*** & ($n=150,471$) & (0.0671) & (0.0558)*** \\ 
 6 & 0.0004 & 0.0046  & 14  & -0.0344 & 0.8920  \\
($n=393,256$) & (0.0009) & (0.0007)*** & ($n=29,849$) & (0.0720) & (0.0650)*** \\ 
 7 & 0.0120 & 0.0582  & 15  & 0.0555 & 0.4518  \\
($n=449,963$) & (0.0063)* & (0.0056)*** & ($n=77,525$) & (0.0228)** & (0.0324)*** \\ 
 8 & 0.0056 & 0.1973  & 16  & 0.0004 & 0.0019  \\
($n=677,576$) & (0.0106) & (0.0090)*** & ($n=339,019$) & (0.0007) & (0.0005)*** \\ 
 \hline \hline
    \end{tabular}
\begin{tablenotes}
      \small
	Note: The unit of observation is a user. The dependent variable is the number of visits to the SKU detail page during the experiment. Standard errors robust to heteroskedasticity are shown between parentheses. \\
	*** \textit{p}<0.01, ** \textit{p}<0.05, * \textit{p}<0.1
    \end{tablenotes}
  \end{threeparttable}
\end{table} 

Out of the 16 estimates, seven are statistically significant at the 10\% level. There is substantial variation across the lift estimates: among the significant effects, they range from 8.73\% (campaign 15) to 98.95\% (campaign 11), with an average of 36.40\%. Finally, to account for multiple testing across the 16 campaigns we estimate a SUR model and perform a joint test for whether all $ATE$s are equal to zero. We obtain a \textit{p}-value lower than 0.00001, which indicates that we can reject this hypothesis. 

Overall, this establishes that front-focused ads have a measurable effect on visitation for several campaigns. However, care must be taken when interpreting these results. They were obtained from a sample of users who, absent the experiment, would have been exposed to both the focal campaign and to its main competitor, that is, those who lie in the overlapping target population, which, as Tables \ref{tab:overlap} and \ref{tab:overlap2} show, can be limited. 

\subsection{Assessing the existence of competitive interference effects}\label{sec:int_std}

\noindent We now assess the existence of competitive interference effects. We begin with an analysis of how each campaign's effect varies with the presence of its main competitor. This is only a first-cut: to more comprehensively assess competitive interference we need to assess how each campaign's effect varies over \textit{all} the $2^{F-1}=2^{15}=32,768$ possible scenarios depending on each of its competitors' presence. As discussed in Section \ref{sec:full_analysis}, several competitors may interfere with a given advertiser's campaign, so such a comprehensive analysis is needed to reveal the full picture regarding competition. 

We estimate the following regression to implement this first-cut analysis:
\begin{align}\label{eq:inter}
	Y_{if} = \alpha_f + \beta_f D_{if} + \gamma_f D_{ig} + \delta_f D_{if}D_{ig} + C_i'\psi_f + \epsilon_{if},
\end{align}
where: $g$ is firm $f$'s main competitor as defined above; $i$ denotes a user; $Y$ is the number of visits to the SKU detail page during the experiment; $D_{\ell}$ is an indicator for whether $i$ was in advertiser $\ell$'s treatment group, where $\ell\in\{f,g\}$; $C_i$ is a vector that contains outcomes from three days prior to the experiment, namely numbers of adds to cart, order and page visits, and gross merchandising value (GMV), a metric of money spent---these variables are independent from treatment assignments and we include them to increase precision; and $[\alpha_f,\beta_f,\gamma_f,\delta_f,\psi_f]'$ is the vector of coefficients to be estimated. We can detect whether competitive interference can alter the focal firm's ad's effect by assessing the significance of the parameter $\delta_f$. Table \ref{tab:inter1} displays the estimates for all campaigns. For ease of exposition, we omit the estimates of $\psi_f$. 

\begin{table}[htp]
\resizebox{\textwidth}{!}{\begin{threeparttable}
\caption{Results from equation (\ref{eq:inter})}
\label{tab:inter1}
\begin{tabular}{c|cccc||c|cccc}
    \hline \hline 
\multirow{2}{*}{Campaign} & Intercept & Ad's effect & Competitor ad's effect & Interaction & \multirow{2}{*}{Campaign} & Intercept & Ad's effect & Competitor ad's effect & Interaction   \\
     & $\alpha_f$ & $\beta_f$ & $\gamma_f$ & $\delta_f$ &  & $\alpha_f$ & $\beta_f$ & $\gamma_f$ & $\delta_f$  \\
\hline 
1 & 0.0796 & 0.0516 & 0.0314 & -0.0319 & 9 & 0.1099 & 0.0427 & 0.0162 & -0.0202 \\
($n=128,576$) & (0.0114)*** & (0.0146)*** & (0.0151)** & (0.0188)* & ($n=677,576$) & (0.0099)*** & (0.0130)*** & (0.0128) & (0.0163)  \\ 
2 & 0.0337 & -0.0124 & -0.0100 & 0.0153 & 10 & 0.1301 & 0.0458 & 0.0233 & -0.0231 \\
($n=535,426$) & (0.0055)*** & (0.0053)** & (0.0058)* & (0.0065)** & ($n=68,454$) & (0.0351)*** & (0.0426) & (0.0414) & (0.0526)  \\ 
3 & 0.0325  & 0.0938 & 0.0212 & -0.0718 & 11 & 0.0206 & 0.0151 & -0.0042 & 0.0052 \\
($n=24,109$) & (0.0131)*** & (0.0267)*** & (0.0167) & (0.0298)** & ($n=161,701$) & (0.0043)*** & (0.0062)** & (0.0049) & (0.0070)  \\ 
4 & 0.0380 & -0.0268 & -0.0344 & 0.0334 & 12 & 1.1854 & -0.0671 & 0.6186 & -0.5597  \\
($n=27,883$) & (0.0278) & (0.0280) & (0.0278) & (0.0281) & ($n=17,347$) & (0.1591)*** & (0.1802) & (0.5946) & (0.6085)  \\ 
5 & 0.0487 & -0.0041 & -0.0132 & 0.0187 & 13 & 1.4252 & -0.0428 & -0.0049 & 0.1018 \\
($n=231,047$) & (0.0095)*** & (0.0104) & (0.0103) & (0.0113)* & ($n=150,471$) & (0.0928)*** & (0.1060) & (0.1151) & (0.1353)  \\ 
6 & 0.0043 & -0.0002 & 0.0005 & 0.0008 & 14 & 0.8806 & -0.0470 & 0.0163 & 0.0181 \\
($n=393,256$) & (0.0011)*** & (0.0014) & (0.0014) & (0.0018) & ($n=29,849$) & (0.0865)*** & (0.1037) & (0.1197) & (0.1392)  \\ 
7 & 0.0518 & 0.0270 & 0.0092 & -0.0215 & 15 & 0.4539 & 0.0438 & -0.0030 & 0.0168 \\
($n=449,963$) & (0.0080)*** & (0.0106)** & (0.0102) & (0.0130)* & ($n=77,525$) & (0.0413)*** & (0.0354) & (0.0351) & (0.0435)  \\ 
8 & 0.1901 & 0.0268 & 0.0102 & -0.0303 & 16 & 0.0011 & 0.0022 & 0.0012 & -0.0026 \\
($n=677,576$) & (0.0161)*** & (0.0184) & (0.0191) & (0.0226) & ($n=339,019$) & (0.0006)* & (0.0013)* & (0.0010) & (0.0015)*  \\ 
 \hline \hline
    \end{tabular}
\begin{tablenotes}
      \small
	Note: The unit of observation is a user. The dependent variable is the number of visits to the SKU detail page during the experiment. Standard errors robust to heteroskedasticity are shown between parentheses. \\
	*** \textit{p}<0.01, ** \textit{p}<0.05, * \textit{p}<0.1
    \end{tablenotes}
  \end{threeparttable}}
\end{table} 

\paragraph{Statistical significance} \quad Out of the 16 interaction coefficients, $\delta_f$, six are statistically significant at the 10\% level. To guard against multiple testing, we estimate a SUR model and perform a joint test of whether all the $\delta_f$s are zero clustering at the user level (2,632,499 users). We obtain a \textit{p}-value = 0.0192, which indicates the existence of competitive interference effects. 

\paragraph{Economic significance} \quad To assess the economic significance of the interference, we calculate the average of $\frac{|\delta_f|}{|\beta_f|}$ across campaigns. Across all 16 campaigns, this ratio is 0.6226, and across the campaigns that had a statistically significant $\beta_f$, the ratio is 0.6733. This demonstrates that, by point estimates, interference equals a substantial fraction of a campaign's main effect. 

A takeaway from these results is that, in this marketplace, competitive interference can substantively change the effect of a focal campaign's advertising and that competitive behavior can significantly change the relevant treatment effect of advertising. However, once again we note that these results were obtained from a sample of individuals who, absent the experiment, would have been exposed to the focal campaign and its main competitor. These individuals lie in the overlapping target population, which, as Tables \ref{tab:overlap} and \ref{tab:overlap2} show, can be limited. We assess the overall impact of competitive behavior in Section \ref{sec:full_analysis}. 

\subsection{Assessing the sources of competitive interference effects}\label{sec:inter_sep}

\noindent The next step is to unpack the sources of such interference effects, which can help us understand the role of competition in driving the interactions we uncovered. In particular, we will look at the two effects we outlined above, ad allocation change and cross-campaign externalities. 

To test for the presence of ad allocation change, we assess whether the presence of the focal advertiser's main competitor changes the average number of impressions per user the focal advertiser obtains. We proceed as follows. First, we take only auctions for display ads to users who belong to the focal advertiser's treatment group, so that the focal advertiser is always able to factually show their ad. Second, we separate this group in two: (1) in which users belong to the main competitor's treatment group (those eligible to see the main competitor's ad) and (2) in which users belong to the main competitor's control group. Third, we use the the numbers of auctions the focal advertiser won and participated in and the number of users in each firm's experiment, as given in Tables \ref{tab:base1} and \ref{tab:inter1}, to compute estimates of the average number of impressions the focal advertiser obtained per user for each of the two groups. Finally, we test whether these averages are the same. Table \ref{tab:avg_change} shows the $p$-values associated with such tests along with the average impressions per users in the main competitor's control group and report the change in average impressions, in terms of percentage points, from users in the control group.

\begin{table}[H]
\resizebox{\textwidth}{!}{\begin{threeparttable}
\caption{Average impressions per user}
\label{tab:avg_change}
\begin{tabular}{c|ccc|| c|ccc}
    \hline \hline 
 \multirow{2}{*}{Campaign} & Avg. impressions in  & \% Change in & \multirow{2}{*}{\textit{p}-value} & \multirow{2}{*}{Campaign} & Avg. impressions in  & \% Change in & \multirow{2}{*}{\textit{p}-value}   \\ 
  & Competitor's control & treatment (percent points) & & & Competitor's control & treatment (percent points)    \\ 
 \hline
 1 & 21.52 & -0.804 & $1.05\times 10^{-15}$ &  9 & 33.66  & 0.584 &  $2.217\times 10^{-61}$    \\
 2 & 28.15 & 0.062 & $0.1578$  &  10 & 66.06  &  0.363 &  $5.603\times 10^{-6}$     \\
 3 & 192.17   &  -0.388  &    $9.986\times 10^{-7}$ &  11 &   100.06 &  -0.075 & $0.073$   \\
 4 & 61.45 & -0.115   &  $0.37275$  &  12 & 186.25  & -0.621  &  $2.139\times 10^{-11}$   \\
 5 & 28.95 &  -1.054  & $5.712\times 10^{-60}$  &  13 & 55.92  & 0.322  &   $2.829\times 10^{-8}$     \\
 6 &    25.65  & 0.155 &    $0.0005$   &  14 & 220.87 &  -0.206  & $0.0002$    \\
 7 & 74.18  & 0.558 &  $6.582\times 10^{-82}$  &  15 & 68.2 & -0.340 &   $2.673\times 10^{-6}$    \\
 8 & 65.78  & -0.001 &  $0.9574$  &  16 &   44.98  & -0.088 &   $0.044$   \\
 \hline \hline
    \end{tabular}
\begin{tablenotes}
      \small Due to data limitations, the $p$-values in this table were computed under the assumptions that auctions are i.i.d. Thus, these $p$-values are possibly deflated and should be interpreted with caution.
    \end{tablenotes}
\label{tab:avg_numb}
  \end{threeparttable}}
\end{table}  

We detect a statistically significant ad allocation effect---the changes in the average number of impressions delivered due to the competitor's campaign are statistically significant at the 5\% level in 12 out of the 16 cases, although the percentage changes are small on average. Thus, ad allocation change due to the main competitor has an impact on the intensity with which users are, on average, exposed to the focal campaign's ad.

Combined with the results from Table \ref{tab:inter1}, these results also demonstrate the presence of cross-campaign externalities. Take, for example, campaign 2. From Table \ref{tab:avg_numb} we detect no change in the average number of impressions delivered per user as a function of their main competitor's presence, but Table \ref{tab:inter1} indicates that there is a significant increase in the average effect of the ad campaign from this competitor's presence. Other examples, such as campaign 5---for which average number of impressions delivered decreased while the treatment effect on consumer visits increased--- and campaign 7---for which we see the opposite pattern---show that the treatment effects on consumer behavior and ad impressions delivered can go in opposite directions. These instances are indicative of the presence of cross-campaign externalities.

Overall, we have evidence that the competitive interference effects we detected in Section \ref{sec:int_std} are a result not only of ad allocation change but also of cross-campaign externalities, which demonstrates the complexity and richness of this competitive advertising environment. The effect of a firm's advertising is not only impacted by the number of competitors it has since this number directly impacts its ability to show its ad; in addition, the identities of the competitors that are able to show their ads may also be relevant. Furthermore, it is noteworthy  that the interaction effects can be either positive or negative.

\subsection{Recovering all possible \pmb{$CATE$}s}\label{sec:full_analysis}
 
We now unpack competitive interference effects in more detail by recovering all possible $CATE$s as defined in equation (\ref{eq:simp_ate}). As we mentioned in Section \ref{sec:estimation}, with such a large number of coefficients OLS is known not to have good properties. Thus, instead of OLS we use a kernel-based estimator, whose details we give below, that directly accounts for the large number of coefficients via regularization. 

We illustrate this approach by focusing on campaign number 3. We chose this campaign for two reasons. First, as Table \ref{tab:base1} shows, its estimated unconditional $ATE$ is significant. Second, its low number of observations (relative to the other campaigns) facilitates the use of the kernel-based estimator, whose implementation utilizes via leave-one-out cross-validation and is thus computationally intensive. 

Importantly, the overall impact on the focal advertiser's treatment effect parameters and, consequently, on their decisions, depends both on the number of competitors with whom the focal firm's target population overlaps \textit{and} the degree of interference due to these competitors campaigns. Hence, we use \textit{all} users who belonged to campaign 3's experiment to perform this analysis, and not just those who overlapped with its main competitor's experiment. Consequently, unlike in Sections \ref{sec:int_std} and \ref{sec:inter_sep}, the results we obtain here apply to the overall target population.

\subsubsection{A kernel-based estimator}\label{sec:kernel}

We employ the kernel-based estimator introduced by \cite{lor2013}. Although typical kernel estimators are designed to smooth over continuous variables, there are more flexible approaches that rely on generalized kernel functions that smooth over both continuous \textit{and} categorical variables to estimate densities \citep{lr2003} and regression functions \citep{rl2004}. In this regard, the method proposed by \cite{lor2013} can be seen as special case of this generalized kernel approach: one in which all variables that are smoothed over are categorical.

To the best of our knowledge, this is the first implementation of this estimator within the context of a factorial design. Since, this estimator smooths over discrete variables, highlighting that data from a factorial experiment are conducive to its use may be of independent interest. In the Appendix, we provide extensive simulation exercises that showcase the performance of this estimator, which we describe below.

We roughly follow the presentation from \cite{lor2013}. Based on regression equation (\ref{eq:est_tau_f_regr}), consider the regression equation for a given partial treatment assignment, $D_{i,-f}=d$, where $D_{i,-f} \equiv \left [D_{i1},\dots, D_{i,f-1}, D_{i,f+1},\dots,D_{iF} \right ]'$:
\begin{align*}
	Y_{if}&= \alpha_{f,d} + \tau_{f,d} D_{if}+\epsilon_{if} \nonumber \\
	 	&= X_{if}'\theta_{f,d} +\epsilon_{if}, 	
\end{align*}
where: $i$ indicates a user; $Y$ is the outcome variable; $\epsilon$ is an error term; $D_{if}$ is an indicator for whether $i$ was in $f$'s treatment group; $\alpha_{f,d}$ and $\tau_{f,d}$ are parameters to be estimated; $X_{if}= \left [1, D_{if} \right ]'$; and $\theta_{f,d}(\cdot)= \left [\alpha_{f,d}, \tau_{f,d} \right ]'$.

Let $D_{i,-f,v}$ be the $v^{\text{th}}$ coordinate of the vector $D_{i,-f}$. The kernel function is:
\begin{align*}
	l \left ( D_{i,-f,v}; d_v ; \lambda_v \right ) = \begin{cases} 1, \text{ if } D_{i,-f,v}=d_v \\   \lambda_v, \text{ otherwise}        \end{cases},
\end{align*}
where $d_v$ is one of the values $D_{i,-f,v}$ can take (0 or 1 in this case) and $\lambda_v$ is the bandwidth associated with $D_{i,-f,v}$, where $\lambda_v \in [0,1]$ for all $v$. The overall kernel is given by:
\begin{align}\label{eq:kernel_total}
	L \left ( D_{i,-f}; d ; \lambda \right ) = \prod_v l \left ( D_{i,-f,v} ; d_v; \lambda_v \right ) = \prod_v \lambda_v^{\mathbbm{1} \left \{ D_{i,-f,v} \neq d_v \right \}}, 
\end{align}
where $d$ is a vector collecting all the $d_v$s and $\lambda$ is a vector containing all bandwidths. The estimator, which resembles a local linear estimator, is:
\begin{align}\label{eq:lor}
	\hat{\theta}_f\left (d \right ) = \left ( \frac{1}{n} \sum_{i=1}^n X_{if} X_{if} ' L \left ( D_{i,-f} ; d; \lambda \right )  \right )^{-1} \left ( \frac{1}{n} \sum_{i=1}^n X_{if} Y_{if}  L \left ( D_{i,-f} ; d ; \lambda \right )  \right ).
\end{align}

\cite{lor2013} propose computing $\lambda$ via cross-validation to minimize the leave-one-out mean squared error (MSE). Through their Theorems 1 and 2, the authors formally prove that the estimator is $\sqrt{n}-$consistent with a limiting normal distribution. Importantly, the rate of convergence of this estimator is square root, which is an attractive property because it is faster than typical kernel-based and machine learning estimators. Hence, this is the approach we follow. Note that these results assume that the number of variables that are smoothed over, which is the dimension of $D_{i,-f}$ in our case, is fixed and does not increase with the sample size. This assumption holds in our case. However, if it is violated, a potential regularization bias must be dealt with as in \cite{ccddhnr2018}.

\paragraph{Intuition} \quad First, we discuss the interpretation of the bandwidths, $\lambda$. The estimator given in (\ref{eq:lor}) is identical to the OLS estimator from equation (\ref{eq:est_tau_f_regr}) when $\lambda=0$. This is because $\lambda=0$ implies that $L\left (D_{i,-f};d;\lambda \right) =1 $ when $D_{i,-f}=d$ and 0 otherwise. Hence, to estimate the treatment effect given the competitor experimental assignment $d$ the procedure uses only the data from observations where $D_{i,-f}=d$. 

Now consider the case where one bandwidth, say, $\lambda_v$, increases from zero. When $\lambda_v\ne0$, the estimator in equation (\ref{eq:lor}) uses additional data from $D_{i,-f}$ combinations that differ from $d$ at the $v$th element to estimate $\hat{\theta}_f(d)$. When $\lambda_v=1$, then $\hat{\theta}_f(d)$ will be identical for values of $d$ that differ only in terms of $d_v$. At the extreme, when all bandwidths are one, that is, $\lambda=1$, then $L\left (D_{i,-f};d;\lambda \right) =1 $ for all $d$ and so $\hat{\theta}_f(d)=\hat{\theta}_f$, which implies pooling across all combinations of $D_{i,-f}$'s.

The intuition behind the estimation of bandwidths is as follows. If the $v$th competitor does not impact the effectiveness of $f$'s ads, increasing its bandwidth will decrease the out-of-sample MSE. This is because a larger bandwidth improves the accuracy of the estimated effects by allowing for more data pooling. Hence, the procedure will yield a larger bandwidth corresponding to a competitor that does not interact with the focal ad's effect.

Overall, in our setting, bandwidths given by this procedure can serve as a first data-driven indication of competitive interference effects and also of which competitors are associated with stronger such effects. The more $CATE$s associated with different competitor treatment assignments are different from one another, the more the cross-validation procedure penalizes higher values of $\lambda$ because they increase the MSE more. Consequently, lower values of the bandwidth associated with the treatment indicator for a given competitor indicate that this competitor induces larger competitive interference effects. In turn, in the extreme case in which there is no competitive interference at all, so that the $CATE$s do not vary with whether competitors are advertising, then the associated bandwidths equal one, implying that the estimate is the same across all partial treatment assignments.
 
\subsubsection{Cross-validated bandwidth estimates}

\noindent Before we present the main results, we first display the cross-validated bandwidths we compute for campaign number 3 with respect to each competitor. Bandwidths with higher (lower) values, that is, values that are closer to one (zero), are associated with competitors who induce small (big) competitive interference effects. Importantly, these bandwidths also capture the extent of overlap between the focal advertiser and their competitors. The more limited the overlap, the less scope there is for overall interference, which drives the values of the bandwidth towards one.

\begin{figure}[H]
    \centering
        {\includegraphics[width=0.6\textwidth]{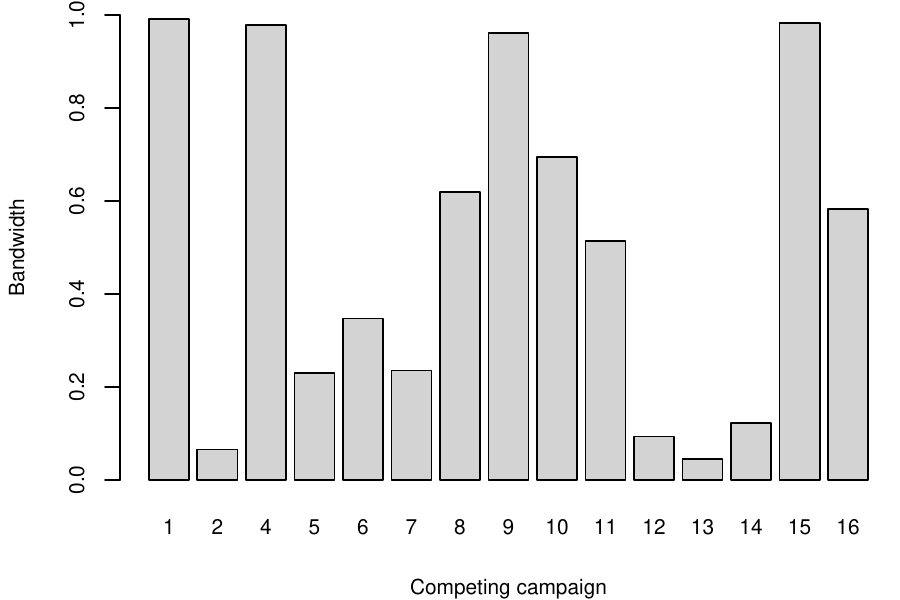}}
     \caption{Cross-validated bandwidths}
     \floatfoot{\scriptsize Note: The graph shows the bandwidths associated with each of firm 3's competitors. The bandwidths were obtained via leave-one-out cross-validation as described in Section \ref{sec:kernel}.}
    \label{fig:bws}%
\end{figure} 

The cross-validated bandwidths computed for campaign 3 with respect to its 15 competitors are displayed in Figure \ref{fig:bws}. The bandwidths associated with competing firms 1, 4, 9 and 15 are very close to one, indicating that firm 3 does not suffer significant competitive interference effects from these competitors relative to others. In turn, firms 2, 12, 13 and 14 display the smallest bandwidths, suggesting that these are the competitors that cause the strongest  competitive interference effects on firm 3.\footnote{Interestingly, firm 10, which we previously assessed as firm 3's main competitor based on our simple ``overlap'' metric, has a relatively high bandwidth. This implies that it is not necessarily firm 3's closest competitor and suggests that our previous test for interference based on the overlap metric was conservative.} Figure \ref{fig:bws} once again reiterates the presence of substantial competitive interference in the data.

\subsubsection{Estimates, competitive interference, and comparison to naive estimate of \pmb{$ATE$}}

\noindent We now present estimates of the $2^{15}=32,768$ different $CATE$s of firm 3's advertising depending on the advertising policies of its rivals. The goal is to assess the extent to which the $CATE$s are heterogeneous, so that taking rivals' policies into account is crucial. Figure \ref{fig:dots} displays the estimated $CATE$s. It distinguishes between statistically significant and non-significant estimates by displaying the former in black and the latter in gray. Significance is at the 5\% level and is determined using nonparametric bootstrap based on 50 replications.

\begin{figure}[H]
    \centering
        {\includegraphics[width=0.6\textwidth]{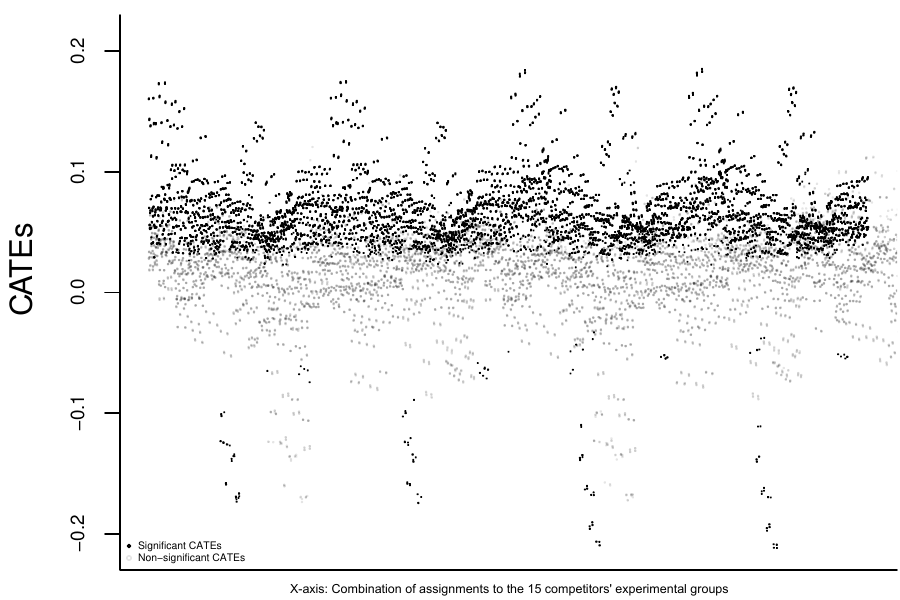}}
     \caption{Estimates of $CATE$s}
     \floatfoot{\scriptsize Note: The graph shows the estimates of the $CATE$s across all assignment combinations of competitors' experiments. Black dots indicate statistically significant estimates at the 5\% level while gray dots indicate non-significant ones, both based on nonparametric bootstrap with 50 replications.}
    \label{fig:dots}%
\end{figure}  

Out of the 32,768 estimates, 11,616, or 35\%, are statistically significant at the 5\% level, indicating that the variation we see in \textit{CATE}s by competitor policies are not driven simply by sampling and estimation noise. It is interesting to note that most negative estimates, which are arguably more counterintuitive, are not statistically significant.\footnote{Out of the 32,768 estimates, 5,317, or 16.23\%, were negative, and out of those only 164 were statistically different from 0, corresponding to only 1.41\% of all significant results.}

One concern with this analysis is multiple testing, which we account for by repeating this exercise using a simple Bonferroni correction in which we divide the significance level, 0.05, by the number of tests. We find that instead of 11,616 only 7,980 effects, or 24.35\%, are significant. Furthermore, out of these only 43 results are negative, corresponding to 0.54\% of all significant results. Put together, these results provide further evidence that sampling and estimation noise are not the sole drivers of the quantities we observe.

To visualize the spread in $CATE$s caused by competitive interference, Figure \ref{fig:pdfs} displays the distribution of the estimated $CATE$s computed using a Gaussian kernel and Silverman's rule-of-thumb bandwidth. It ranges from -0.2408 to 0.1853, with a mean of 0.032. Compared to this mean, the distribution has a standard deviation of 0.0411, indicating the presence of meaningful heterogeneity across $CATE$s with respect to the advertising policies of the focal advertiser's competitors. Finally, with a median of 0.0348 the distribution is fairly symmetric.

The analysis of dispersion in the distribution of $CATE$s is relevant to assess the importance of what we refer to as ``competitive environmental uncertainty'': it is a reflection of a firm's uncertainty over which of its competitors will be advertising. Because $CATE$s are estimated to aid firms in their decision-making, high dispersion across them indicates that consideration of environmental uncertainty is important.

Nevertheless, environmental uncertainty is ignored by the usual approach that focuses only on the unconditional $ATE$, obtained typically by the difference in means estimator. If environmental uncertainty is relevant, following this approach can hurt decision-making, as we study in more detail below. As a precursor to this analysis, we now consider how much of the competitive environmental uncertainty is captured by the statistical uncertainty around the estimate of the unconditional $ATE$ obtained from the typical difference in means approach. If statistical uncertainty around the unconditional $ATE$ captures very little of the environmental uncertainty, typical decision-making that incorporates the former, but ignores the latter, can be sub-optimal.

The $ATE$'s point estimate is 0.03 with a 95\% confidence interval ranging from 0.0147 to 0.0453. This confidence interval covers 7.18\% of the full range of possible $CATE$s in Figure \ref{fig:pdfs}. From another perspective, we find that only 36.56\% of the area under the curve in Figure \ref{fig:pdfs} is captured by the confidence interval, demonstrating that the typical measure of statistical uncertainty underestimates the amount of uncertainty faced by advertisers.

\begin{figure}[H]
    \centering
        {\includegraphics[width=0.6\textwidth]{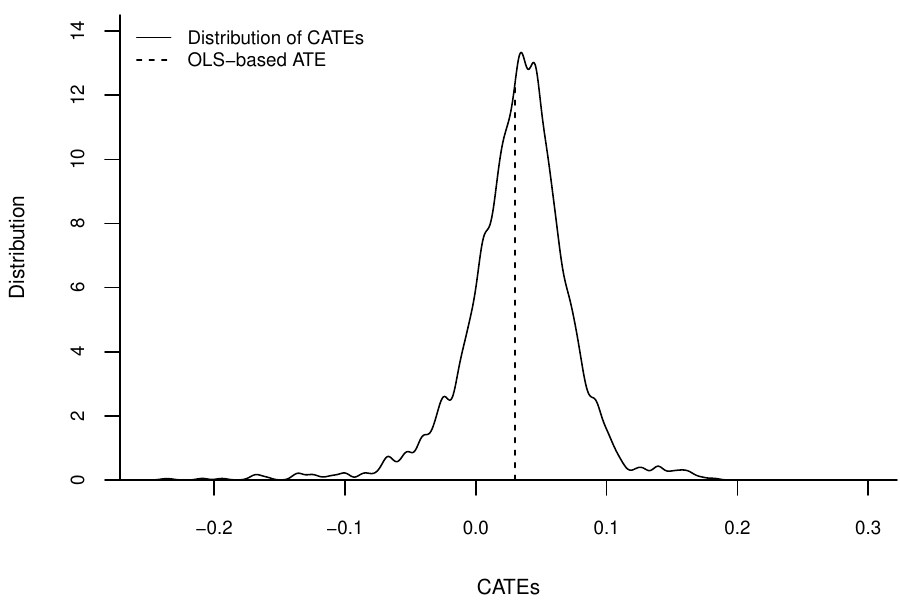}}
     \caption{Distribution of $CATE$s}
     \floatfoot{\scriptsize Note: The graph shows the distribution of estimated $CATE$s (solid line) and the estimate of the $ATE$ based on a simple linear regression (vertical dashed line). The distribution is computed using a Gaussian kernel with Silverman's rule-of-thumb bandwidth.}
    \label{fig:pdfs}%
\end{figure}

This suggests that even if decision-making accounts for statistical uncertainty around the unconditional $ATE$ it can still perform poorly by ignoring a substantial portion of the competitive environmental uncertainty. We investigate this in more detail below.

\subsubsection{Implications for firm's decision-making}

\noindent So far we have described the results we obtained when we recovered all possible $CATE$s and how they compare to the unconditional $ATE$ recovered via difference in means. We now assess how the variation in competitive states can impact the firm's post experiment decisions. We assume that the status quo decision-making by firms is based on the $ATE$, which is the average change in outcomes going from the control to the treatment group.

\paragraph{Advertising lift} \quad
A managerially relevant measure of the advertising effect is its lift, which refers to the incremental business outcomes caused by advertising relative to the baseline that occurs in the absence of advertising \citep{gordon2023close}. We ask the question: how different could the actual advertising lift be from the lift the firm estimates if it ignores the competitive states?

To make this comparison, we first calculate the firm's baseline lift, ignoring competitive states, as $\ubar{l}=\frac{ \ubar{Y}_{T} - \ubar{Y}_{C} }{\ubar{Y}_{C}}$ where $\ubar{Y}_T$ and $\ubar{Y}_C$ are the average outcomes (visits in our application) in the focal firm's treatment and control groups, respectively. Next, we calculate the lift corresponding to each competitive state $d$ using the $CATE$ estimates from the previous section as $l_d=\frac{CATE_d}{\ubar{Y}_{C,d}}$, where $\ubar{Y}_{C,d}$ is the average outcome across experimental units belonging to the control group that experience the state $d$ and $CATE_d$ is the $CATE$ at this state. 

To visualize how different the firm's actual lift could be relative to the baseline, we present the empirical CDF of $\Delta l_d \equiv |\frac{l_d-\ubar{l}}{\ubar{l}}|$ in Figure \ref{fig:dist_lift}. It shows that, depending on the actual state of competitors' advertising, the firm's estimated lift using the baseline method can be off by a large factor. In 60\% of the states, $\Delta l_d \ge 100\%$. In 40\% of the states, the firm might actually get a lift that differs from its $ATE$-based estimate by a factor of two. This could be consequential for a firm's decision of whether to advertise if such decision is based on the lift.

\begin{figure}[H]
    \centering
        {\includegraphics[width=0.6\textwidth]{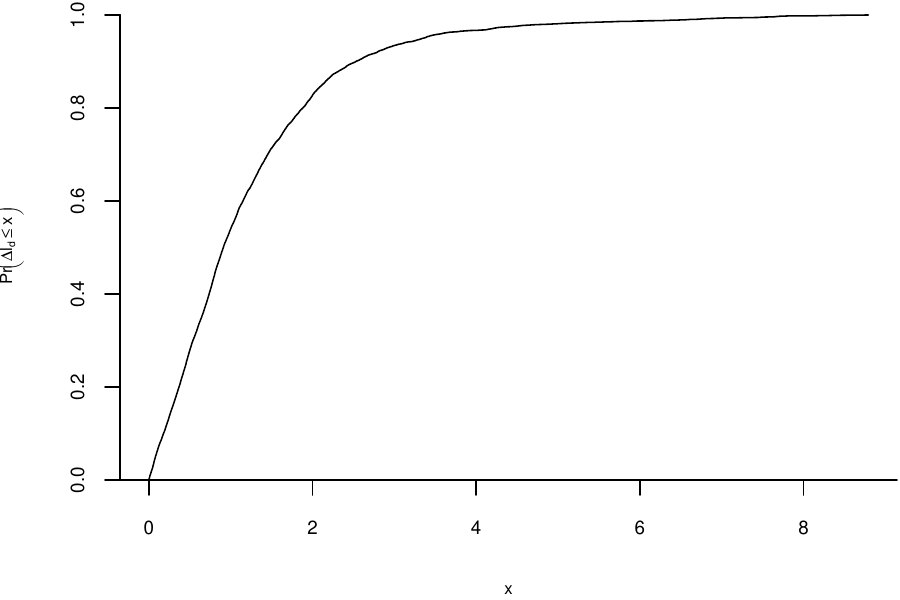}}
     \caption{Distribution of changes in actual lift relative to the $ATE$-based lift}
     \floatfoot{\scriptsize Note: The graph shows the empirical cumulative distribution of $\Delta l_d = |\frac{l_d-\ubar{l}}{\ubar{l}}|$. It implies that in about 40\% of the states the actual lift the firm might get differs by a factor larger than 2 from the lift the firm would estimate based on the $ATE$.}
    \label{fig:dist_lift}%
\end{figure} 

\paragraph{Firm 3's simulated advertising decision} \quad
Assuming the status quo scenario where firms make their advertising decisions based on the $ATE$ ignoring competitive states, we ask: what is the economic impact of a firm deviating from the status quo? Answering this question sheds light on the significance of ignoring competitive interference for decision-making. 

We consider a simple decision problem in which firm 3 has to choose whether to advertise on the platform. This decision is made to maximize its expected payoff from advertising, so that the optimal decision is to advertise if and only if the impact of advertising exceeds its cost, which we denote by a constant $\kappa$. 
 
In the status quo, when the firm $f$ ignores competitive interference, the optimal advertising policy, $A_f$, is based on the unconditional $ATE$ and is given by:
\begin{align}\label{eq:opt_pol-SQ}
	A_{f} = \mathbbm{1} \left \{ ATE\geq \kappa \right \}.
\end{align}

What would the firm do if it knew its competition and the $CATE$? Given the $ATE$ estimates from Table \ref{tab:base1}, we know that advertisers 1, 5, 9, 11, and 15 would potentially advertise given the status quo policy rule because their estimated $ATE$s are positive and significant at the 5\% significance level. Thus, the expected state of the world is $d^*$ such that $d^*_1=d^*_5=d^*_9=d^*_{11}=d^*_{15}=1$, and the remaining elements are 0. In this state, firm 3's optimal policy is
\begin{align}\label{eq:opt_pol}
	A_{f}^* = \mathbbm{1} \left \{ CATE_f(d^*)\geq \kappa \right \},
\end{align}
which would yield an expected incremental profit of $\pi_f^*(\kappa)=\left [ CATE_f(d^*) - \kappa \right ] \times A_f^*$. We denote the expected incremental profit from following the status quo policy by $\pi_f(\kappa)=\left [ CATE_f(d^*) - \kappa \right ] \times A_f$, where $A_f$ is given in (\ref{eq:opt_pol-SQ}), since the actual gains depend on $d^*$ while the firm's policy ignores it under the status quo.

Figure \ref{fig:prof_ex} displays $\pi_f^*(\kappa)$ and $\pi_f(\kappa)$. Unsurprisingly, the firm would make the same decision of not advertising if the cost is high enough and of advertising if the cost is low enough. However, since the $CATE$ is higher than the $ATE$ in this case, the firm is more likely to advertise at intermediate cost levels if it is able to account for competition, and it makes a loss if it does not. Depending on the cost, the firm can lose more than 30\% of its optimal incremental expected profits if it ignores competition.

\begin{figure}[H]
    \centering
        {\includegraphics[width=0.7\textwidth]{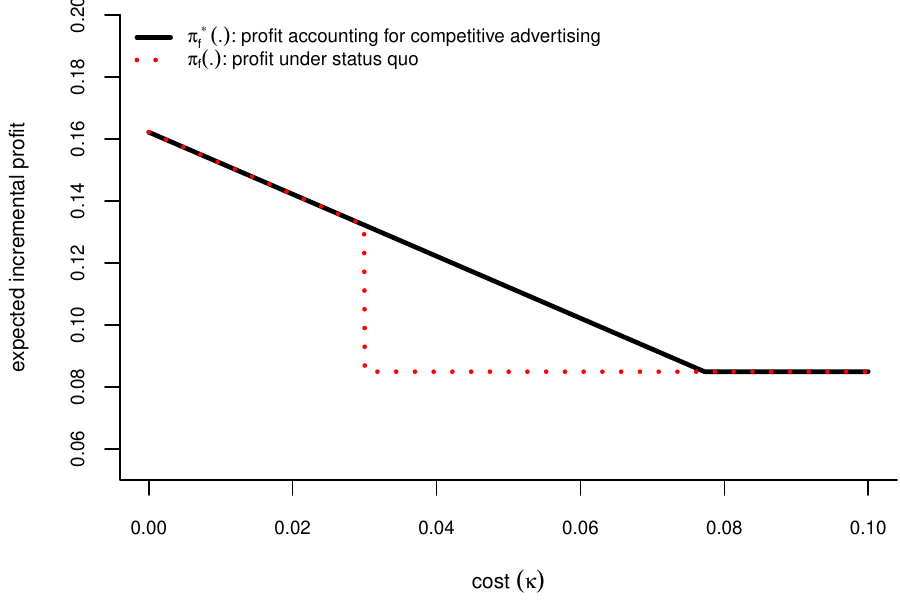}}
     \caption{Expected incremental profit with and without accounting for competitive state}
    \label{fig:prof_ex}%
\end{figure} 

\paragraph{Discussion} \quad
The above simulations show that accounting for the competitive states can be significant for a firm's decision-making, and we expect similar impact for all advertising firms. Without accounting for the actual competitive state, companies can end up losing money, and the value of experimental results is diminished. However, as mentioned earlier, the advertiser will need the support of the platform to identify the most likely states and the relevant $CATE$s. We hope that our analysis will draw attention to competitive interference and encourage platforms to take it into account when designing their experimentation services.

\end{section}

\begin{section}{Conclusions}\label{sec:conclusion}

An assessment of experimentation on platforms when multiple firms experiment in parallel is presented. The core focus is on interactions that are induced by the competition between the firms and the impact of such interactions on the measurement of treatment effects under such \textit{parallel experimentation}. The main point of the paper is that when competition between firms is significant, such as when many advertisers compete for capacity constrained ad-slots on publishers, these interactions can non-trivially affect the value of platform experimentation because the causal effects are sensitive to competitive actions. In our setting, parallel experimentation affects both the allocation of ads and their effectiveness, which, in turn, affect the relevance of the advertisers' experimental results. We discuss how a full factorial experiment can help recover more useful estimands and demonstrate our approach using novel data from large-scale field experiments conducted on \texttt{JD.com}.

Although we find a significant impact of parallel experimentation in our context, future research is needed on the generalizability of our findings and solving the problems caused by competitive interactions. The impact of parallel experimentation depends on how precisely ads are targeted and the overlap in populations targeted by competitors. An increase in a competitor's bid targeting a similar audience might affect the likelihood of an advertiser reaching its audience, which can be consequential in ways similar to what we found. However, this issue might be less significant in a setting less competitive than ours or one where ads are not as precisely targeted. Future research is needed to establish the material extent of such effects in other media. 

Our research also raises methodological issues, particularly on how platforms can estimate competitive interactions at scale. One promising direction would be to identify potentially interfering competitors that have similar and precise targeting criteria, and variability in advertising campaigns. This area requires further investigation to develop reliable estimation methods.

Another topic for future exploration concerns the potential outcomes for a platform that incorporates competitive environmental uncertainty into its reporting. If platforms could accurately report on this uncertainty, it might not only enhance their attractiveness to advertisers by improving decision-making but also minimize inefficient ad spending. However, the broader implications, such as the effect on advertising prices and the overall ad market density, remain uncertain. Understanding how advertisers respond to such information is crucial and warrants further investigation.

While the paper focuses on advertising, similar issues arise in other scenarios, such as in pricing experiments where the effect of a focal firm's price depends on the prices offered by other firms. Therefore, the ideas presented in this paper may be useful in developing experimental approaches to understanding situations with interactions across experiments more broadly. 

\end{section}

\newpage

\appendix
\renewcommand{\thesubsection}{\Alph{subsection}}
\counterwithin{table}{subsection}   
\counterwithin{figure}{subsection} 
\numberwithin{equation}{subsection}

\section*{Appendix}

\subsection{Randomization checks}\label{app:random}

\begin{figure}[htp]
    \centering
    \caption{KS test for \textit{p}-values of fractions of users in treatment groups}
    \label{fig:ks_tests}
    \begin{subfloat}[\scriptsize 70/30 split\label{fig:ks_fracs}]
        {\includegraphics[width=0.32\textwidth]{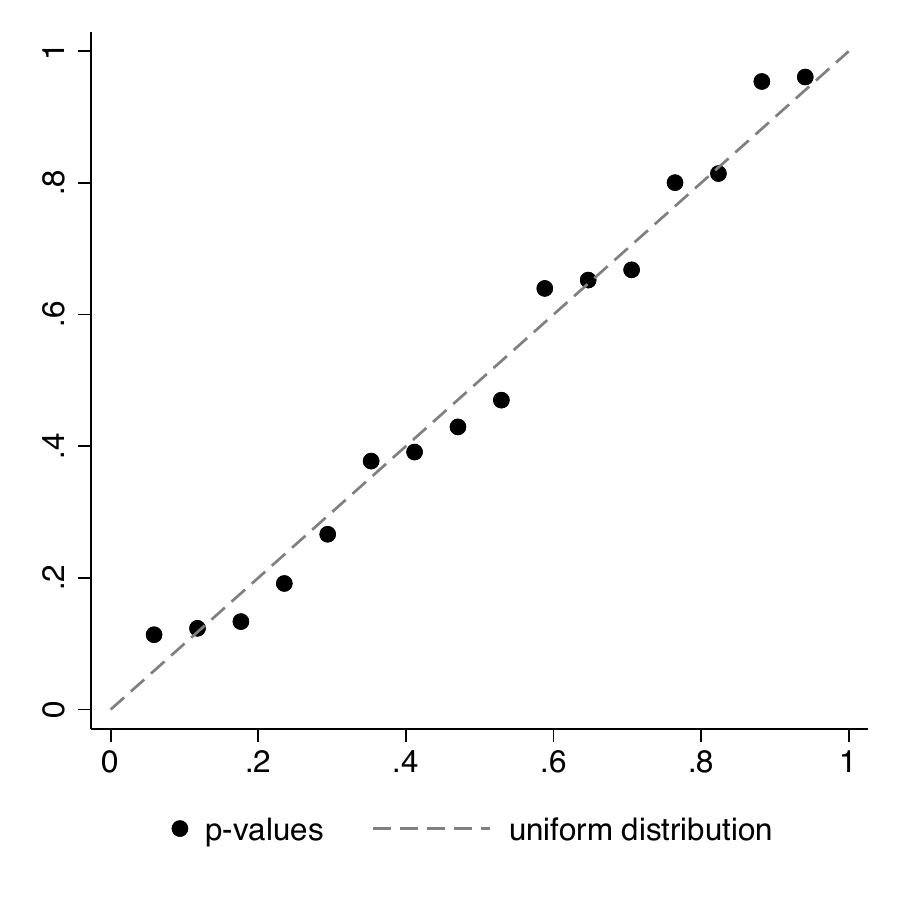}}
    \end{subfloat} 
    \begin{subfloat}[\scriptsize Differences pre-experiment\label{fig:ks_all}]
    {\includegraphics[width=0.32\textwidth]{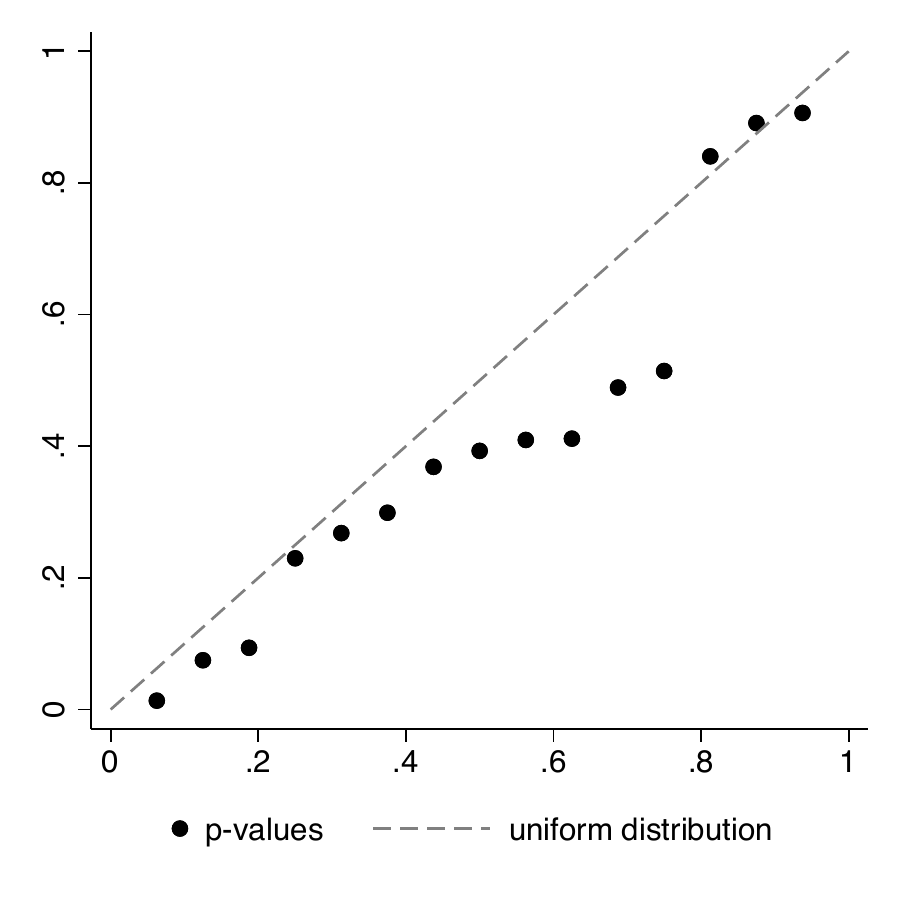}}
     \end{subfloat} 
     \begin{subfloat}[\scriptsize Uncorrelated assignments\label{fig:ks_inds}]
    {\includegraphics[width=0.32\textwidth]{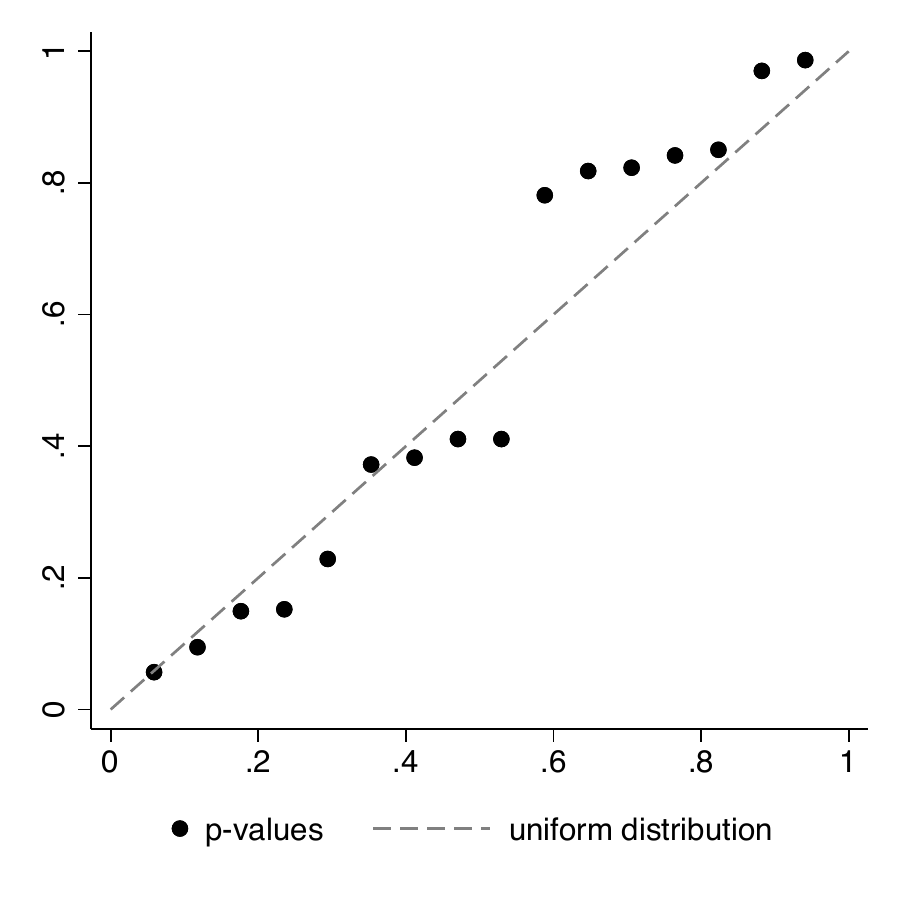}}
     \end{subfloat}
     \floatfoot{\scriptsize Note: Graphs show quantiles plots of 16, 15 and 16 \textit{p}-values, respectively, computed using standard errors robust to heteroskedasticity. The \textit{p}-values of KS tests for whether these \textit{p}-values are drawn from a standard uniform distribution are 0.986, 0.173 and 0.428, respectively.}
\end{figure}

This section shows a series of randomization checks to ensure that the experiment was implemented correctly. First, for all campaigns 70\% of users should have been assigned to the treatment group. To test whether this was the case and handle multiple testing, we perform a proportion test for each campaign, collect the associated \textit{p}-values and use a KS statistic to test whether their distribution corresponds to a standard uniform distribution, as should be the case. We display this comparison in Figure \ref{fig:ks_fracs}. The \textit{p}-value associated with this KS test for this hypothesis is 0.986, indicating that the test/control splits were correct.

We now proceed to investigate possible differences in user behavior before the experiment as a function of whether they belong to the treatment versus the control group of each campaign. If randomization was implemented correctly, there should not be significant differences across groups. To verify whether this is the case, for each campaign, we run regressions of the number of adds to cart, of orders, of visits to the product detail page, and GMV during the three days prior to the start of the experiment on the treatment assignment indicator. The regression coefficients along with standard errors robust to heteroskedasticity are shown in Table \ref{tab:random_check}. Out of the 64 coefficients, 54 are not statistically significant at the 10\% level.

\begin{table}[htp]
\resizebox{\textwidth}{!}{\begin{threeparttable}
\caption{Regressions of behavioral variables three days before experiment on treatment}
\label{tab:random_check}
\begin{tabular}{c|cccc||c|cccc}
    \hline \hline 
Campaign & \# of adds to cart & GMV & \# of orders & \# of visits & Campaign & \# of adds to cart & GMV & \# of orders & \# of visits   \\
\hline
1  & -0.0108  & -0.1035  & 0.0008 & -0.0121 & 9 & 0.0003 & 0.1096 & 0.0001 & -0.0030 \\
($n=128,576$) & (0.0073) & (0.3025) & (0.0014) &  (0.0180) & ($n=539,702$) & (0.0007) & (0.2891) & (0.0001) & (0.0080) \\ 
2 & 0.0002 & -0.0007 & 0.0002 & 0.0001 & 10 & 0.0025 & 0.6120 & 0.0001 & 0.0226 \\
($n=535,343$) & (0.0004) & (0.0046) & (0.0001) & (0.0012) & ($n=66,824$) & (0.0014)* & (0.4016) & (0.0001) & (0.0209) \\ 
3 & 0.0094 & 0.3232 & 0.0021 & 0.0144 & 11 & 0.0007 & -0.0118 & 0.0004 & 0.0041 \\
($n=12,809$) & (0.0059) & (0.2576) & (0.0036) &  (0.0349) & ($n=130,071$) & (0.0009) & (0.0254) & (0.0007) & (0.0046) \\ 
4 & 0.0001 & 0.0010 & 0.0001 & 0.0128 & 12 & -0.0047 & -0.5814 & 0.0009 & -0.0099 \\
($n=14,846$) & (0.0001) & (0.0010) & (0.0001) & (0.0071)* & ($n=14,806$) & (0.0137) & (0.9264) & (0.0052) & (0.0855) \\ 
5 & -0.0006 & -0.0777  & -0.0003 & 0.0024 & 13 & 0.0059 & 1.9995 & 0.0005 & 0.0357 \\
($n=179,516$) & (0.0012) & (0.1393) & (0.0006)  & (0.0052) & ($n=111,307$) & (0.0034)* & (1.6644) & (0.0004) & (0.0856) \\ 
6 & 0.0013 & 0.0131 & 0.0009 & 0.0028 & 14 & 0.0239 & 0.3669 & 0.0032 & 0.0459 \\
($n=184,689$)  & (0.0006)** & (0.0069)* & (0.0004)** & (0.0015)* & ($n=20,907$) & (0.0137)* & (0.4281) & (0.0043) & (0.0660) \\ 
7 & 0.0003 & -0.1681 & -0.0001 & 0.0026 & 15 & 0.0168 & 1.4439 & 0.1330 & -0.0294 \\
($n=302,471$)  & (0.0005) & (0.3101) & (0.0002) & (0.0071) & ($n=66,984$) & (0.0104) & (0.6600)** & (0.0663)*** & (0.0270) \\ 
8 & -0.0068  & -0.1054 & -0.0001 & -0.0804 & 16 &-0.0001 & 0.0025 & 0.0002  & 0.0004 \\
($n=131,743$)  & (0.0044) & (2.0914) & (0.0003) & (0.0564) & ($n=339,018$) & (0.0003) & (0.0018) & (0.0001) & (0.0008) \\ 
 \hline \hline
    \end{tabular}
\begin{tablenotes}
      \normalsize
	Note: Each cell contains the coefficient associated with the treatment indicator from a linear regression.
		The unit of observation is a user. Dependent variables are events three days prior to the experiment. Standard errors robust to heteroskedasticity are shown between parentheses. \\
	*** \textit{p}<0.01, ** \textit{p}<0.05, * \textit{p}<0.1
    \end{tablenotes}
  \end{threeparttable}}
\end{table} 

To account for multiple testing, we estimate a SUR model for each campaign, gather the \textit{p}-values associated with a test for whether all coefficients were equal to zero, and test whether they correspond to draws from a standard uniform distribution using a KS test.\footnote{We only obtain 15 \textit{p}-values because the variance matrix associated with the SUR estimates for campaign 4 was not invertible.} Results are shown in Figure \ref{fig:ks_all}. The \textit{p}-value associated with the KS test is 0.173, so that we cannot reject the null hypothesis that the \textit{p}-values follow a standard uniform distribution at the usual significance level.

Finally, we assess whether users were indeed allocated to treatment groups independently across campaigns. To do so, for each campaign we run a regression of an indicator for whether the user is in its treatment group on an indicator for whether the user is in its main competitor's treatment group as determined in Table 2. We then collect all 16 \textit{p}-values, computed using standard errors robust to heteroskedasticity, and perform a KS test for whether they follow a standard uniform distribution, which should be the case if the treatment assignments are uncorrelated. Results are displayed in Figure \ref{fig:ks_inds}. The \textit{p}-value associated with this KS test is 0.428, indicating that users were indeed allocated to treatment groups independently across campaigns.

\subsection{Alternative estimators}\label{app:ests}

As we noted, our estimands of interest are conditional average treatment effect parameters, so any method that can consistently estimate conditional expectations is applicable in our setting. The use of nonparametric and kernel-based methods to estimate treatment effect parameters is not novel, as attested by \cite{imbens2004}. These methods are often employed to treat continuous variables nonparametrically, which could be incorporated into our framework by including user- and/or target audience-level variables to the analysis. 

Traditional kernel-based methods applied to continuous variables in conjunction with tools that smooth over discrete variables, such as \cite{lor2013}, have also been applied. For example, \cite{lrw2009} smooth over both continuous and categorical variables to estimate the propensity score and then use it to provide an asymptotically efficient estimator of the $ATE$. However, unlike us they do not address estimation of $CATE$s. In a setup where these are the objects of interest and where there are continuous variables over which the econometrician needs to smooth, the approach proposed by \cite{rl2004}, which allows for smoothing over categorical and continuous variables, can be followed. However, the rate of convergence of this estimator is affected by the bandwidths associated with the continuous variables, making it slower.

While the kernel-based estimator we propose has several attractive properties, it also has shortcomings. As noted in page 555 of \cite{lor2013}, situations in which the number of variables that are smoothed increases with the sample size are not considered. This is especially relevant with user- and target audience-level variables, which can be numerous, and when many advertisers experiment in parallel. Thus, alternative machine learning methods can become an attractive option. 

Given the linear structure of our estimation problem, as shown in equation (5), methods that rely on or apply to such structure are especially attractive. For binary dependent variables, \cite{ir2013} provide a method that combines Support Vector Machines with separate LASSO constraints to perform regularization and variable selection, which would be especially appropriate if the outcome of interest was an indicator for purchase, for example. In turn, \cite{tagt2014} combine LASSO constraints with a simple covariate modification; given the correct specification of the regression equation, maximum likelihood estimation can be implemented, ensuring asymptotic efficiency. 

Assuming sparsity, \cite{aiw2018} propose approximate residual balancing as a way to debias penalized regression adjustments so that methods like the LASSO and elastic net can be used to perform inference on $CATE$ parameters in high-dimensional settings. Importantly, this method does not require the propensity score to be known or estimable, even though in settings such as ours treatment assignment is randomly determined according to known probabilities. In turn, \cite{bwz2019} allow the regression equation not be sparse provided that the propensity score is while assuming that it is given by the usual logistic formula. We note that these restrictions on the propensity are usually satisfied in randomized experiments such as the ones we consider. 

Other studies propose methods to deal with potential misspecification of the outcome equation or the propensity score and achieve double robustness in the estimation of conditional average treatment effect parameters, such as \cite{tan2020} and \cite{npi2020}. Misspecification, however, is not a concern in our setting.

\subsection{Simulations}

We now present simulations to illustrate of the performance and properties of the estimator introduced by \cite{lor2013}. We further compare its performance to that of OLS and to the method proposed by \cite{yzzzz2023}.

\subsubsection{Setup}

In these simulation exercises, we consider the following model:
\begin{align}\label{eq:reg_2comps}
    Y_i=\beta_0 + \beta_1 D_f+ \beta_2 D_g + \beta_3 D_h + \beta_4 D_f D_g + \beta_5 D_f D_h + \beta_6 D_g D_h + \beta_7 D_f D_g D_h + \epsilon,
\end{align}
where: $Y$ is the outcomes of interest; the $\beta$s are the coefficients to be estimated, which we collect in a vector, $\beta$; $D_{\ell}$ is an indicator for whether the user was in advertiser $\ell$'s treatment group, in which case they were eligible to be exposed to $\ell$'s ad, and where $\ell\in\{f,g,h\}$; and $\epsilon$ is the error term. For all simulations, $\epsilon \sim N(0,1)$. 

Our objective is to estimate the effects of $f$'s advertising, and whether and how they vary depending on whether $f$'s competitors, $g$ and $h$, are advertising.

\subsubsection{Data generating processes (DGPs)}

We consider five different DGPs, that is, five different vectors $\beta$ in our simulations. The values we use are taken from the results we obtained in our application of the kernel-based estimator, which we present in Section 6.4. We consider situations where only two competitors could be experimenting, so that the remaining 13 indicators $D$ were always set to zero. We chose specific combinations to create different types of competitive advertising environments. The specific values for $\beta$ we used are given in Table \ref{tab:dgp} below.

\begin{table}[htp]
\begin{threeparttable}
\caption{Distribution of users across campaign exposures}
\label{tab:dgp}
\begin{tabular}{c|ccccc}
    \hline \hline 
\multirow{2}{*}{Coefficients} & \multicolumn{5}{c}{DGPs}   \\ \cline{2-6}
 & 1 & 2 & 3 & 4 & 5  \\ 
\hline
 $\beta_0$ & 0.09 & 0.09 & 0.09 & 0.09 & 0.09 \\
 $\beta_1$ & 0.16 & 0.16 & 0.16 & 0.16 & 0.16 \\
 $\beta_2$ & 0.00 & 0.00 & 0.03 & 0.00 & 0.00 \\
 $\beta_3$ & 0.00 & 0.05 & 0.05 & 0.05 & 0.01 \\
 $\beta_4$ & 0.00 & 0.00 & -0.13 & -0.10 & -0.10 \\
 $\beta_5$ & 0.00 & -0.15 & -0.15 & -0.15 & -0.02 \\
 $\beta_6$ & 0.00 & 0.00 & -0.07 & -0.01 & -0.01 \\
 $\beta_7$ & 0.00 & 0.00 & 0.23 & 0.09 & 0.02 \\
 \hline \hline
    \end{tabular}
  \end{threeparttable}
\end{table} 

When $f$ faces no competition, $f$'s $CATE$ is 0.16. The DGPs are such that:
\begin{itemize}
    
    \item DGP 1: $g$ and $h$ do not impact $f$'s $CATE$ either unilaterally or combined.

    \item DGP 2: unilaterally, $g$ does not impact $f$'s $CATE$ but $h$ does as the $CATE$ goes from 0.16 to 0.01. With $g$ and $h$ combined, $f$'s $CATE$ is also 0.01.

    \item DGP 3: $g$ unilaterally reduces $f$'s $CATE$ to 0.03 and $h$ to 0.01. Combined, however, the effects partially cancel each other out, and $f$'s $CATE$ becomes 0.11.

    \item DGP 4: $g$ unilaterally reduces $f$'s $CATE$ to 0.06 and $h$ to 0.01. When combined, the effects act together and $f$'s $CATE$ becomes 0.

    \item DGP 5: $g$ unilaterally reduces $f$'s $CATE$ to 0.06 and $h$ only to 0.14. When combined, $g$'s negative impact is not altered by $h$ and the $CATE$ remains 0.06.

\end{itemize}

\subsubsection{Estimators}

We consider and compare three estimators. First, we consider OLS. Second, we consider the kernel-based estimator from \cite{lor2013}. Third, we consider the estimator proposed by \cite{yzzzz2023}. Given the absence of covariates other than the treatments and the functional form implied by Assumption 1 from \cite{yzzzz2023}, their estimator collapses to nonlinear least squares (NLS) applied to the following model:
\begin{align}\label{eq:reg_2comps_ye}
\begin{split}
    Y_i&=\theta_0 + \theta_0\theta_1 D_f+ \theta_0\theta_2 D_g + \theta_0\theta_3 D_h + \theta_0 \theta_1 \theta_2 D_f D_g + \theta_0 \theta_1 \theta_3 D_f D_h \\
    &\quad\quad + \theta_0 \theta_2 \theta_3 D_g D_h + \theta_0 \theta_1 \theta_2 \theta_3 D_f D_g D_h + \eta.
\end{split}
\end{align}

This estimator can also be interpreted simply as OLS applied to (\ref{eq:reg_2comps}) under specific nonlinear constraints between the parameters. Notice that these constraints are only true under DGP 1.

\subsubsection{Data sets}

In the simulations, the $D$s are randomly assigned with equal probability independently from one another. We consider three sample sizes---80, 800, and 8,000---to assess how the performance of the estimators vis-à-vis the number of coefficients and how it changes as the sample size increases. We choose these specific values so that the number of observations for each of the eight possible full treatment assignments is the same. 

For each DGP-sample size combination, we create 10,000 data sets and implement the three aforementioned estimators on each data set. The results, which we present and detail below, are obtained from these estimates.

We separate our results in two parts. First, we show the performance of the kernel-based in detail because it is the one we use in the main manuscript. Then, we compare its performance to that of OLS and \cite{yzzzz2023}.

\subsubsection{Performance of kernel-based estimator}

\paragraph{Consistency}\quad The first property of the kernel-based estimator we verify is consistency: as the sample size grows, the estimates of the coefficients should approximate their true values more and more. Tables \ref{tab:kern1}-\ref{tab:kern5} display, for each coefficient-DGP-sample size, the average and standard deviation of the estimates over the 10,000 data sets. Expectedly, given Theorem 2 from \cite{lor2013}, the larger the sample size, the closer the estimates are to the true values of the coefficients. This patterns holds for all coefficients and DGPs.

\begin{table}[htp]
\begin{threeparttable}
\caption{Average bias and standard deviation of kernel estimates under DGP 1}
\label{tab:kern1}
\begin{tabular}{c|c|c|c|c}
    \hline \hline 
Coefficients & True Value & $n$ & mean($\hat{\beta}_{k,n}^{s}$) & SD($\hat{\beta}_{k,n}^{s}$)  \\ 
\hline 
 \multirow{3}{*}{$\beta_0$} & \multirow{3}{*}{0.09} & 80 & 0.098 & 0.229 \\
  & & 800  & 0.1 & 0.074 \\
  & & 8,000 & 0.101 & 0.025 \\
 \hline
 \multirow{3}{*}{$\beta_1$}  & \multirow{3}{*}{0.16} & 80 & 0.18 & 0.324 \\
  & & 800 & 0.181 & 0.106 \\
  & & 8,000 & 0.181 & 0.038 \\
 \hline
 \multirow{3}{*}{$\beta_2$} & \multirow{3}{*}{0} & 80 & -0.0004 & 0.222 \\
  & & 800 & 0.0007 & 0.068 \\
  & & 8,000 & -0.0002 & 0.022 \\
 \hline
 \multirow{3}{*}{$\beta_3$} & \multirow{3}{*}{0} & 80 & -0.004 & 0.216 \\
  & & 800 & 0.001 & 0.067 \\
  & & 8,000 & -0.0001 & 0.02 \\
 \hline
 \multirow{3}{*}{$\beta_4$} & \multirow{3}{*}{0} & 80 & 0.004 & 0.303 \\
  & & 800 & -0.0004 & 0.095 \\
  & & 8,000 & -0.0001 & 0.03 \\
 \hline
 \multirow{3}{*}{$\beta_5$} & \multirow{3}{*}{0} & 80 & -0.004 & 0.306 \\
  & & 800 & 0.0001 & 0.094 \\
  & & 8,000 & 0.0001 & 0.029 \\
 \hline
 \multirow{3}{*}{$\beta_6$} & \multirow{3}{*}{0} & 80 & -0.002 & 0.199 \\
  & & 800 & -0.0005 & 0.06 \\
  & & 8,000 & -0.0002 & 0.019 \\
 \hline
 \multirow{3}{*}{$\beta_7$} & \multirow{3}{*}{0} & 80 & 0.003 & 0.406 \\
  & & 800 & 0.001 & 0.124 \\
  & & 8,000 & 0.0003 & 0.039 \\
 \hline \hline
    \end{tabular}
  \end{threeparttable}
\end{table} 

\begin{table}[htp]
\begin{threeparttable}
\caption{Average bias and standard deviation of kernel estimates under DGP 2}
\label{tab:kern2}
\begin{tabular}{c|c|c|c|c}
    \hline \hline 
Coefficients & True Value & $n$ & mean($\hat{\beta}_{k,n}^{s}$) & SD($\hat{\beta}_{k,n}^{s}$)  \\ 
\hline 
 \multirow{3}{*}{$\beta_0$} & \multirow{3}{*}{0.09} & 80 & 0.113 & 0.23  \\
  & & 800  & 0.117 & 0.078  \\
  & & 8,000 & 0.108 & 0.029  \\
 \hline
 \multirow{3}{*}{$\beta_1$}  & \multirow{3}{*}{0.16} & 80 & 0.134 & 0.329 \\
  & & 800 & 0.143 & 0.118  \\
  & & 8,000 & 0.17 & 0.04  \\
 \hline
 \multirow{3}{*}{$\beta_2$}  & \multirow{3}{*}{0} & 80 & -0.0005 & 0.222 \\
  & & 800 & 0.001 & 0.069 \\
  & & 8,000 & -0.0002 & 0.021  \\
 \hline
 \multirow{3}{*}{$\beta_3$}  & \multirow{3}{*}{0.05} & 80 & 0.029 & 0.222  \\
  & & 800 & 0.03 & 0.082  \\
  & & 8,000 & 0.048 & 0.034  \\
 \hline
 \multirow{3}{*}{$\beta_4$}  & \multirow{3}{*}{0} & 80 & 0.003 & 0.303  \\
  & & 800 & -0.001 & 0.097  \\
  & & 8,000 & -0.0001 & 0.031 \\
 \hline
 \multirow{3}{*}{$\beta_5$}  & \multirow{3}{*}{-0.15} & 80 & -0.08 & 0.324  \\
  & & 800 & -0.088 & 0.135  \\
  & & 8,000 & -0.144 & 0.057  \\
 \hline
 \multirow{3}{*}{$\beta_6$}  & \multirow{3}{*}{0} & 80 & -0.001 & 0.202  \\
  & & 800 & -0.001 & 0.066  \\
  & & 8,000 & -0.0003 & 0.026  \\
 \hline
 \multirow{3}{*}{$\beta_7$}  & \multirow{3}{*}{0} & 80 & 0.029 & 0.413 \\
  & & 800 & 0.031 & 0.141 \\
  & & 8,000 & 0.048 & 0.057  \\
 \hline \hline
    \end{tabular}
  \end{threeparttable}
\end{table} 

\begin{table}[htp]
\begin{threeparttable}
\caption{Average bias and standard deviation of kernel estimates under DGP 3}
\label{tab:kern3}
\begin{tabular}{c|c|c|c|c}
    \hline \hline 
Coefficients & True Value & $n$ & mean($\hat{\beta}_{k,n}^{s}$) & SD($\hat{\beta}_{k,n}^{s}$)  \\ 
\hline 
 \multirow{3}{*}{$\beta_0$}  & \multirow{3}{*}{0.09} & 80 & 0.118 & 0.23  \\
  & & 800  & 0.121 &  0.078 \\
  & & 8,000 & 0.118 & 0.033 \\
 \hline
 \multirow{3}{*}{$\beta_1$}  & \multirow{3}{*}{0.16} & 80 & 0.11 & 0.328 \\
  & & 800 & 0.113 & 0.116  \\
  & & 8,000 & 0.152 & 0.053  \\
 \hline
 \multirow{3}{*}{$\beta_2$}  & \multirow{3}{*}{0.03} & 80 & 0.003 & 0.225 \\
  & & 800 & 0.005 & 0.076 \\
  & & 8,000 & 0.018 & 0.039 \\
 \hline
 \multirow{3}{*}{$\beta_3$}  & \multirow{3}{*}{0.05} & 80 & 0.017 & 0.22 \\
  & & 800 & 0.016 & 0.077 \\
  & & 8,000 & 0.036 & 0.041  \\
 \hline
 \multirow{3}{*}{$\beta_4$}  & \multirow{3}{*}{-0.13} & 80 & -0.023 & 0.311  \\
  & & 800 & -0.032 & 0.119 \\
  & & 8,000 & -0.086 & 0.073 \\
 \hline
 \multirow{3}{*}{$\beta_5$}  & \multirow{3}{*}{-0.15} & 80 & -0.041 & 0.316 \\
  & & 800 & -0.042 & 0.121 \\
  & & 8,000 & -0.104 & 0.078 \\
 \hline
 \multirow{3}{*}{$\beta_6$}  & \multirow{3}{*}{-0.07} & 80 & -0.013 & 0.206 \\
  & & 800 & -0.015 & 0.078 \\
  & & 8,000 & -0.045 & 0.053  \\
 \hline
 \multirow{3}{*}{$\beta_7$}  &\multirow{3}{*}{0.23} & 80 & 0.055 & 0.431 \\
  & & 800 & 0.064 & 0.188 \\
  & & 8,000 & 0.183 & 0.141 \\
 \hline \hline
    \end{tabular}
  \end{threeparttable}
\end{table}

\begin{table}[htp]
\begin{threeparttable}
\caption{Average bias and standard deviation of kernel estimates under DGP 4}
\label{tab:kern4}
\begin{tabular}{c|c|c|c|c}
    \hline \hline 
Coefficients & True Value & $n$ & mean($\hat{\beta}_{k,n}^{s}$) & SD($\hat{\beta}_{k,n}^{s}$)  \\ 
\hline 
 \multirow{3}{*}{$\beta_0$}  & \multirow{3}{*}{0.09} & 80 & 0.113 & 0.23 \\
  & & 800  & 0.117  & 0.078  \\
  & & 8,000 & 0.121 & 0.033 \\
 \hline
 \multirow{3}{*}{$\beta_1$}  & \multirow{3}{*}{0.16} & 80 & 0.108 & 0.329 \\
  & & 800 & 0.113 & 0.118 \\
  & & 8,000 & 0.152 & 0.052 \\
 \hline
 \multirow{3}{*}{$\beta_2$}  & \multirow{3}{*}{0} & 80 & -0.002 & 0.223 \\
  & & 800 & -0.001 & 0.073 \\
  & & 8,000 & -0.001 & 0.033  \\
 \hline
 \multirow{3}{*}{$\beta_3$}  & \multirow{3}{*}{0.05} & 80 & 0.027 & 0.221  \\
  & & 800 & 0.026 & 0.079 \\
  & & 8,000 & 0.042 & 0.039 \\
 \hline
 \multirow{3}{*}{$\beta_4$} & \multirow{3}{*}{-0.1} & 80 & -0.031 & 0.309  \\
  & & 800 & -0.037 & 0.112 \\
  & & 8,000 & -0.067 & 0.061 \\
 \hline
 \multirow{3}{*}{$\beta_5$} & \multirow{3}{*}{-0.15} & 80 & -0.064 & 0.319 \\
  & & 800 & -0.065 & 0.125 \\
  & & 8,000 & -0.116 & 0.072 \\
 \hline
 \multirow{3}{*}{$\beta_6$} & \multirow{3}{*}{-0.01} & 80 & -0.003 & 0.203 \\
  & & 800 & -0.002 & 0.067 \\
  & & 8,000 & -0.005 & 0.038 \\
 \hline
 \multirow{3}{*}{$\beta_7$} & \multirow{3}{*}{0.09} & 80 & 0.041 & 0.416 \\
  & & 800 & 0.044 & 0.15 \\
  & & 8,000 & 0.09 &  0.09 \\
 \hline \hline
    \end{tabular}
  \end{threeparttable}
\end{table}

\begin{table}[htp]
\begin{threeparttable}
\caption{Average bias and standard deviation of kernel estimates under DGP 5}
\label{tab:kern5}
\begin{tabular}{c|c|c|c|c}
    \hline \hline 
Coefficients & True Value & $n$ & mean($\hat{\beta}_{k,n}^{s}$) & SD($\hat{\beta}_{k,n}^{s}$)  \\ 
\hline 
 \multirow{3}{*}{$\beta_0$} & \multirow{3}{*}{0.09} & 80 & 0.1 & 0.23 \\
  & & 800  & 0.105 & 0.077 \\
  & & 8,000 & 0.108 & 0.027 \\
 \hline
 \multirow{3}{*}{$\beta_1$} & \multirow{3}{*}{0.16} & 80 & 0.144 & 0.327 \\
  & & 800 & 0.149 & 0.113 \\
  & & 8,000 & 0.168 & 0.041 \\
 \hline
 \multirow{3}{*}{$\beta_2$} & \multirow{3}{*}{0} & 80 & -0.002 & 0.224 \\
  & & 800 & -0.001 & 0.076 \\
  & & 8,000 & -0.003 & 0.033 \\
 \hline
 \multirow{3}{*}{$\beta_3$} & \multirow{3}{*}{0.01} & 80 & 0.007 & 0.217 \\
  & & 800 & 0.004 & 0.069 \\
  & & 8,000 & 0.004 & 0.022 \\
 \hline
 \multirow{3}{*}{$\beta_4$} & \multirow{3}{*}{-0.1} & 80 & -0.042 & 0.311 \\
  & & 800 & -0.052 & 0.118 \\
  & & 8,000 & -0.085 & 0.054 \\
 \hline
 \multirow{3}{*}{$\beta_5$} & \multirow{3}{*}{-0.02} & 80 & -0.011 & 0.307 \\
  & & 800 & -0.006 & 0.096 \\
  & & 8,000 & -0.007 & 0.032 \\
 \hline
 \multirow{3}{*}{$\beta_6$} & \multirow{3}{*}{-0.01} & 80 & -0.003 & 0.2 \\
  & & 800 & -0.002 & 0.065 \\
  & & 8,000 & -0.003 & 0.026 \\
 \hline
 \multirow{3}{*}{$\beta_7$} & \multirow{3}{*}{0.02} & 80 & 0.009 & 0.408 \\
  & & 800 & 0.008 & 0.132 \\
  & & 8,000 & 0.01 & 0.052 \\
 \hline \hline
    \end{tabular}
  \end{threeparttable}
\end{table}

\paragraph{Bandwidths} \quad We now assess the behavior of the bandwidths. As explained in \cite{lor2013}, whenever a variable is irrelevant, the bandwidth associated with it should be one; otherwise, the bandwidth should converge to zero as the sample size increases. Thus, we should observe the following:
\begin{itemize}
    
    \item DGP 1: both bandwidths should be one.

    \item DGP 2: the bandwidths associated with $g$ and $h$ should be one and zero, respectively. 

    \item DGP 3: both bandwidths should be zero.

    \item DGP 4: both bandwidths should be zero.

    \item DGP 5: the bandwidths associated with $g$ and $h$ should be zero and one, respectively. 

\end{itemize}

\begin{figure}[H]
	\centering
	\caption{Bandwidths for DGP 1}
	\subfloat
	{\includegraphics[width=0.7\textwidth,page=1] {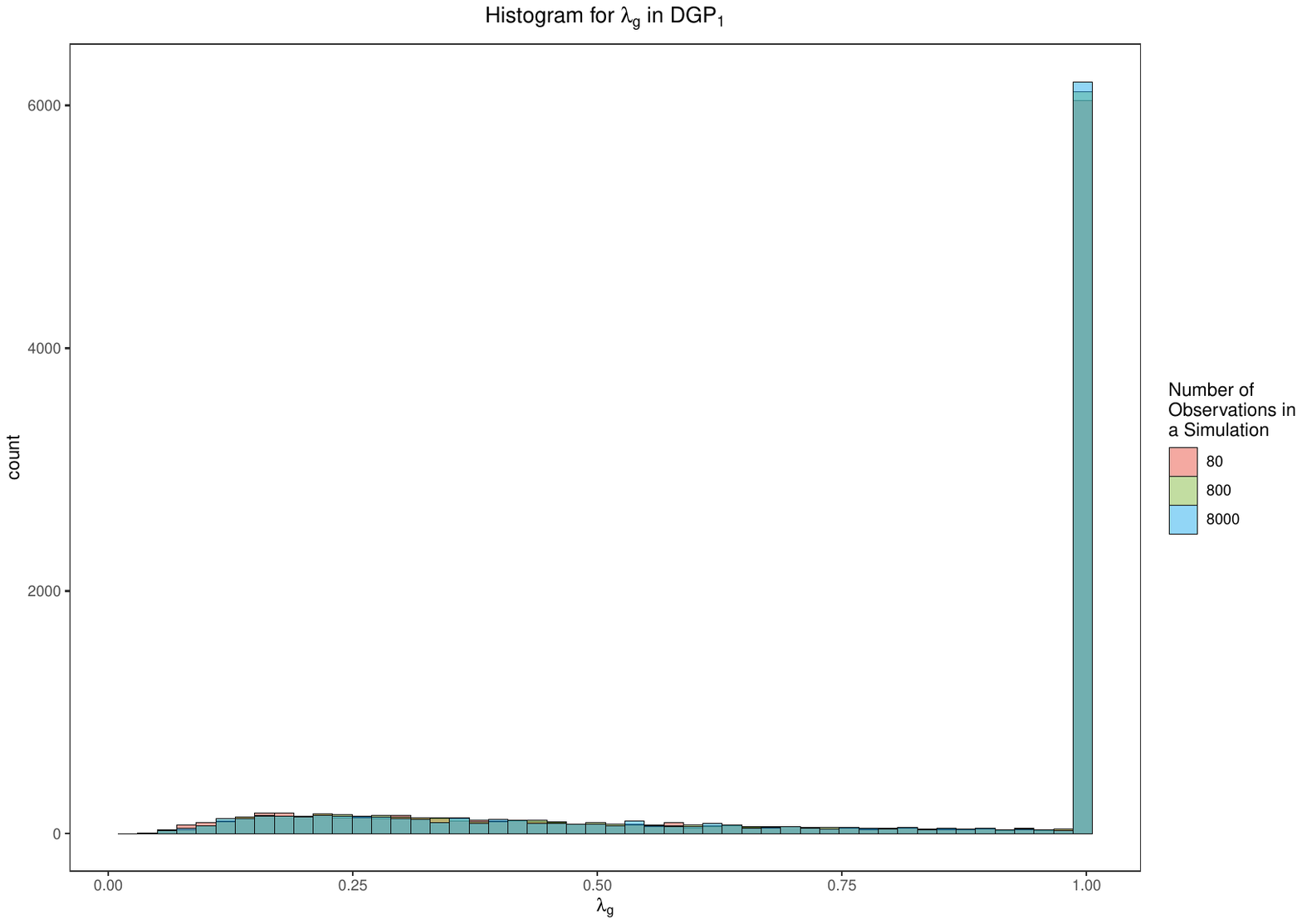}} \\
 \subfloat
	{\includegraphics[width=0.7\textwidth,page=2] {graphs_bandwidths.pdf}} 
	\label{fig:bws1}
\end{figure}

\begin{figure}[H]
	\centering
	\caption{Bandwidths for DGP 2}
	\subfloat
	{\includegraphics[width=0.7\textwidth,page=3] {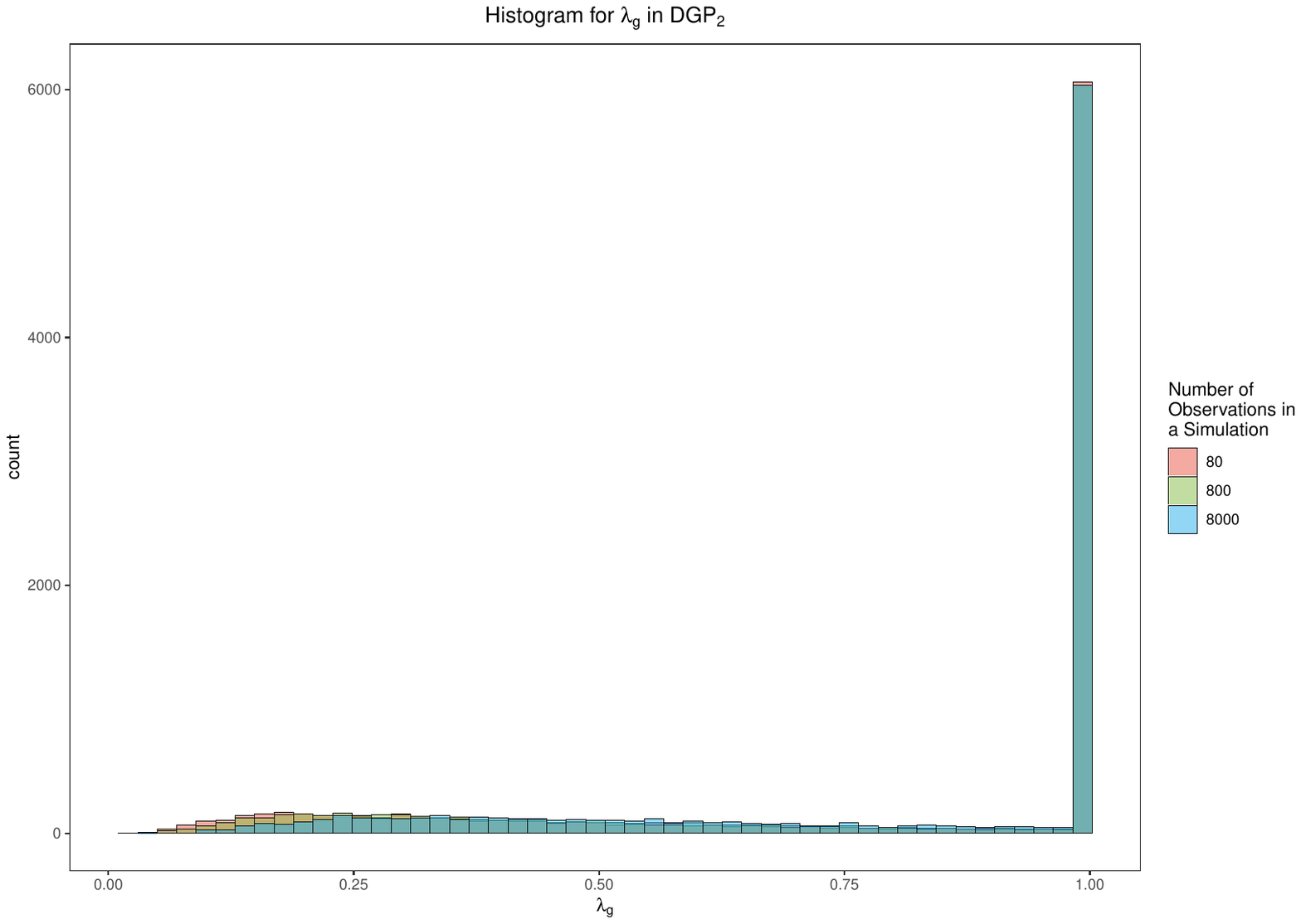}} \\
 \subfloat
	{\includegraphics[width=0.7\textwidth,page=4] {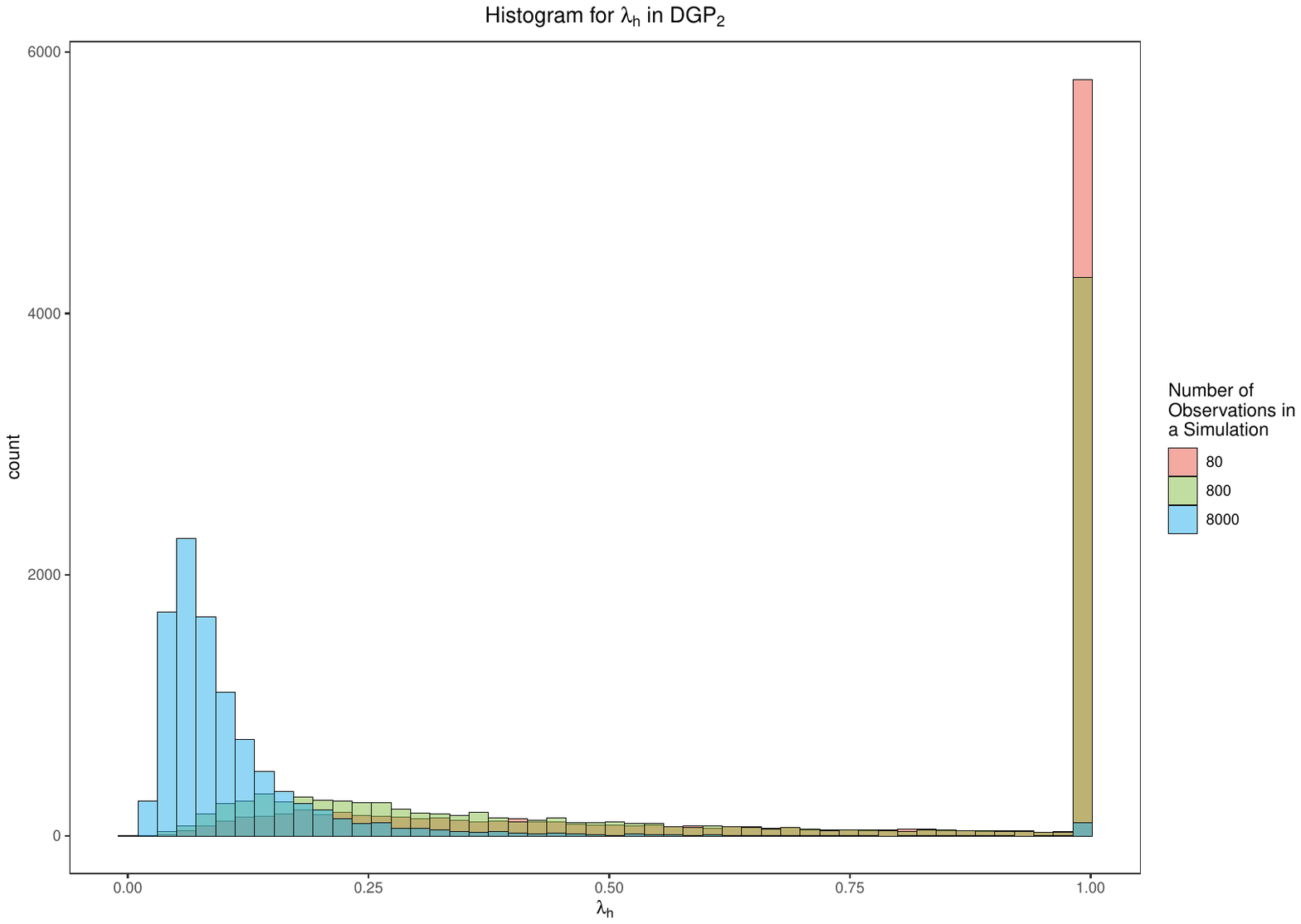}} 
	\label{fig:bws2}
\end{figure}

\begin{figure}[H]
	\centering
	\caption{Bandwidths for DGP 3}
	\subfloat
	{\includegraphics[width=0.7\textwidth,page=5] {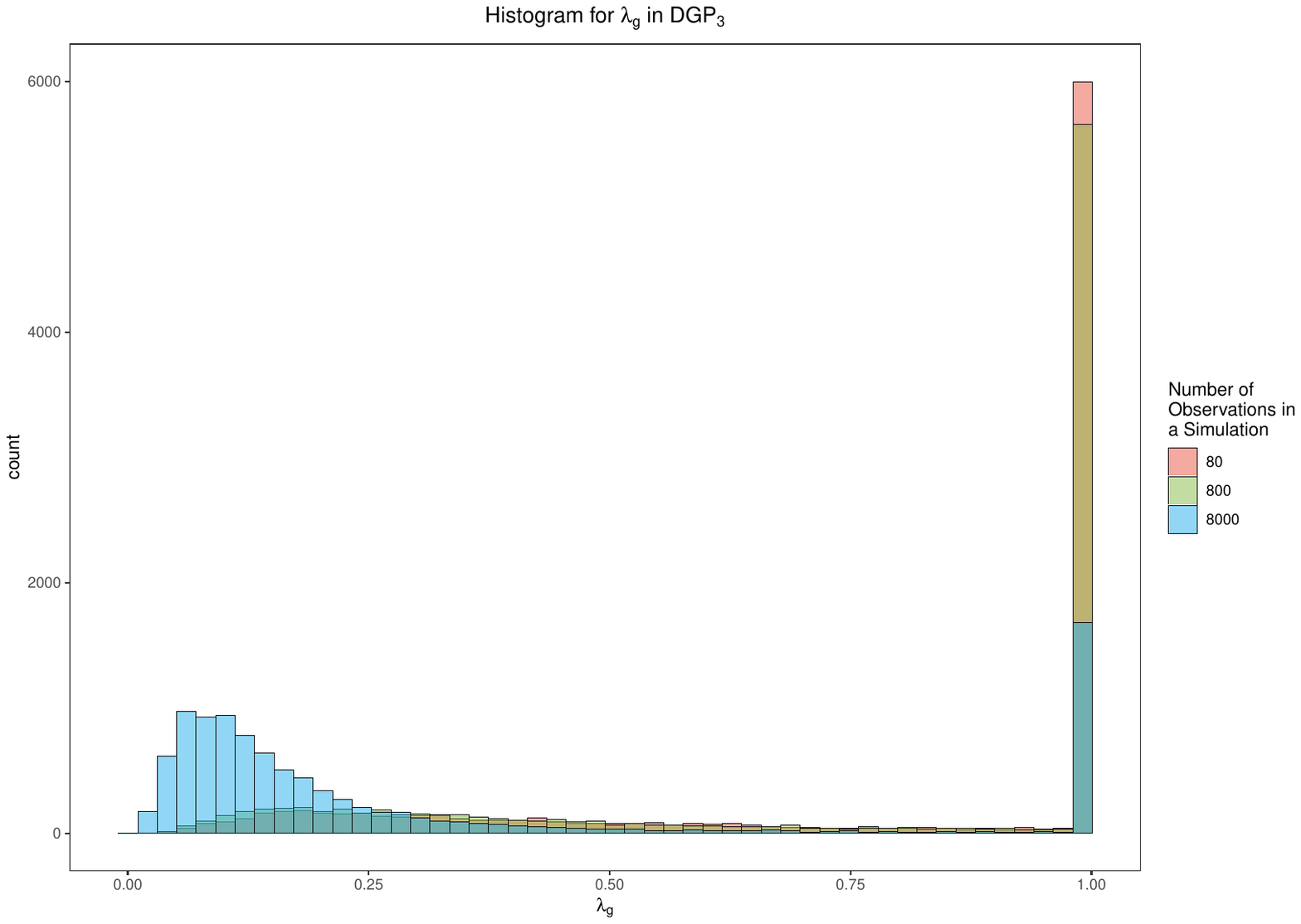}} \\
 \subfloat
	{\includegraphics[width=0.7\textwidth,page=6] {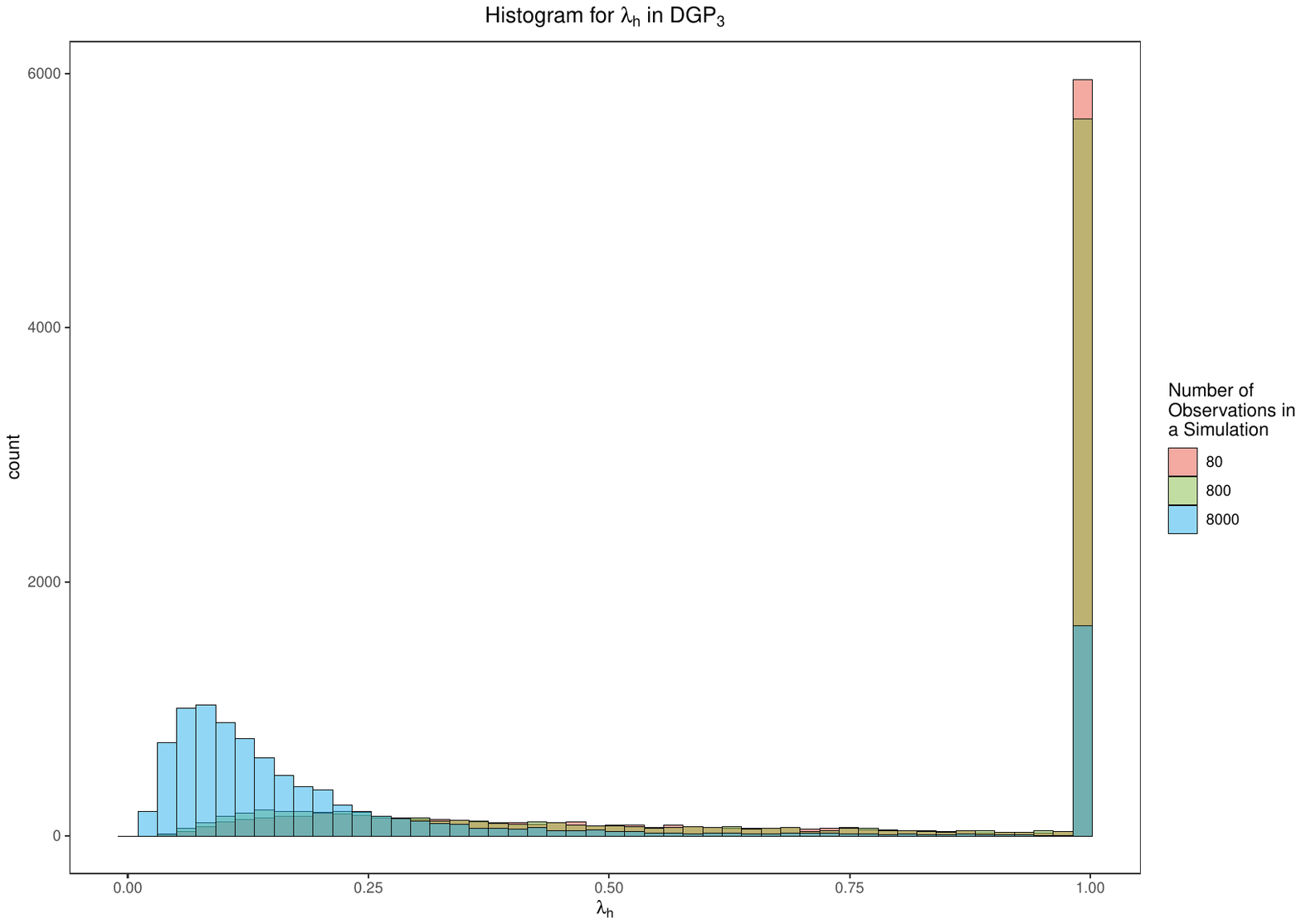}} 
	\label{fig:bws3}
\end{figure}

\begin{figure}[H]
	\centering
	\caption{Bandwidths for DGP 4}
	\subfloat
	{\includegraphics[width=0.7\textwidth,page=7] {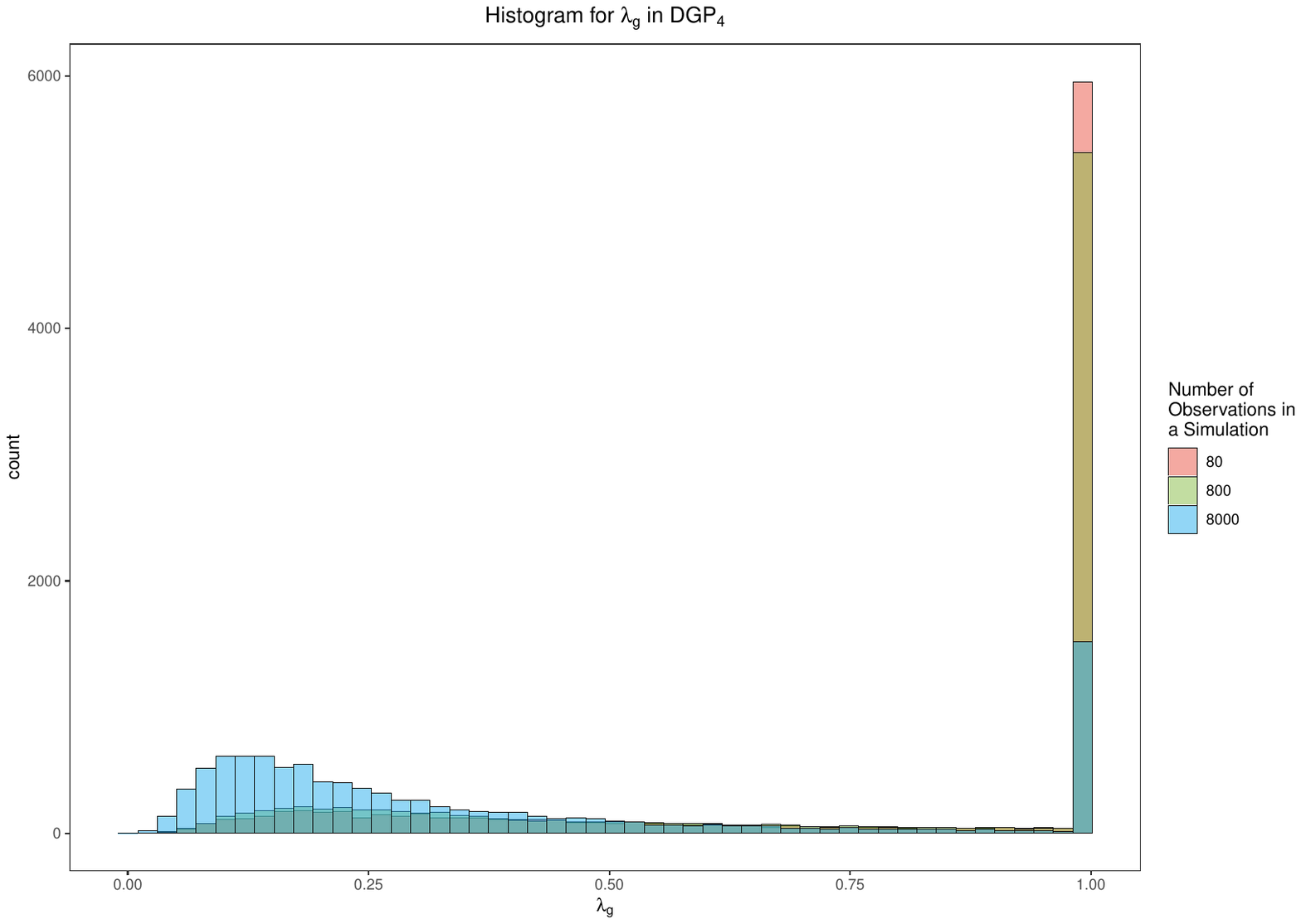}} \\
 \subfloat
	{\includegraphics[width=0.7\textwidth,page=8] {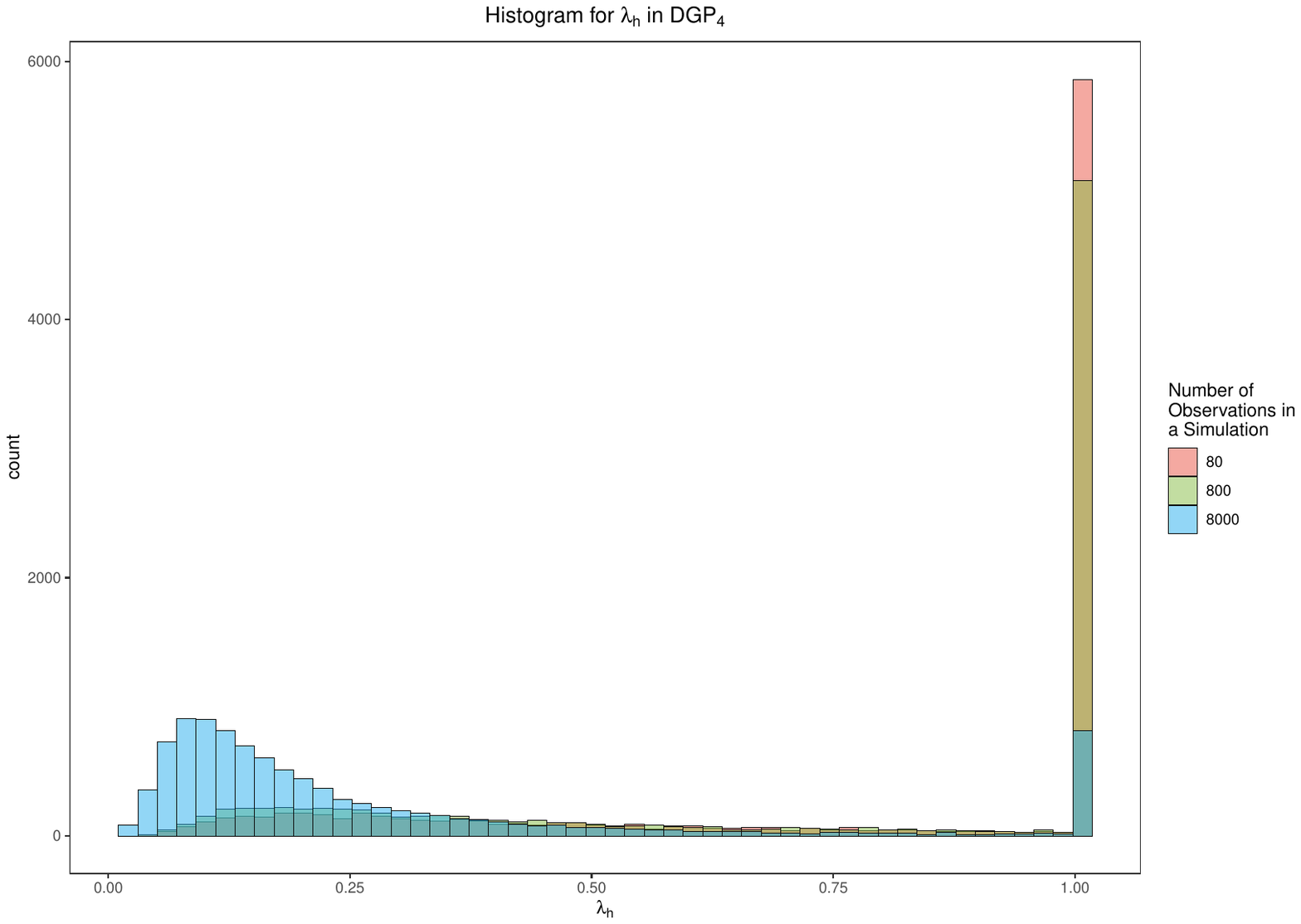}} 
	\label{fig:bws4}
\end{figure}

\begin{figure}[H]
	\centering
	\caption{Bandwidths for DGP 5}
	\subfloat
	{\includegraphics[width=0.7\textwidth,page=9] {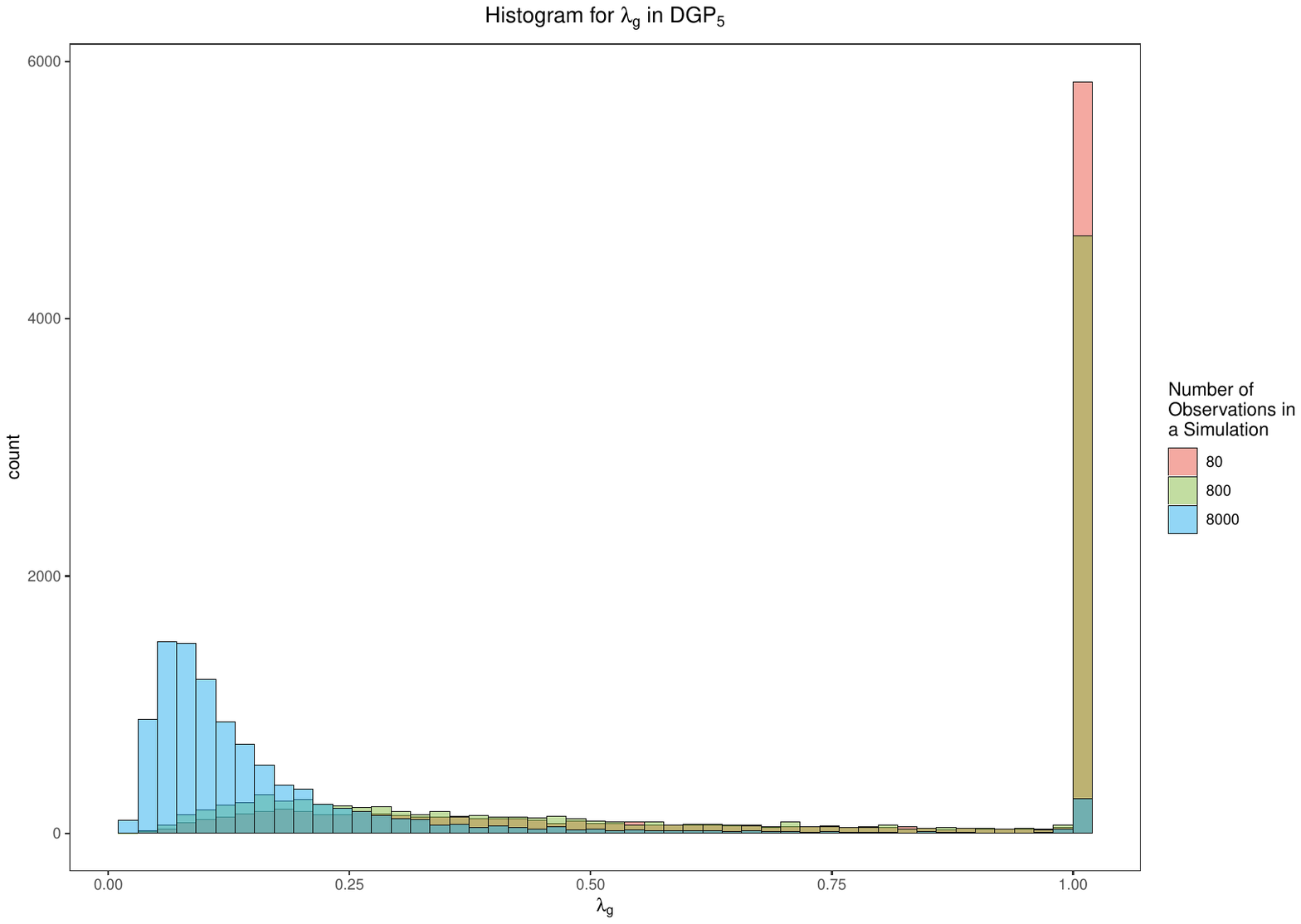}} \\
 \subfloat
	{\includegraphics[width=0.7\textwidth,page=10] {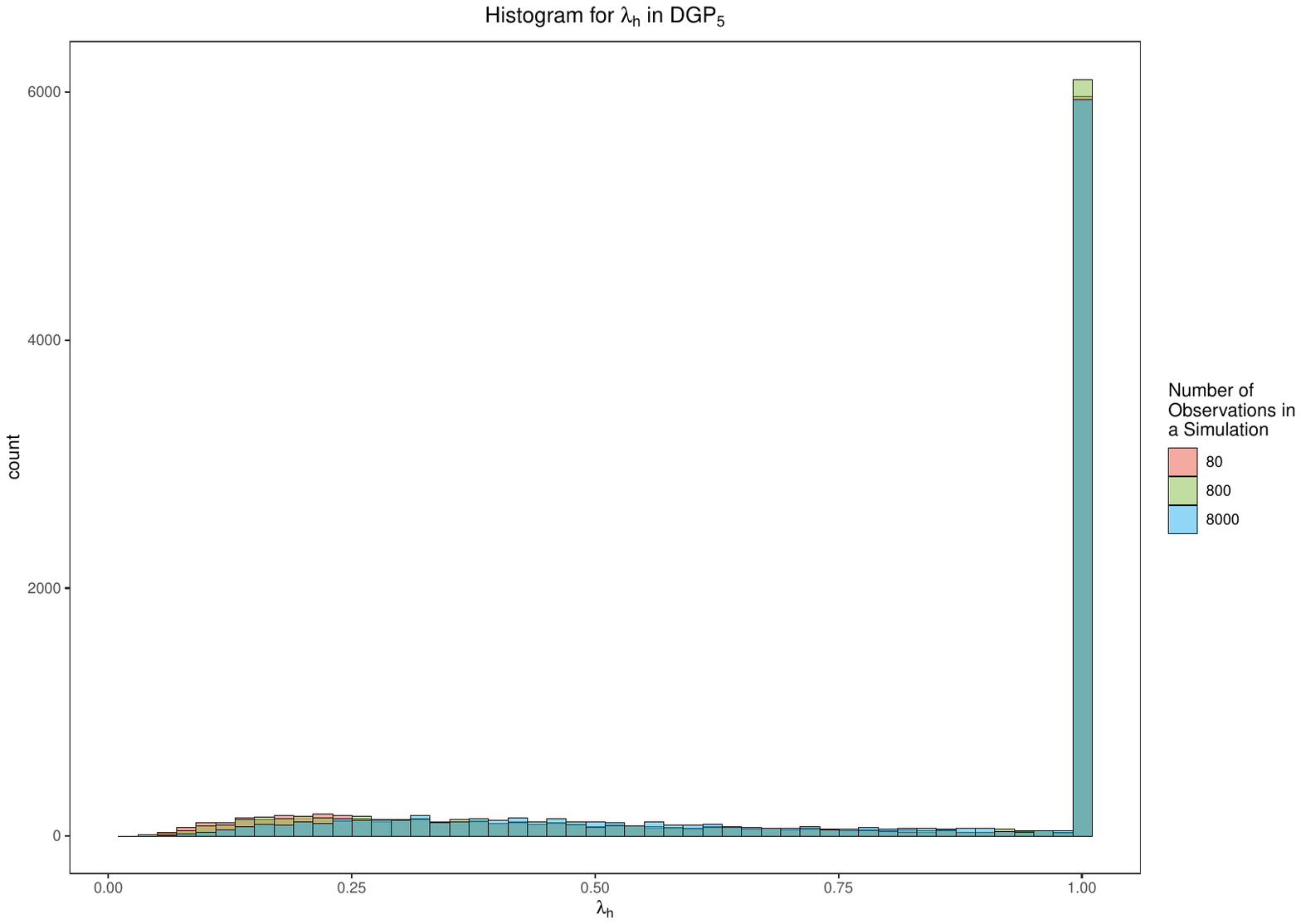}} 
	\label{fig:bws5}
\end{figure}

Figures \ref{fig:bws1}-\ref{fig:bws5} display the bandwidths across the 10,000 simulations for all DGPs, highlighting the ones from data sets with 80 observations in red, with 800 in green, and with 8,000 in blue. We verify that all expected patterns hold.

Furthermore, we observe that it is not straightforward for the estimator to pick up on the interactions. Notably, when the sample size is just 80 it is very difficult for the estimator to detect the interactions as the majority of bandwidths equal one even when these interactions are significant. This is an important because with only 80 observations there are only 10 per full treatment assignment, a number that is comparable to what our experiments features. This reinforces our confidence that the interactions we were able to detect are indeed significant. 

Notice that these results also illustrate the necessity to observe all treatment combinations in the data. To see this, consider DGPs 3 and 4. Under DGP 3, $g$ and $h$ unilaterally reduce $f$'s treatment effect from 0.16 to 0.03 and 0.01, respectively. Together, however, these effects seem to cancel each other out and $f$'s treatment effect falls to 0.11 only. In turn, under DGP 4, $g$ and $h$ unilaterally reduce $f$'s treatment effect to 0.06 and 0.01, respectively, and together they reduce it to zero. Hence, while $h$ always interferes with $f$'s advertising, its interference is somewhat mitigated under DGP 3 relative to DGP 4, so we would expect $\lambda_h$ to have lower values under DGP 4. The results confirm this intuition. 

Now suppose that a partial factorial design was used instead and that this design did not consider the case where $D_g=D_h=1$. Under both DGPs 3 and 4, $h$'s unilateral effect on $f$'s treatment effect is to reduce it from 0.16 to 0.01. Consequently, using the kernel-based estimator on such data would yield the same value for $\lambda_h$. This would potentially lead to a misleading extrapolation of the effect on $f$'s treatment effect when both $g$ and $h$ were present, which highlights that to obtain estimates that account for the different treatment interactions it is necessary to observe all treatment combinations in the data.

\subsubsection{Comparison between methods}

\paragraph{Criteria of comparison} \quad We compare the performance of the three estimators using three criteria: bias, mean squared error (MSE), and median absolute error (MAE).

We compute MSE as follows. Let $\hat{\beta}_{k,n}^{s,m}$ be the estimate of the $k-$th coefficient from sample $s$ of size $n$ using method $m$. We compute the MSE as:
\begin{align}\label{eq:mse}
    \text{MSE}_{k,n}^m = \frac{1}{10,000}\sum_{s=1}^{10,000} \left (\hat{\beta}_{k,n}^{s,m} - \beta_k \right)^2.
\end{align}

To assess the bias-variance tradeoff, which is especially important because of the regularization used by the kernel-based estimator, we assess how much of the MSE is comprised by bias by considering the ratio of bias-squared to MSE where bias squared is:
\begin{align}\label{eq:bias_sq}
    \left (\text{bias}_{k,n}^m \right )^2 = \left [ \frac{1}{10,000}\sum_{s=1}^{10,000} \left (\hat{\beta}_{k,n}^{s,m} - \beta_k \right) \right ]^2.
\end{align}

Finally, we also consider the median absolute error (MAE). The absolute error of the estimate of the $k-$th coefficient from sample $s$ of size $n$ using method $m$ is $\left |\hat{\beta}_{k,n}^{s,m} - \beta_k \right|$. Using order statistics notation, let the $s$-th highest absolute error be $\left |\hat{\beta}_{k,n}^{(s),m} - \beta_k \right|$. Then:
\begin{align}\label{eq:mae}
    \text{MAE}_{k,n}^m =  \frac{\left |\hat{\beta}_{k,n}^{(5,000),m} - \beta_k \right| + \left |\hat{\beta}_{k,n}^{(5,001),m} - \beta_k \right|}{2}.
\end{align}

\begin{figure}[H]
	\centering
	\caption{MSEs under DGP 1: $\beta_0-\beta_3$}
	\includegraphics[width=0.87\textwidth,page=1] {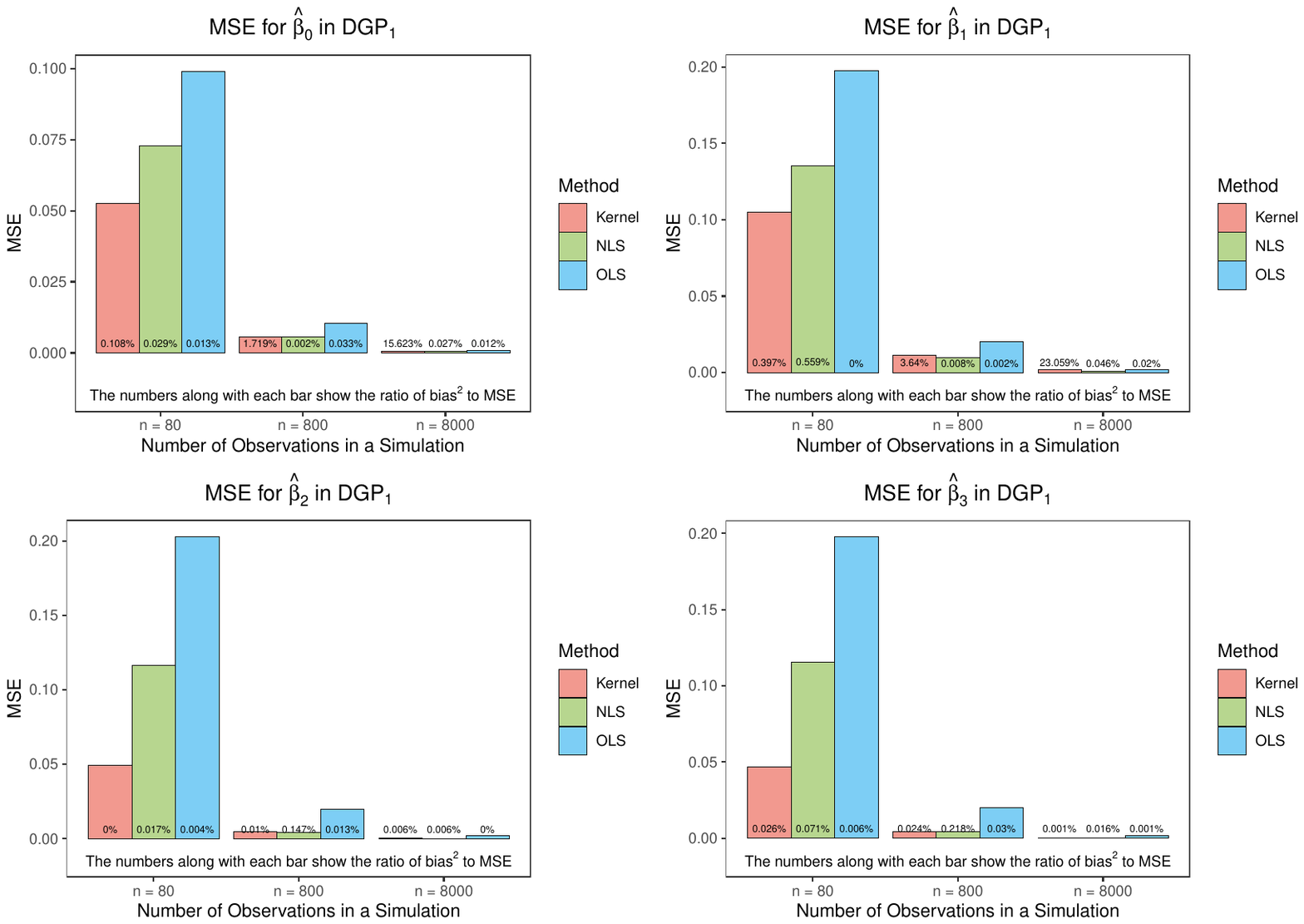}
	\label{fig:mse1a}
\end{figure}

\begin{figure}[H]
	\centering
	\caption{MSEs under DGP 1: $\beta_4-\beta_7$}
	\includegraphics[width=0.87\textwidth,page=2] {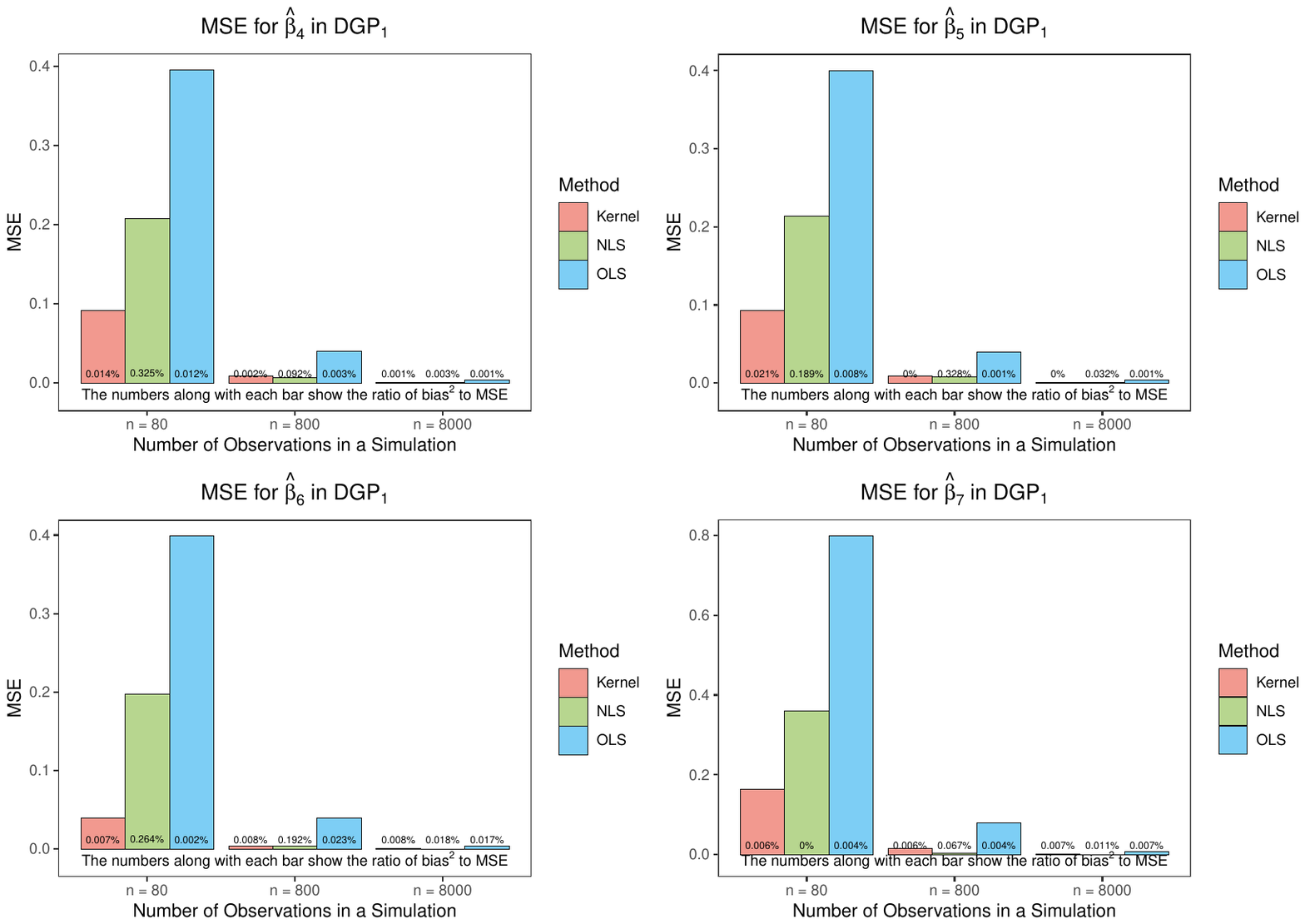}
	\label{fig:mse1b}
\end{figure}

\begin{figure}[H]
	\centering
	\caption{MSEs under DGP 2: $\beta_0-\beta_3$}
	\includegraphics[width=0.87\textwidth,page=3] {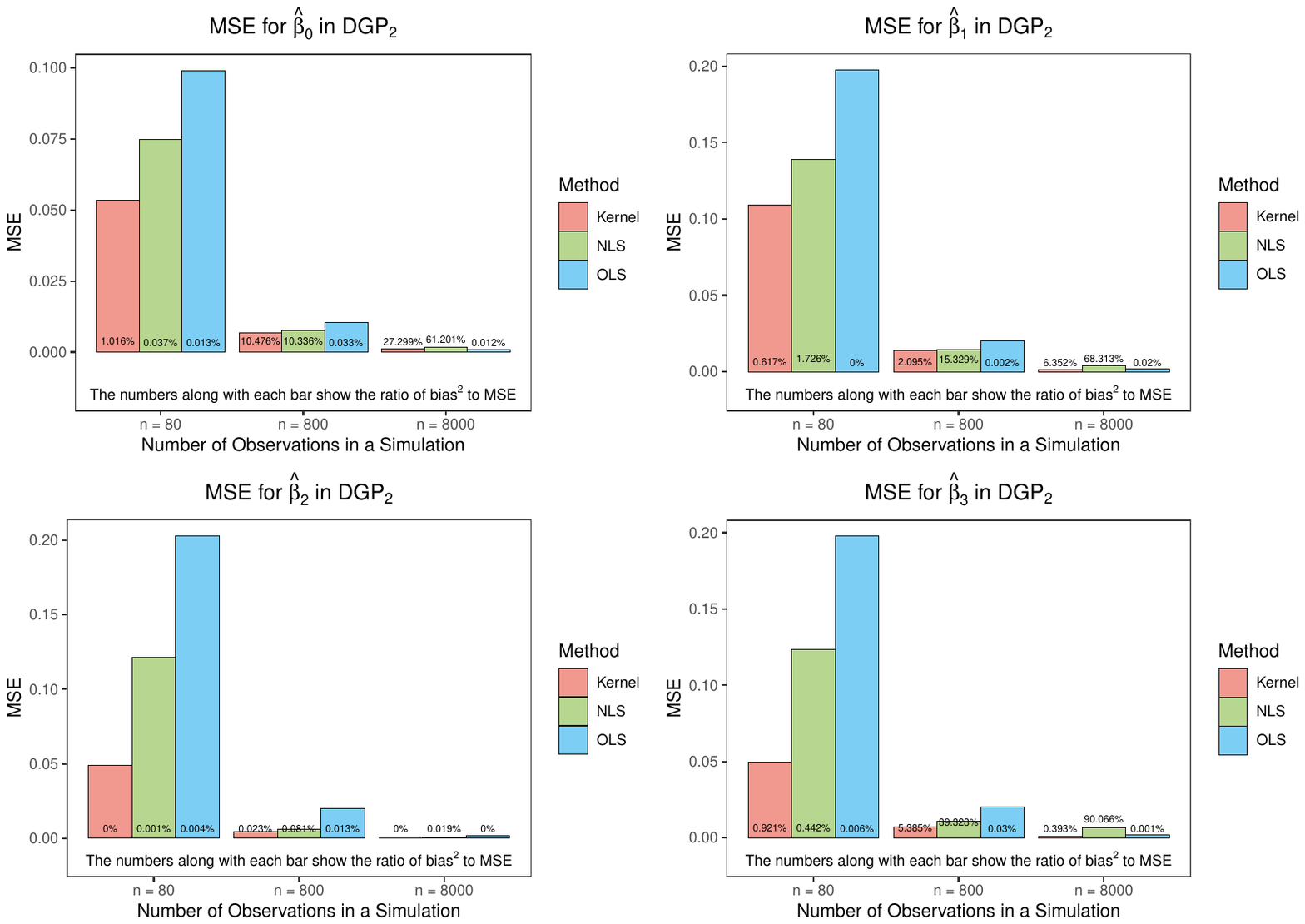}
	\label{fig:mse2a}
\end{figure}

\begin{figure}[H]
	\centering
	\caption{MSEs under DGP 2: $\beta_4-\beta_7$}
	\includegraphics[width=0.87\textwidth,page=4] {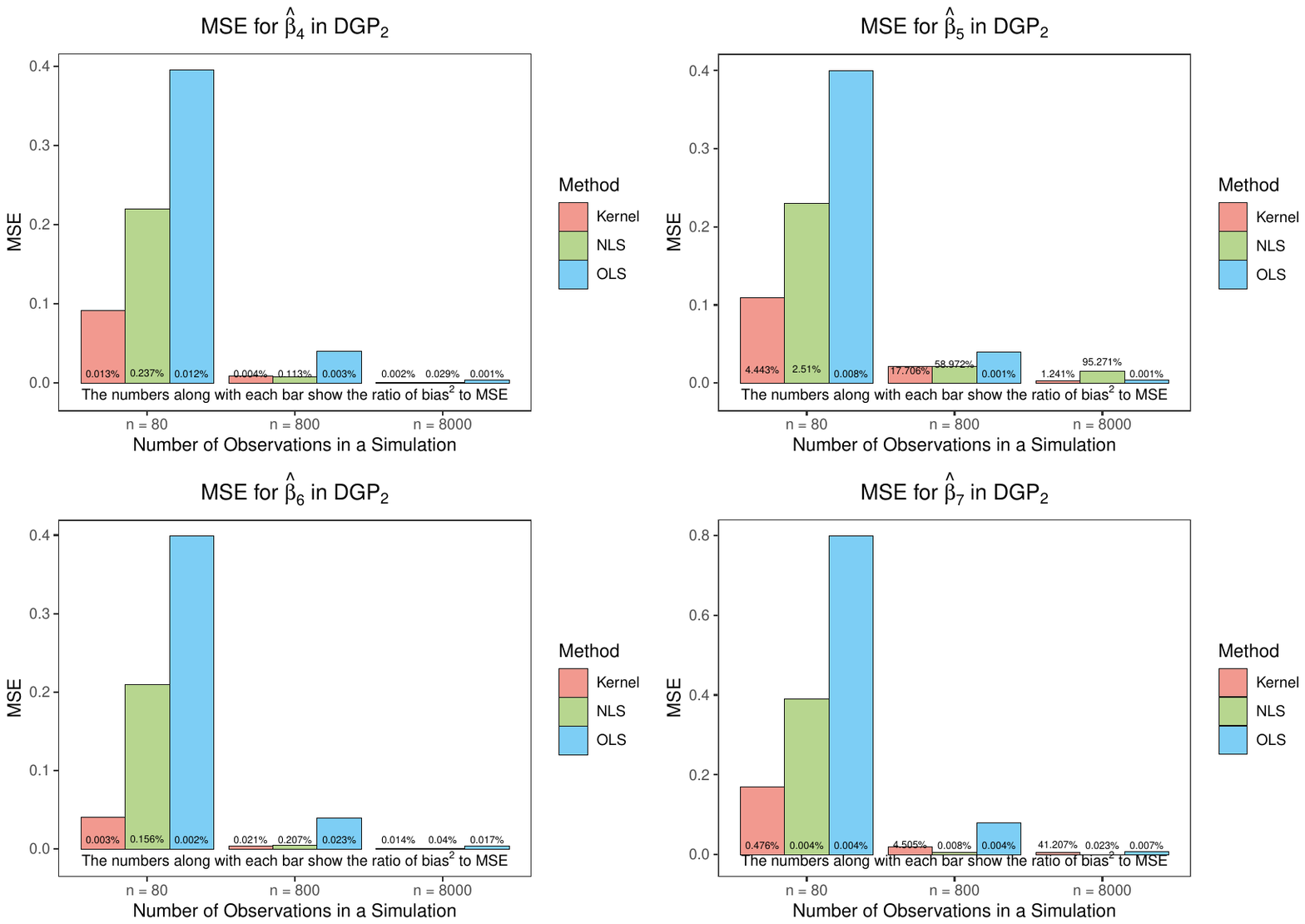}
	\label{fig:mse2b}
\end{figure}

\begin{figure}[H]
	\centering
	\caption{MSEs under DGP 3: $\beta_0-\beta_3$}
	\includegraphics[width=0.87\textwidth,page=5] {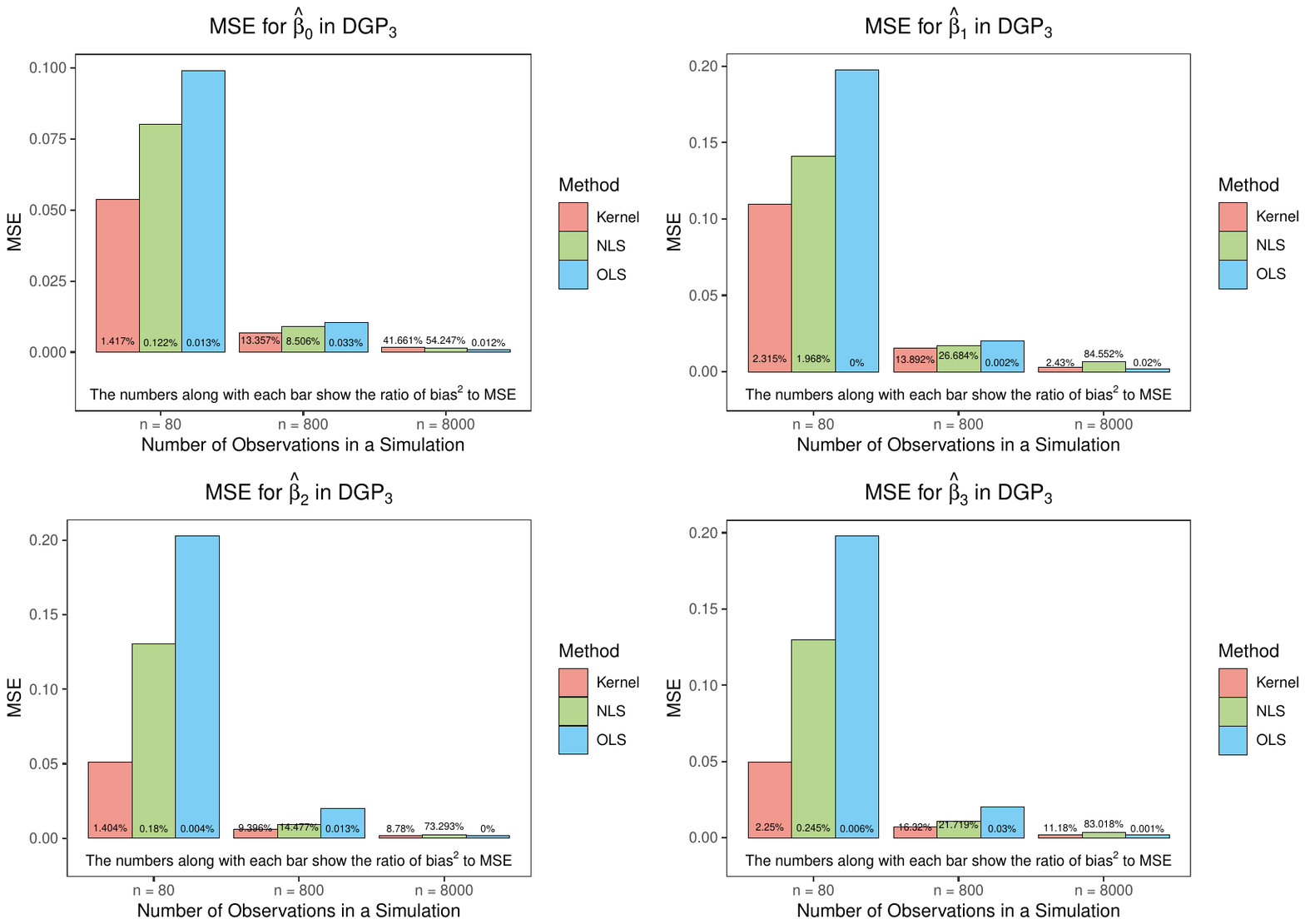}
	\label{fig:mse3a}
\end{figure}

\begin{figure}[H]
	\centering
	\caption{MSEs under DGP 3: $\beta_4-\beta_7$}
	\includegraphics[width=0.87\textwidth,page=6] {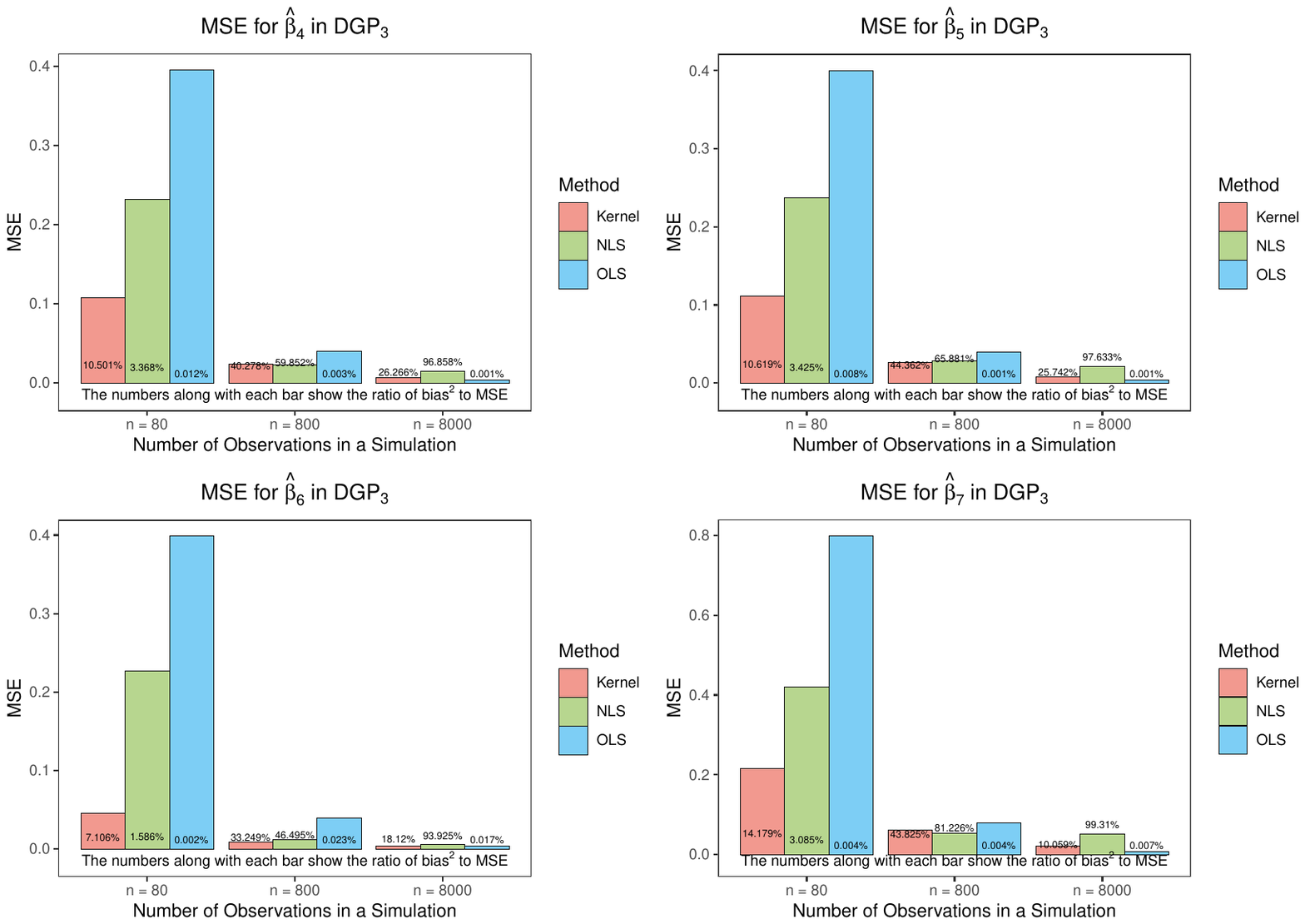}
	\label{fig:mse3b}
\end{figure}

\begin{figure}[H]
	\centering
	\caption{MSEs under DGP 4: $\beta_0-\beta_3$}
	\includegraphics[width=0.87\textwidth,page=7] {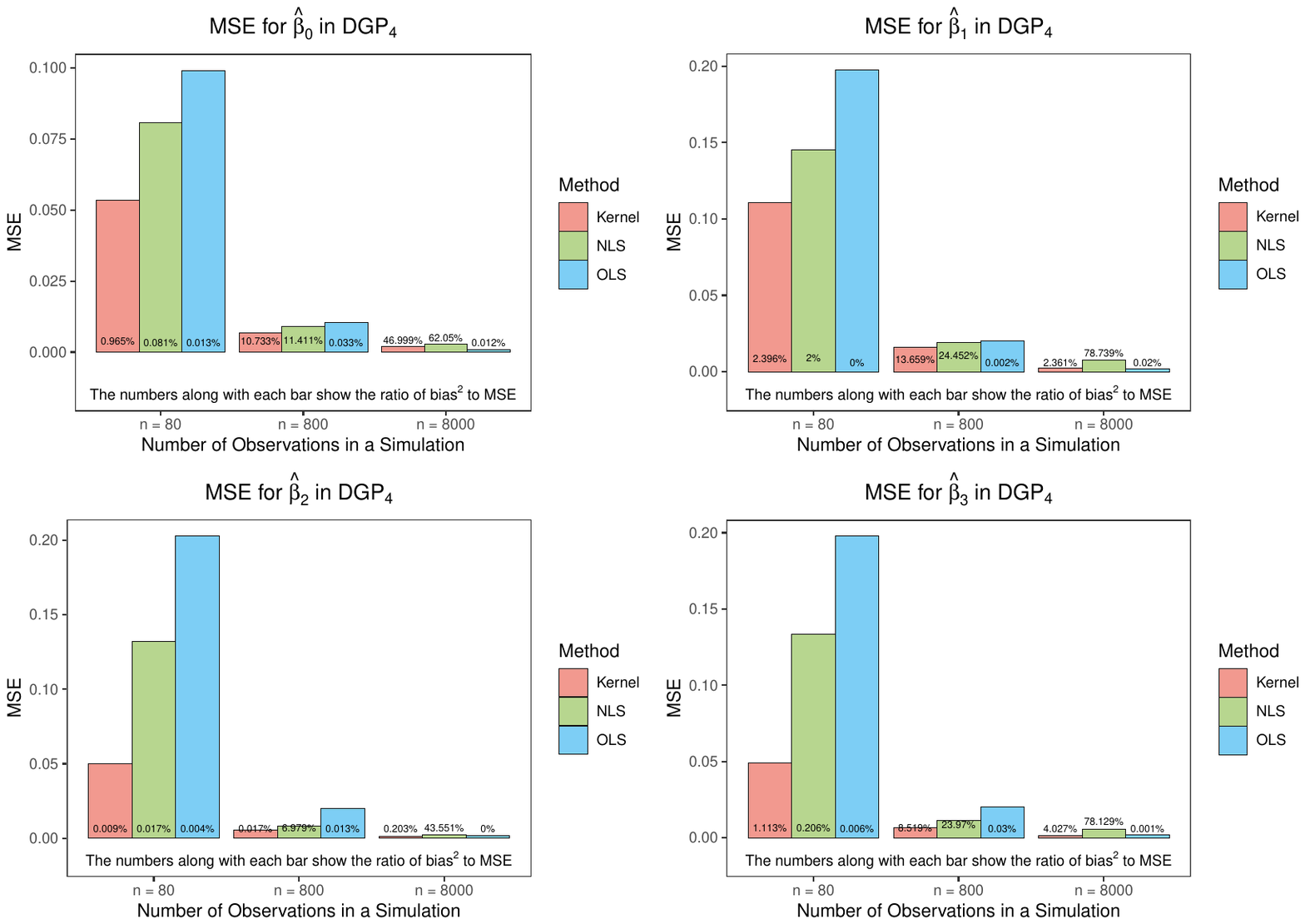}
	\label{fig:mse4a}
\end{figure}

\begin{figure}[H]
	\centering
	\caption{MSEs under DGP 4: $\beta_4-\beta_7$}
	\includegraphics[width=0.87\textwidth,page=8] {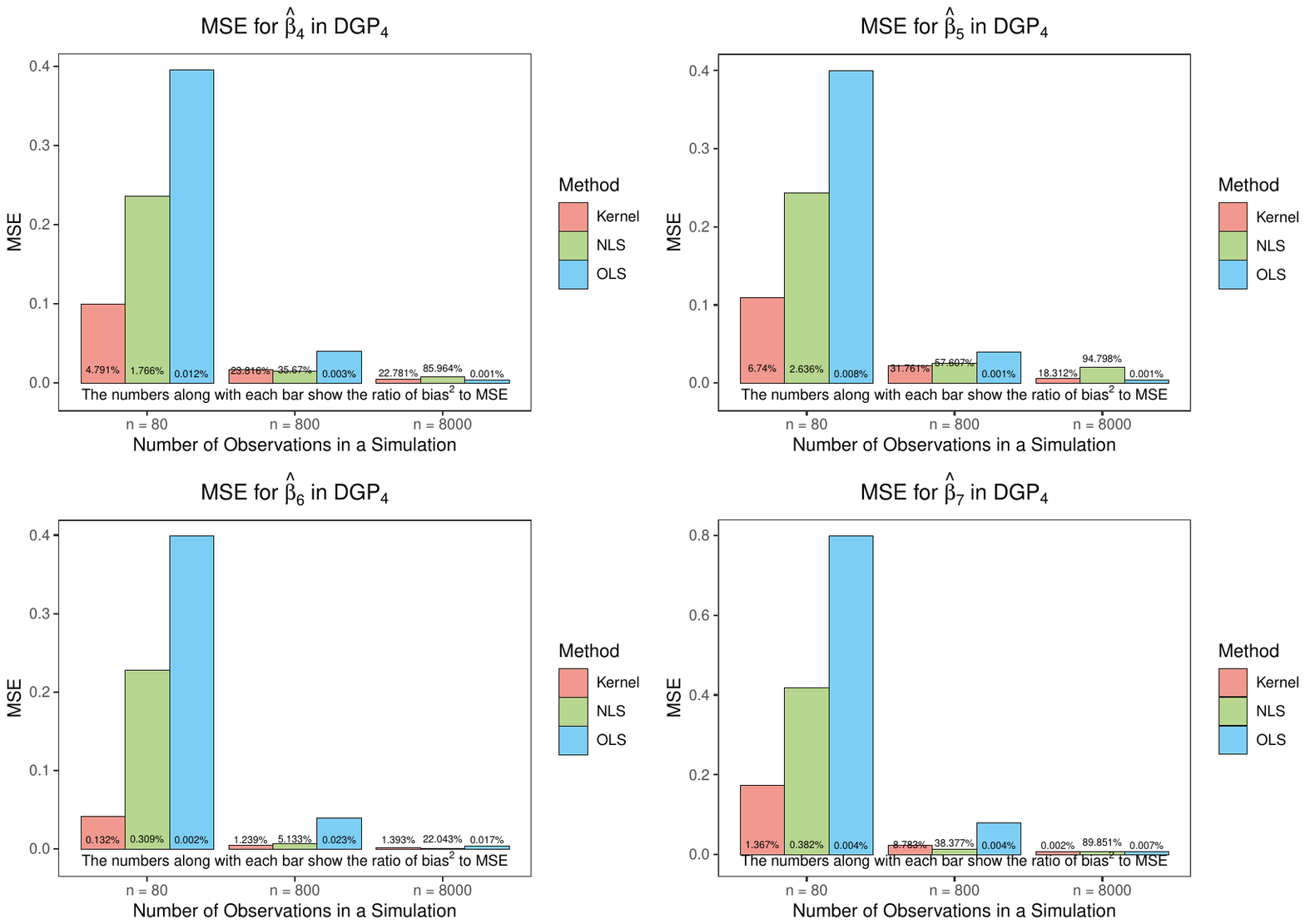}
	\label{fig:mse4b}
\end{figure}

\begin{figure}[H]
	\centering
	\caption{MSEs under DGP 5: $\beta_0-\beta_3$}
	\includegraphics[width=0.87\textwidth,page=9] {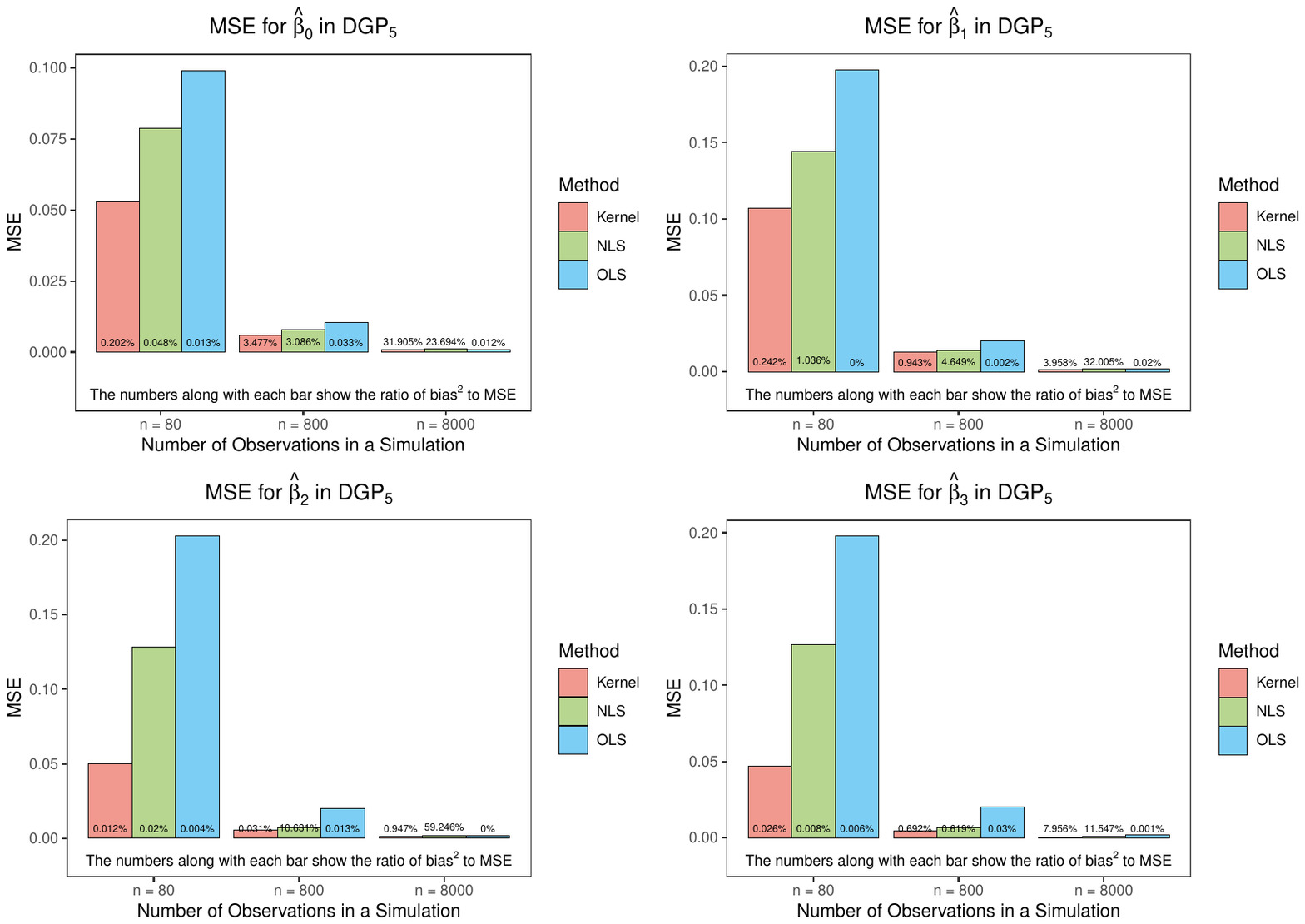}
	\label{fig:mse5a}
\end{figure}

\begin{figure}[H]
	\centering
	\caption{MSEs under DGP 5: $\beta_4-\beta_7$}
	\includegraphics[width=0.87\textwidth,page=10] {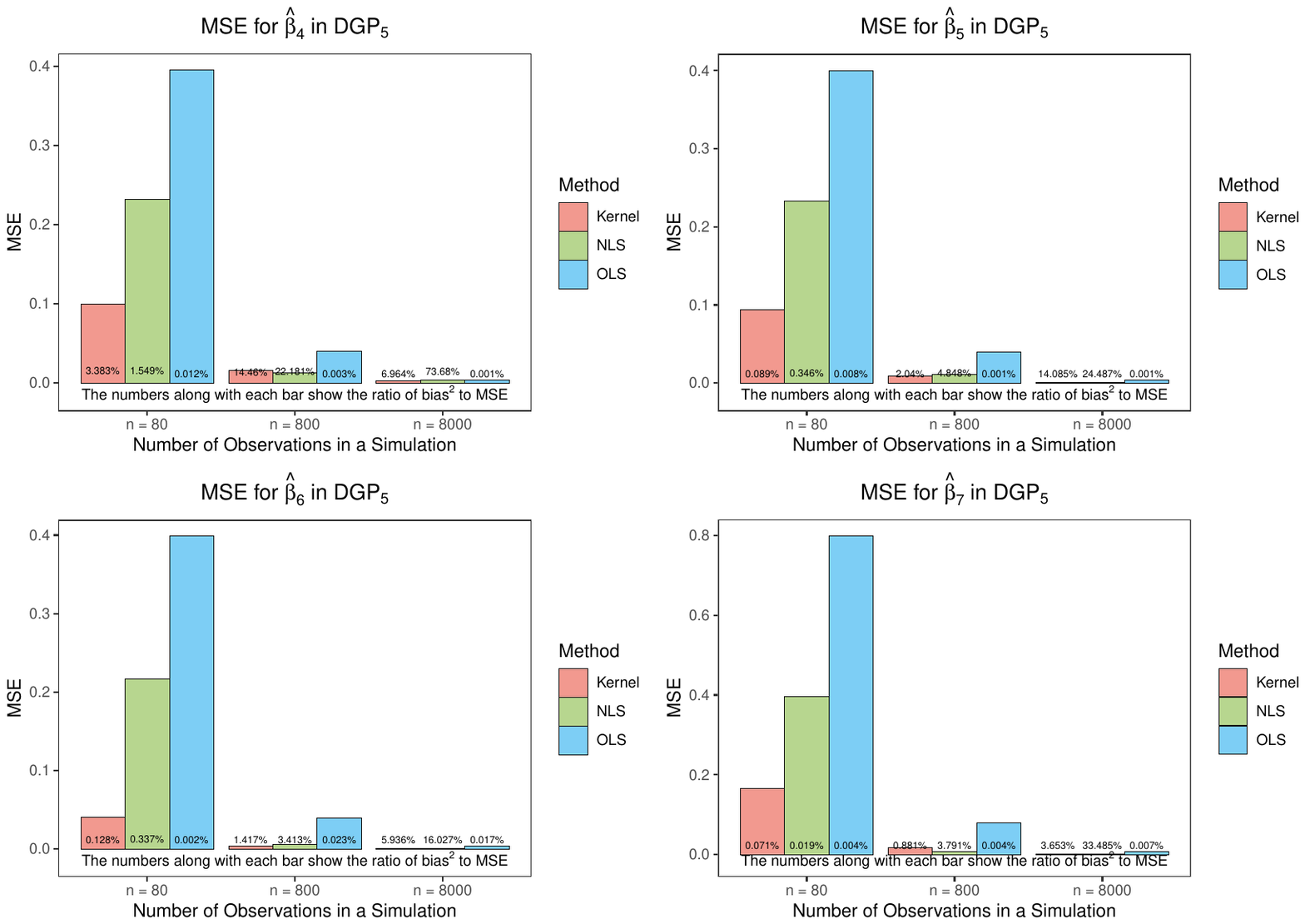}
	\label{fig:mse5b}
\end{figure}

\paragraph{Results} \quad We begin by comparing the MSE of the methods and the fraction of them that consist of bias squared. Figures \ref{fig:mse1a}-\ref{fig:mse5b} display these quantities for the all coefficients, DGPs, and sample sizes. Results from the kernel-based estimator are shown in red, from NLS in green, and from OLS in blue.

The results are as expected. In smaller samples, the kernel-based estimator performs better in minimizing MSE due to the regularization. However, OLS performs best in minimizing bias, as it yields the smallest fractions of the MSEs corresponding to bias squared. Expectedly, as the sample size increases the differences between the methods vanishes. 

It is also worth noting that often NLS outperforms OLS in minimizing MSE when the sample sizes are smaller and that it performs best amongst the three methods under DGP 1. This is unsurprising as NLS imposes restrictions between parameters that are true for this DGP. For other DGPs, however, not all such restrictions hold, and so as the sample size increases NLS's performance worsens relative to that of the other two methods.

\begin{figure}[H]
	\centering
	\caption{MAEs under DGP 1: $\beta_0-\beta_3$}
	\includegraphics[width=0.85\textwidth,page=1] {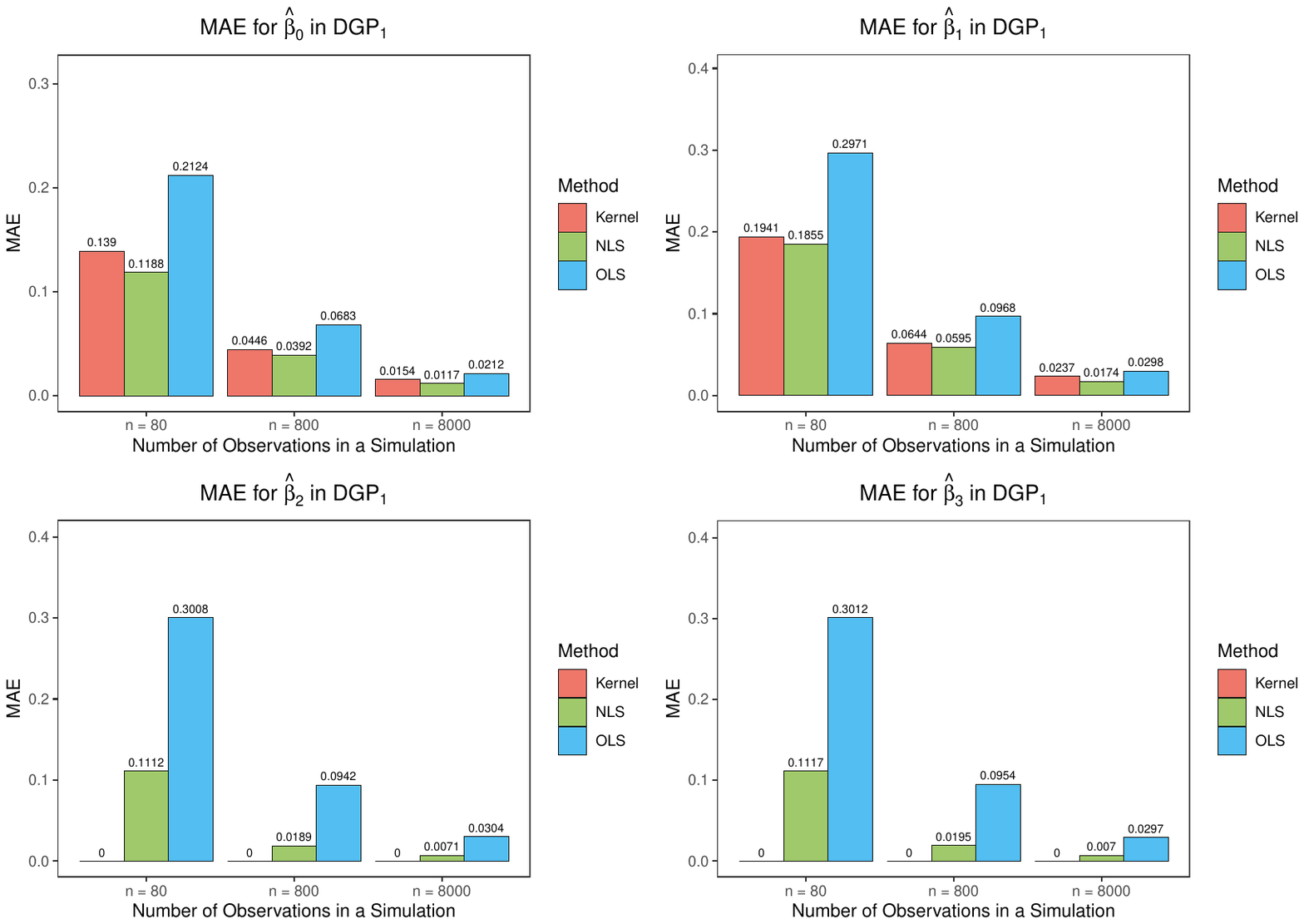}
	\label{fig:mae1a}
\end{figure}

\begin{figure}[H]
	\centering
	\caption{MAEs under DGP 1: $\beta_4-\beta_7$}
	\includegraphics[width=0.85\textwidth,page=2] {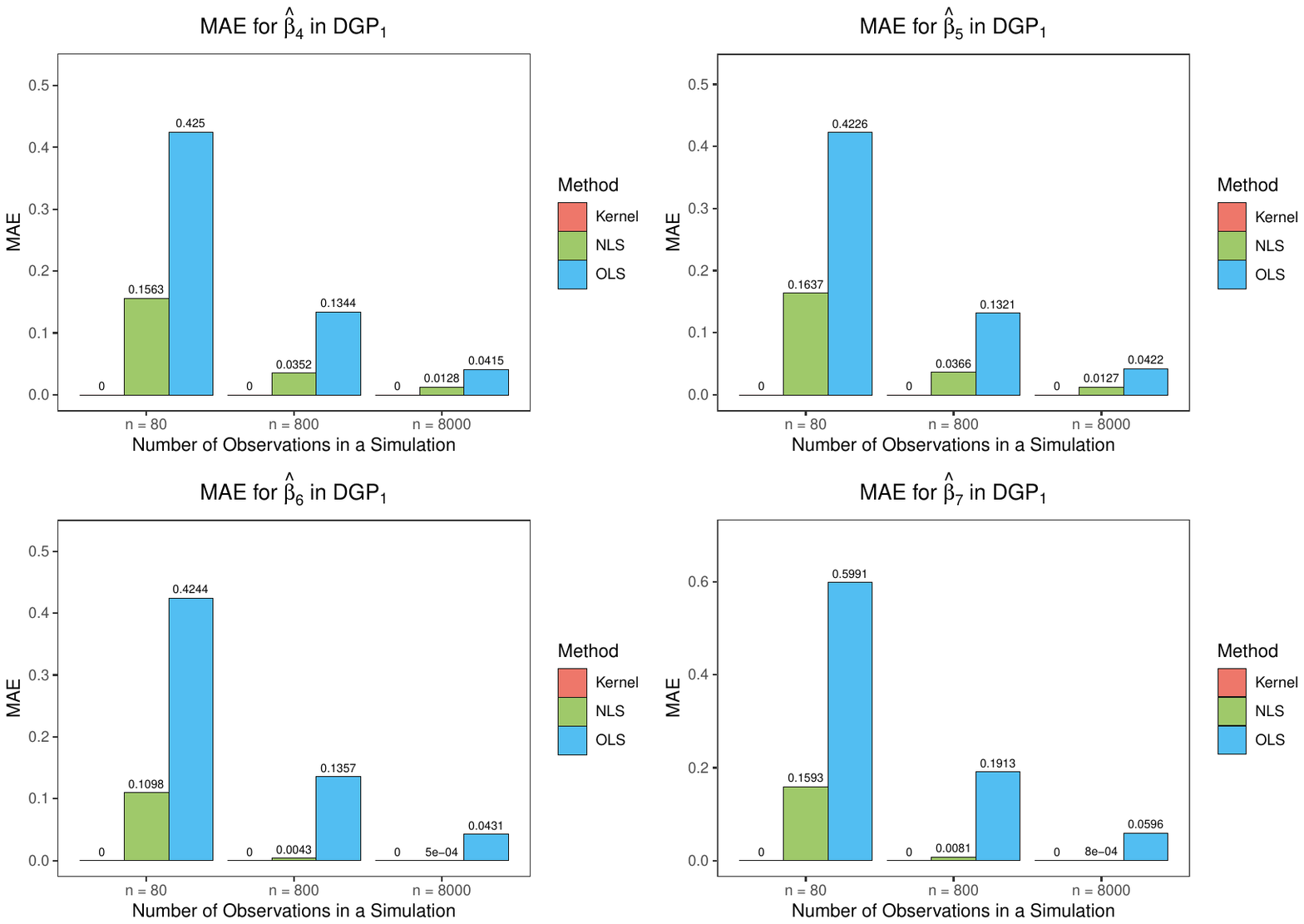}
	\label{fig:mae1b}
\end{figure}

\begin{figure}[H]
	\centering
	\caption{MAEs under DGP 2: $\beta_0-\beta_3$}
	\includegraphics[width=0.85\textwidth,page=3] {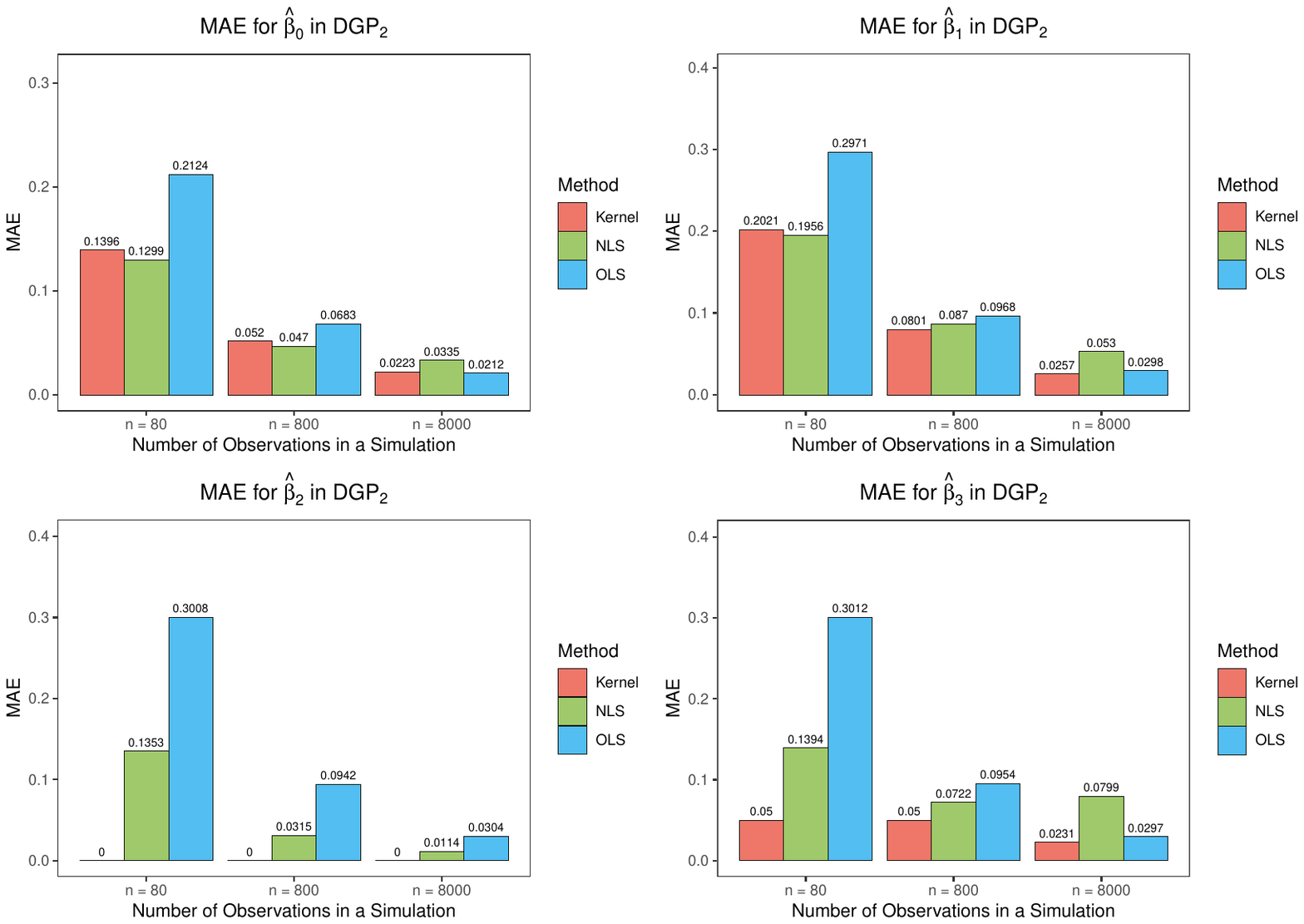}
	\label{fig:mae2a}
\end{figure}

\begin{figure}[H]
	\centering
	\caption{MAEs under DGP 2: $\beta_4-\beta_7$}
	\includegraphics[width=0.85\textwidth,page=4] {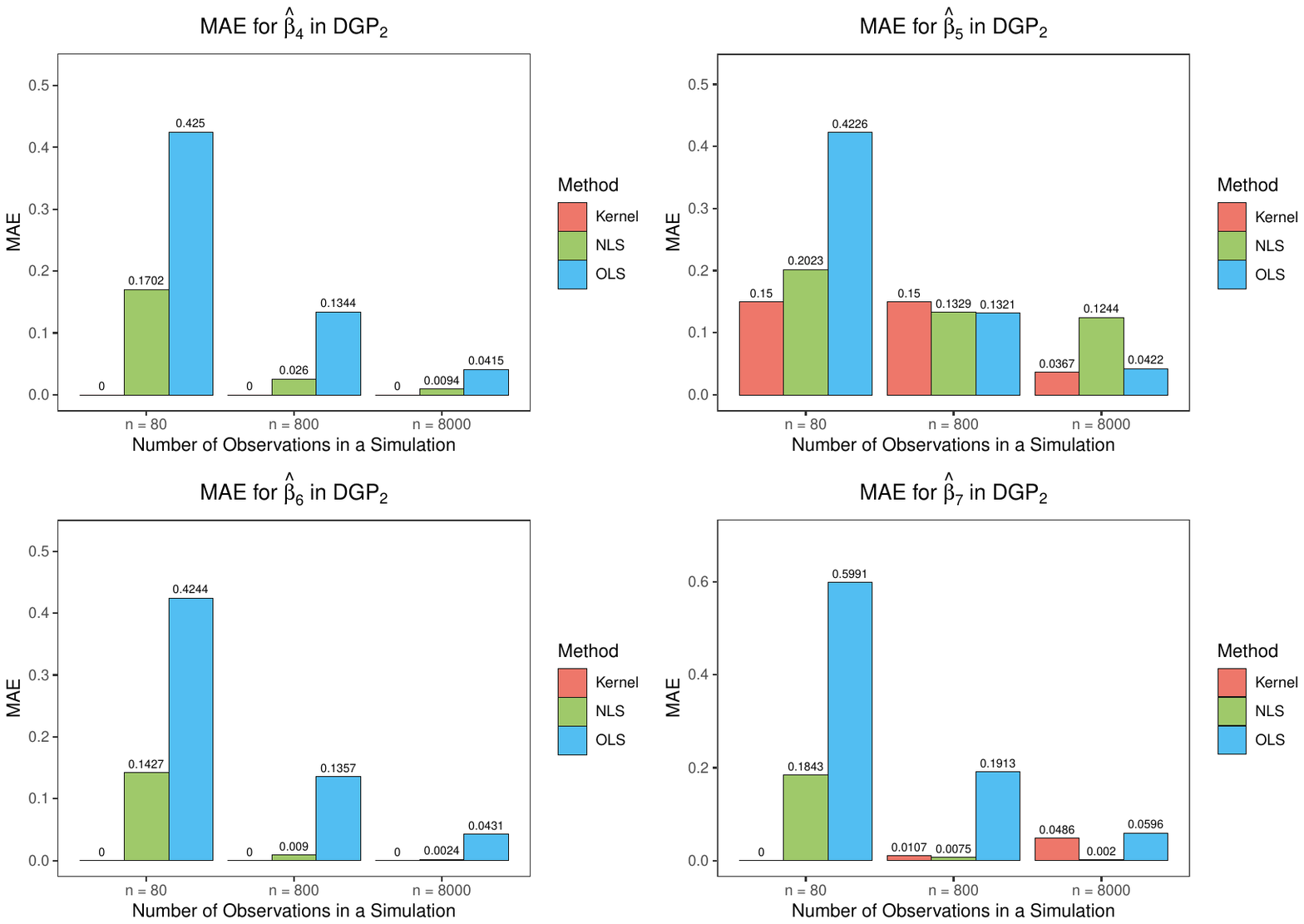}
	\label{fig:mae2b}
\end{figure}

\begin{figure}[H]
	\centering
	\caption{MAEs under DGP 3: $\beta_0-\beta_3$}
	\includegraphics[width=0.85\textwidth,page=5] {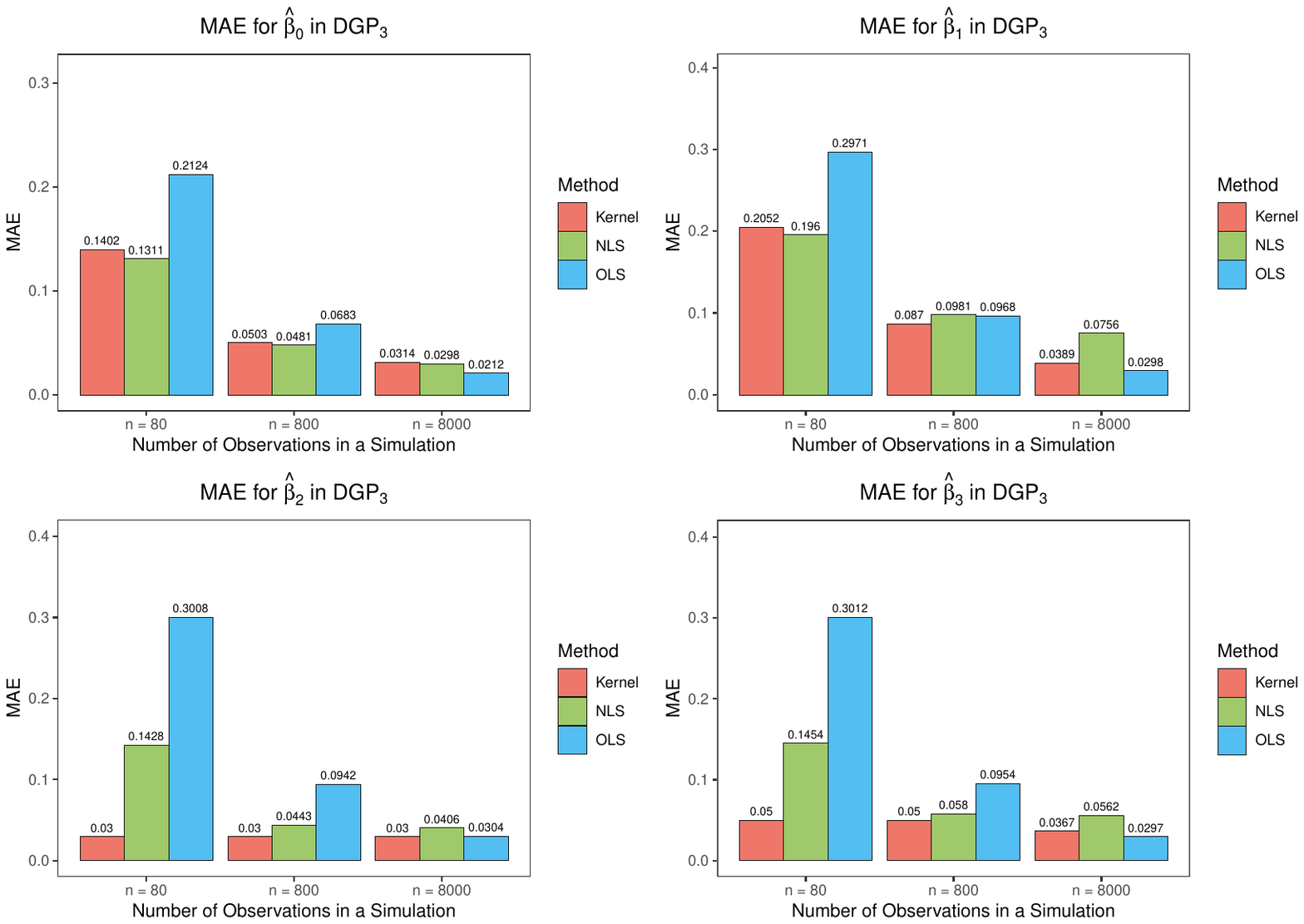}
	\label{fig:mae3a}
\end{figure}

\begin{figure}[H]
	\centering
	\caption{MAEs under DGP 3: $\beta_4-\beta_7$}
	\includegraphics[width=0.85\textwidth,page=6] {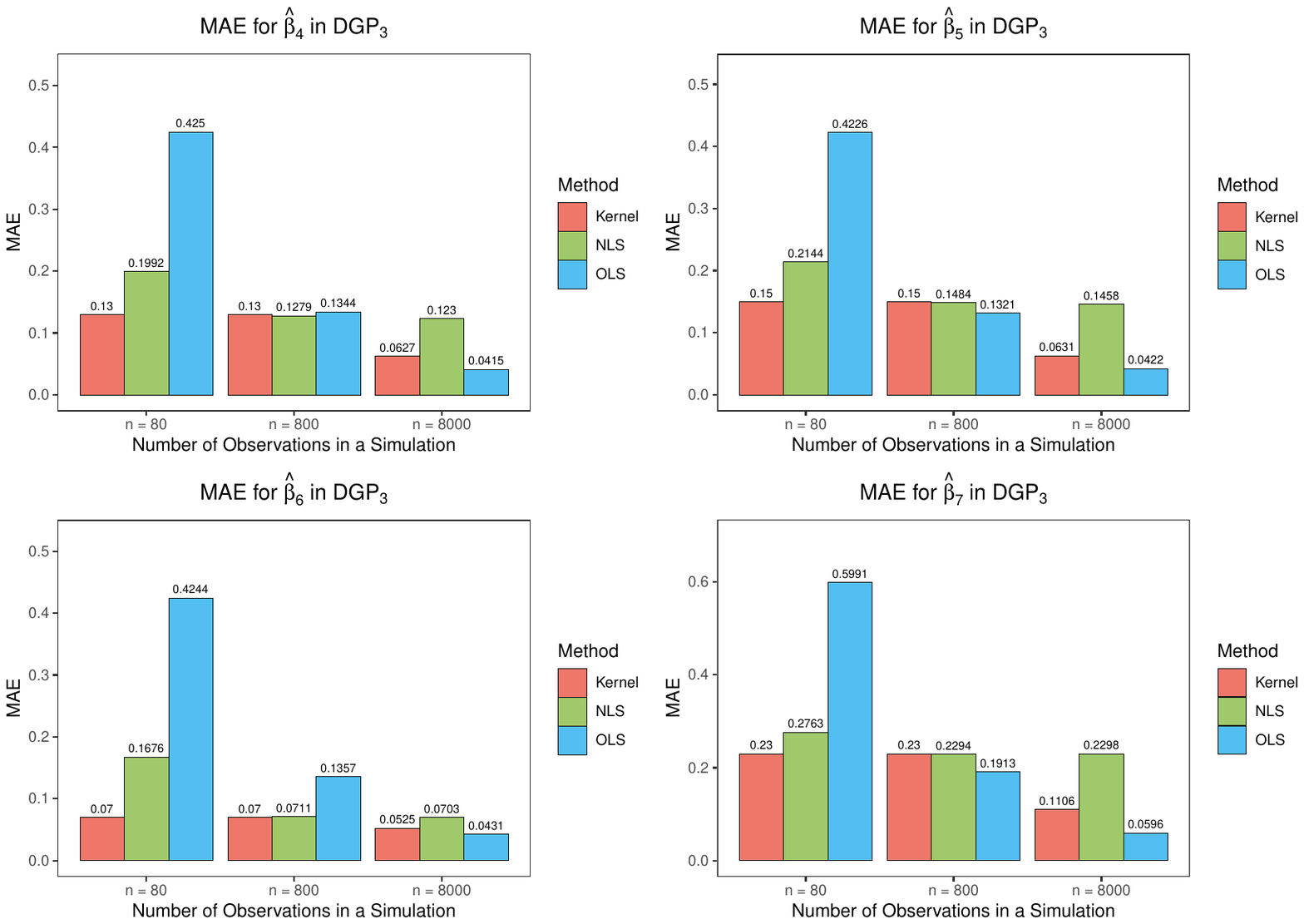}
	\label{fig:mae3b}
\end{figure}

\begin{figure}[H]
	\centering
	\caption{MAEs under DGP 4: $\beta_0-\beta_3$}
	\includegraphics[width=0.85\textwidth,page=7] {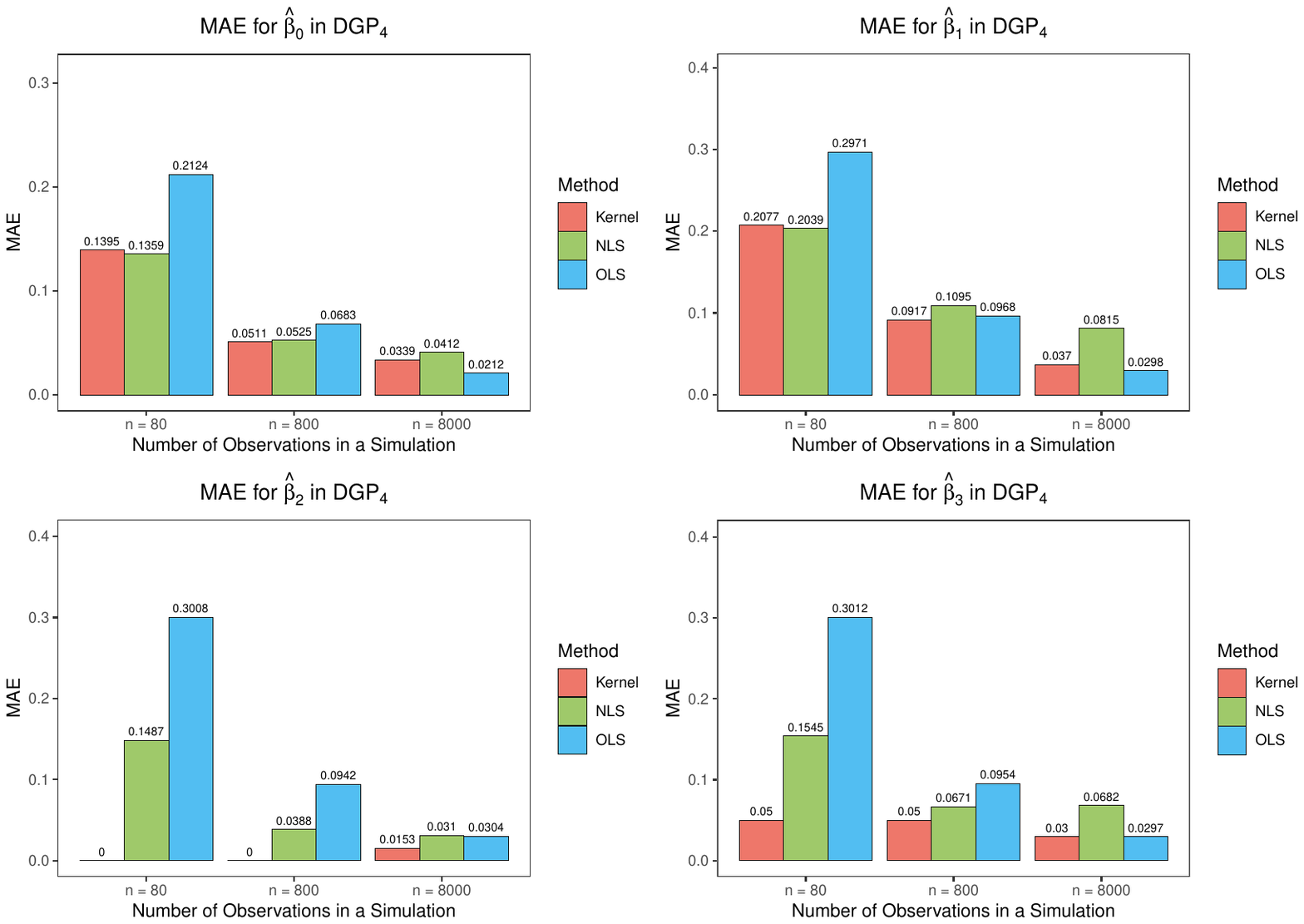}
	\label{fig:mae4a}
\end{figure}

\begin{figure}[H]
	\centering
	\caption{MAEs under DGP 4: $\beta_4-\beta_7$}
	\includegraphics[width=0.85\textwidth,page=8] {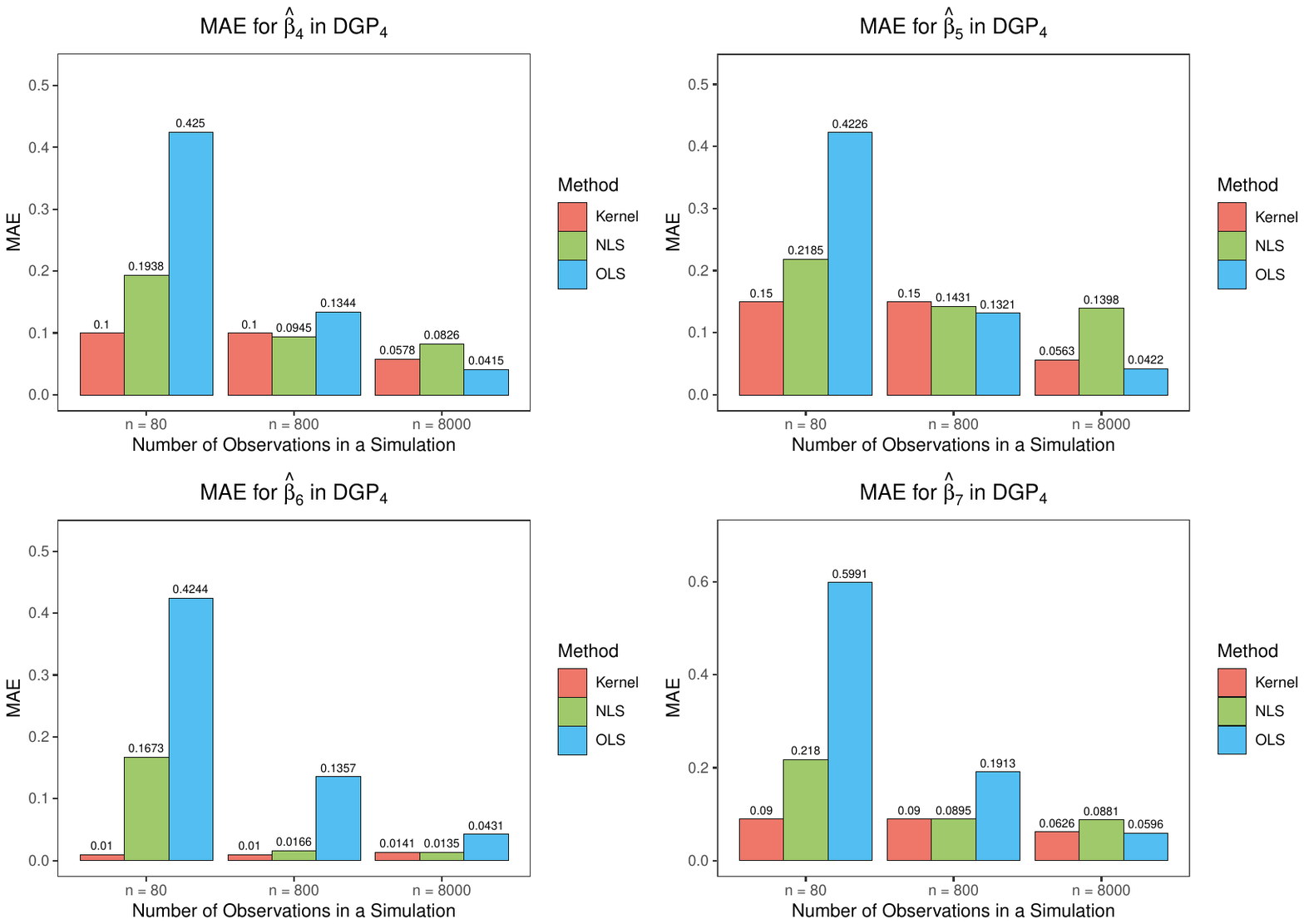}
	\label{fig:mae4b}
\end{figure}

\begin{figure}[H]
	\centering
	\caption{MAEs under DGP 5: $\beta_0-\beta_3$}
	\includegraphics[width=0.85\textwidth,page=9] {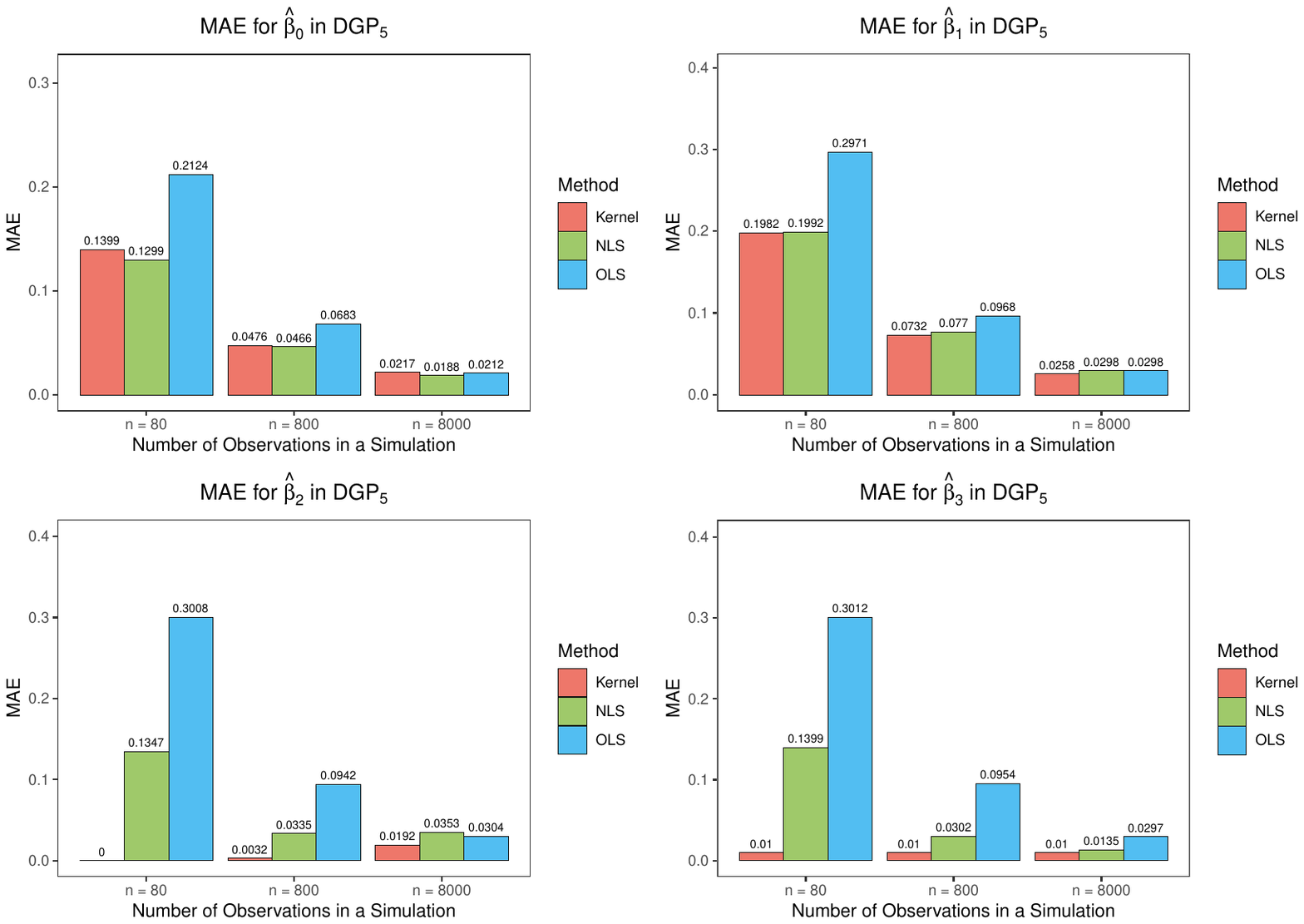}
	\label{fig:mae5a}
\end{figure}

\begin{figure}[H]
	\centering
	\caption{MAEs under DGP 5: $\beta_4-\beta_7$}
	\includegraphics[width=0.85\textwidth,page=10] {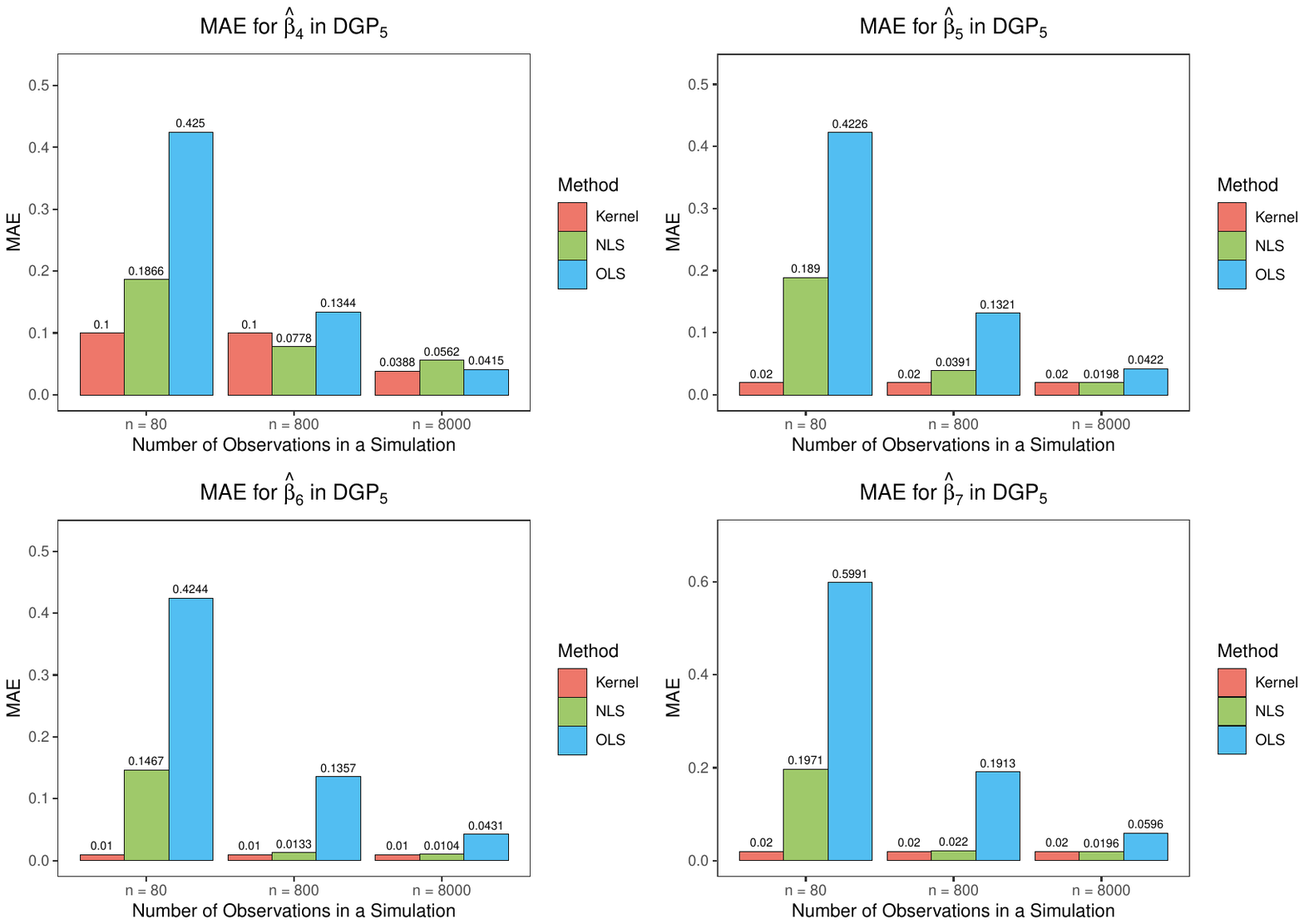}
	\label{fig:mae5b}
\end{figure}

For completeness, Figures \ref{fig:mae1a}-\ref{fig:mae5b} display the MAEs for the all coefficients, DGPs, and sample sizes. Once again, results from the kernel-based estimator are shown in red, from NLS in green, and from OLS in blue.

As expected, results improve as the sample size increases. We can see that the kernel-based estimator performs very well as it often obtains an MAE of zero. This is not entirely surprising given the results shown in Figures \ref{fig:bws1}-\ref{fig:bws5}. This estimator has a propensity to set the bandwidths to one, implying that it does not detect interference between advertisers. When if fact there is no such interference, the coefficients associated with the indicators for the ``non-intefering'' advertisers are zero, which the kernel-based estimator will impose, finally yielding no error at all. When the coefficients are non-zero, there is no chance any estimator can often obtain no error. In such cases, the performance of the different methods varies.

\newpage

{\footnotesize\bibliography{par_exp_ref}}

\end{document}